\documentclass[twoside,twocolumn]{aa}
\usepackage{times}
\usepackage{graphicx}
\usepackage{amsmath}
\usepackage{longtable}
\makeatletter

\fontfamily{ptmcm}\selectfont

\newcommand{\noun}[1]{\textsc{#1}}

\def\gradip{\hbox{\rlap{\hbox{.}}\raise 5.truept \hbox{{\small $\circ$}}}}

\catcode`\@=11
\def\gsim{\ifmmode{\mathrel{\mathpalette\@versim>}}
    \else{$\mathrel{\mathpalette\@versim>}$}\fi}
\def\lsim{\ifmmode{\mathrel{\mathpalette\@versim<}}
    \else{$\mathrel{\mathpalette\@versim<}$}\fi}
\def\@versim#1#2{\lower 2.9truept \vbox{\baselineskip 0pt \lineskip
    0.5truept \ialign{$\m@th#1\hfil##\hfil$\crcr#2\crcr\sim\crcr}}}
\catcode`\@=12
\makeatother

\begin{document}

\title{HST color-magnitude diagrams of 74 galactic globular \\
clusters in the HST F439W and F555W bands 
\thanks{Based on observations with the NASA/ESA \textit{Hubble 
Space Telescope}, obtained at the Space Telescope Science Institute, which 
is operated by AURA, Inc., under NASA contract NAS5-26555, and on observations
 retrieved from the ESO ST-ECF Archive.}}

\author{G. Piotto \inst{1}, I.R. King \inst{2}, S.G. Djorgovski \inst{3},\\ 
C. Sosin \inst{2}, M. Zoccali \inst{4}, I. Saviane \inst{5}, F. De Angeli \inst{1,2},
M. Riello \inst{1,6}, A. Recio-Blanco \inst{1},\\ R. M. Rich \inst{7}, 
G. Meylan \inst{8}, A. Renzini \inst{4}}

\offprints{G. Piotto \email{piotto@pd.astro.it}}

\institute{Dipartimento di Astronomia -- Universit\`a di Padova, 
	I-35122 Padova, Italy; piotto@pd.astro.it, deangeli@pd.astro.it, 
	recio@pd.astro.it
	\and Astronomy Department, University of California,
	Berkeley, CA 94720, USA; king@glob.berkeley.edu, craig@sosin.org
	\and California Institute of Technology, MS 105-24,
	Pasadena, CA 91125, USA; george@deimos.caltech
	\and European Southern Observatory, D-85748 Garching bei Munich, 
	Germany; mzoccali@eso.org, gmeylan@eso.org, arenzini@eso.org
	\and European Southern Observatory, Casilla 19001, Santiago, Chile
	\and Osservatorio Astronomico, I-35122 Padova, Italy, riello@pd.astro.it
	\and Physics and Astronomy Department, University of California,
	Los Angeles, CA 90024, USA; rmr@astro.ucla.edu
        \and Space Telescope Science Institute, Baltimore, MD 21218, USA  }

\date{Received xxx/ accepted xxx}

\titlerunning{Globular cluster HST color-magnitude diagrams}
\authorrunning{Piotto et al.}
\markboth{Piotto et al.}{GGC CMDs}

\abstract{We present the complete photometric database and the color-magnitude
diagrams for 74 Galactic globular clusters observed with the HST/WFPC2
camera in the F439W and F555W bands. A detailed discussion of the
various reduction steps is also presented, and of the procedures to transform
instrumental magnitudes into both the HST F439W and F555W flight
system and the standard Johnson \( B \) and \( V \) systems. We
also describe the artificial star experiments which have been
performed to derive the star count completeness in all the relevant
branches of the color magnitude diagram. The entire photometric database
and the completeness function will be made available on the Web immediately
after the publication of the present paper.
}

\maketitle

\section{Introduction}

\label{intro}

Because of its excellent resolving power, the \textit{Hubble Space
Telescope} offers an exceptional opportunity to study the crowded
centers of Galactic globular clusters (GGC). We therefore began
several years ago a study of color--magnitude diagrams (CMD) of
clusters, using HST's WFPC2 camera.

Our first program, GO-6095, examined 10 clusters in the HST \( B \)
(F439W) and \( V \) (F555W) bands (Sosin et al.\ 1997a, Piotto et al.\
1997), with accompanying ultraviolet images for better study of the
extended blue tails that occur on the horizontal branches (HB) of a
number of clusters. That program gave some quite interesting results,
like the first discovery of extended horizontal branches (EHB) in the
metal-rich clusters NGC~6388 and NGC~6441 (Rich et al.\ 1997), and the
discovery in NGC~2808 of an EHB extending down the helium-burning
main sequence, with a multimodal distribution of stars along it (Sosin
et al.\ 1997b). GO-6095 also persuaded us that it would be more
profitable to explore a larger number of clusters more rapidly in \( B
\) and \( V
\) alone, in order to delineate the upper parts of their CMDs, and
especially their HBs, with the intention of returning to the more
interesting clusters for more intensive follow-up studies.  We
therefore changed our program to a snapshot study (GO-7470) aimed at
producing CMDs down to a little below the main-sequence (MS) turnoff
for all Galactic globular clusters with apparent \( B \) distance
modulus \( \leq18 .0 \) and whose centers had not yet been observed by
HST in a comparable way---53 clusters in all. Since snapshot programs
have no guarantee of being completed in a given year, we have
resubmitted each year the list of clusters not yet observed (GO-8118,
GO-8723). By the time we are writing this paper, only one of the GGCs
in our original list remains to be observed: NGC~6779. There were also
in the HST archive images of a number of clusters that could be
treated similarly. We have measured these in the same way as our own
images, and present here CMDs of a total of 74 GGCs, measured and
reduced in a uniform way, all observed with WFPC2 with the same filter
set, and with the PC centered on the cluster center. For all of the 74
GGCs, we ran artificial star experiments to measure the completeness
of the star counts in all the relevant branches of the CMD. In this
paper we describe the observations, the reduction procedures, the
artificial star tests, and the steps followed to transform the
instrumental magnitudes into magnitudes in both the HST flight and
standard Johnson photometric systems.

The data set that we present has already been shown to be
extremely valuable in attacking a number of still-open topics on evolved
stars in GGCs.  In particular, these data have been used by Piotto et
al.\ (1999a) to investigate the problem of the EHBs and by Raimondo et
al.\ 2002) to study the properties of the red HB in metal-rich clusters;
in Zoccali et al.\ (1999) and Bono et al.\ (2001) we have throughly
discussed the red giant branch (RGB) bump, and compared the predicted
position and dimension of this feature with the observed ones; in
Zoccali et al.\ (2000) we have used the star counts on the HB and on the
RGB to gather information on the helium content and on the dependence of
the helium content on the cluster metallicity; in Zoccali \& Piotto
(2000) we have made the most extensive comparison so far available
between the model evolutionary times away from the main sequence and the
actual star counts on the subgiant branch (SGB) and RGB; in Cassisi et
al.\ (2001) we have compared the observed and theoretical properties of
the asymptotic giant branches; in Piotto et al.\ (2000) we have used the
CMDs in our database for the study of the GGC relative ages, and,
finally, in Piotto et al.\ (1999b), we have started to investigate the
GGC blue straggler (BS) population, and have shown how the BS CMD and
luminosity function can differ in clusters with rather different
morphologies.  Future papers will carry out other detailed studies based
on the same data.  Among these, we are presently working on the
derivation of the relative ages, following the strategy already
delineated in Rosenberg et al.\ (1999) on a similarly photometrically
homogeneous set of CMDs, but from ground-based data, and by Piotto et
al.\ (2000) on a small subsample of the present HST database. We are
also working on the large BS database that came from the CMDs
presented in the following Sections (Piotto et al.\ 2002, in
preparation).

As better described in Section 3, all the CMDs, the star positions, the
magnitudes in both the F439W and F555W flight system, and the \( B \) and $V$
standard Johnson system will be made available to the astronomical community
immediately after the publication of the present paper.

\section{Observations, reductions and calibrations}

\subsection {Pre-reduction of the images}

\begin{figure}
\resizebox{1.6\columnwidth}{!}{\includegraphics{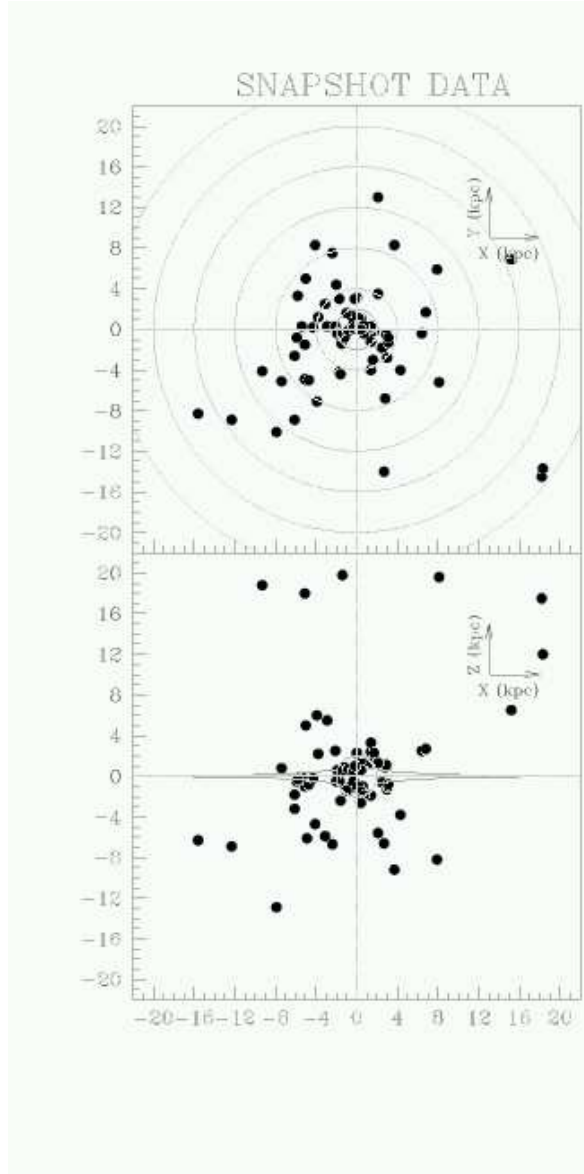}}
\caption{Spatial distribution of the target clusters within the Galaxy.}
\label{distr}
\end{figure}

All the photometric data for the 74 GGCs presented in this paper come
from HST/WFPC2 observations in the F439W and F555W bands; in all
%GPAdded the next sentence and a new Table.
cases, the PC camera was centered on the cluster center. Tables \ref{obs1} and \ref{obs2} list
the observed clusters (Col.~1), the origin of the observations
(Col.~2), and the exposure times in F555W (Col.~3) and F439W (Col.~4)
bands. Tables \ref{par1} and \ref{par2}  give a few relevant parameters (from the Harris
\cite{harris-catalog} compilation) 
%to help the reader to compare the CMDs of clusters with different properties. 
Fig.~\ref{distr} shows
the spatial distribution of the target clusters within the Galaxy.

\begin{table}
\begin{tabular}{llcc}
\hline  
%\multicolumn{1}{l}{ID} & Program ID & \multicolumn{2}{c}{Exposure Times (F555W, F439W)} \\ 
Cluster & Program ID & Exp. Time & Exp. Time \\
        &            &   F555W  [s]     &  F439W [s]       \\             
\hline
 NGC 104 & GO6095  & 1, 7        & 7, 2x(50)     \\ 
 NGC 362 & GO6095  & 3, 18       & 18, 2x(160)   \\
 NGC 1261$^*$& GO7470  & 10, 40      & 40, 2x(160)   \\
 NGC 1851& GO6095  & 6, 40       & 40, 2x(160)   \\
 NGC 1904& GO6095  & 7, 40       & 40, 2x(160)   \\
 NGC 2419$^*$& archive & 10, 2x(100) & 2x(260)        \\
 NGC 2808& GO6095  & 7, 50       & 50, 2x(230)   \\
 NGC 3201$^*$& GO8118  & 3, 2x(30)   & 40, 2x(100)   \\
 NGC 4147$^*$& GO7470  & 10, 40      & 40, 2x(160)   \\
 NGC 4372& GO8118  & 3, 2x(30)   & 40, 2x(100)   \\
 NGC 4590$^*$& GO7470  & 5, 40       & 40, 2x(100)   \\
 NGC 4833& GO8118  & 3, 2x(30)   & 40, 2x(100)   \\
 NGC 5024& GO8118  & 5, 40       & 40, 2x(160)   \\
 NGC 5634& GO7470  & 10, 40      & 40, 2x(160)   \\
 NGC 5694& archive & 10, 4x(60)  & 120, 3x(500)  \\
 IC 4499$^*$& GO8723  & 10, 40      & 40, 2x(160)   \\
 NGC 5824& archive & 2x(10), 5x(60) & 120, 3x(500) \\
 NGC 5904& GO8118  & 3, 2x(30)   & 40, 2x(100)   \\
 NGC 5927& GO6095  & 10, 50      & 50, 2x(160)   \\
 NGC 5946$^*$& GO7470  & 10, 40      & 40, 2x(160)   \\
 NGC 5986$^*$& GO7470  & 10, 40 ec   & 40, 2x(160)   \\
 NGC 6093$^*$& archive & 2x(2), 4x(23) & 2x(30)       \\
 NGC 6139& GO7470  & 10, 40      & 40, 2x(160)   \\
 NGC 6171$^*$& GO7470  & 5, 40       & 40, 2x(100)   \\
 NGC 6205$^*$& archive & 4x(1), 4x(8) & 2x(14)        \\
 NGC 6229& GO8118  & 10, 40      & 40, 2x(160)   \\
 NGC 6218& GO8118  & 3, 2x(30)   & 40, 2x(100)   \\
 NGC 6235& GO7470  & 10, 40      & 40, 2x(160)   \\
 NGC 6256$^*$& GO7470  & 10, 40      & 40, 2x(160)   \\
 NGC 6266& GO8118  & 3, 2x(30)   & 40, 2x(100)   \\
 NGC 6273& GO7470  & 10, 40      & 40, 2x(160)   \\
 NGC 6284$^*$& archive & 4, 4x(40)   & 30, 6x(160)   \\
 NGC 6287$^*$& archive & 2x(5), 3x(50) & 60, 230 \\
&& 3x(1000) & \\
 NGC 6293& archive & 2, 5x(40)   & 20, 6x(160)   \\
 NGC 6304$^*$& GO7470  & 10, 40      & 40, 2x(160)   \\
 NGC 6316$^*$& GO7470  & 10, 40      & 40, 2x(160)   \\
 NGC 6325$^*$& GO7470  & 10, 40      & 40, 2x(160)   \\
 NGC 6342& GO7470  & 10, 40      & 40, 2x(160)   \\
 NGC 6356& GO7470  & 10, 40      & 40, 2x(160)   \\
 NGC 6355$^*$& GO7470  & 10, 40      & 40, 2x(160)   \\
 IC 1257$^*$ & GO8723  & 30, 100     & 160, 2x(300)  \\
 NGC 6362$^*$& GO7470  & 5, 40       & 40, 2x(100)   \\
 NGC 6380& GO7470  & 10, 40      & 40, 2x(160)   \\
 NGC 6388& GO6095  & 12, 50      & 50, 2x(160)   \\
 NGC 6402& GO8118  & 7, 40       & 40, 2x(160)   \\
 NGC 6401& GO7470  & 10, 40      & 40, 2x(160)   \\
 NGC 6397$^*$& archive & 1, 8, 6x(40) & 2x(10), 2x(80) \\
&&& 16x(400), 4x(500) \\
 NGC 6440$^*$& GO8723  & 10, 40      & 40, 2x(160)   \\
 NGC 6441& GO6095  & 14, 50      & 50, 2x(160)   \\
 NGC 6453& GO8118  & 10, 40      & 40, 2x(160)   \\
 NGC 6517$^*$& GO8723  & 10, 40      & 40, 2x(160)   \\
 NGC 6522& GO6095  & 10, 50      & 50, 2x(160)   \\
 NGC 6539& GO8118  & 10, 40      & 40, 2x(160)   \\
 NGC 6540& GO8118  & 5, 40       & 40, 2x(160)   \\
 NGC 6544& GO8118  & 3, 2x(30)   & 40, 2x(100)   \\
 NGC 6569& GO8118  & 5, 40       & 40, 2x(160)   \\
 NGC 6584& GO8118  & 5, 2x(30)    & 40, 2x(100)   \\
\hline 
\end{tabular}
\caption{Observation log. Asterisks indicate clusters reduced
by using Stetsons PSFs.\label{obs1}}
\end{table}

\begin{table}
\begin{tabular}{llcc}
\hline  
%\multicolumn{1}{l}{ID} & Program ID & \multicolumn{2}{c}{Exposure Times (F555W, F439W)} \\
Cluster & Program ID & Exp. Time  & Exp. Time  \\
        &            &   F555W [s]      &  F439W [s]       \\             
\hline
 NGC 6624& archive & 4x(10)       & 2x(50)         \\
 NGC 6638& GO8118  & 5, 40       & 40, 2x(160)   \\
 NGC 6637& GO8118  & 3, 2x(30)   & 40, 2x(100)   \\
 NGC 6642& GO8118  & 3, 40       & 40, 2x(160)   \\
 NGC 6652$^*$& GO6095  & 10, 50      & 50, 2x(160)   \\
 NGC 6681$^*$& GO8723  & 3, 2x(30)   & 40, 2x(100)   \\
 NGC 6712$^*$& archive & 60           & 160            \\
 NGC 6717& GO8118  & 3, 2x(30)   & 40, 2x(100)   \\
 NGC 6723& GO8118  & 3, 2x(30)   & 40, 2x(100)   \\
 NGC 6760$^*$& GO8723  & 7, 40       & 40, 2x(160)   \\
 NGC 6838& GO8118  & 3, 2x(30)   & 40, 2x(100)   \\
 NGC 6864$^*$& GO8723  & 7, 40       & 40, 2x(160)   \\
 NGC 6934$^*$& GO7470  & 10, 40      & 40, 2x(160)   \\
 NGC 6981& GO7470  & 10, 40      & 40, 2x(160)   \\
 NGC 7078& archive & 4x(8)        & 2x(40)         \\
 NGC 7089& GO8118  & 3, 2x(30)   & 40, 2x(100)   \\
 NGC 7099& archive & 4x(4)        & 2x(40)         \\
\hline 
\end{tabular}
\caption{Observation log (cont.). \label{obs2}}
\end{table}

After the snapshot observations (or, for 12 clusters, after the public
release date), the images were retrieved via ftp from the HST archive
in Baltimore, and they were furtherly processed partially following
the recipe in Silbermann et al.\ (\cite{silbermann}). As in Silbermann
et al., the vignetted pixels, and bad pixels and columns, were masked
out using a vignetting frame created by P.\ B.\ Stetson, together with
the appropriate data-quality file for each frame. However, we did
\emph{not} correct for the pixel area map, since this correction is
included in the calibration process described below. Finally the
single-chip frames were extracted from the 4-chip-stack files, and
analyzed separately. Frames obtained both at gain = \( 7\rm e^{-}/ADU
\) and at \( 15\rm e^{-}/ADU \) were available, so care was taken to
adjust the various parameters to the
applicable gain value for each image.

\begin{table*}
\begin{tabular}{llrrrcccclrccc}
\hline  \multicolumn{1}{l}{~~~ID} & Name & $l$~~ & $b$~~ & $R_{\rm GC}$
& $E(B-V)$ & $(m-M)_V$ & $M_{Vt}$ & [Fe/H] & ~$c$ & $r_t$~  &$\log
t_c$&$\log t_h$& $\log \rho_0$ \\ \hline
 NGC 104 & 47 Tuc &   305.90 & $-$44.89 &   7.4 &   0.04 &   13.37 &   $-$9.42 &  $-$0.76 &   2.03  &  47.25&  8.06&  9.48 &   4.77\\
 NGC 362 &        &   301.53 & $-$46.25 &   9.3 &   0.05 &   14.80 &   $-$8.40 &  $-$1.16 &   1.94c:&  16.11&  7.76&  8.92 &   4.70\\
 NGC 1261&        &   270.54 & $-$52.13 &  18.2 &   0.01 &   16.10 &   $-$7.81 &  $-$1.35 &   1.27  &   7.28&  8.74&  9.20 &   2.96\\
 NGC 1851&        &   244.51 & $-$35.04 &  16.7 &   0.02 &   15.47 &   $-$8.33 &  $-$1.22 &   2.32  &  11.70&  6.98&  8.85 &   5.32\\
 NGC 1904& M 79   &   227.23 & $-$29.35 &  18.8 &   0.01 &   15.59 &   $-$7.86 &  $-$1.57 &   1.72  &   8.34&  7.78&  9.10 &   4.00\\
 NGC 2419&        &   180.37 &  25.24 &  91.5 &   0.11 &   19.97 &   $-$9.58 &  $-$2.12 &   1.40  &   8.74&  9.96& 10.55 &   1.54\\
 NGC 2808&        &   282.19 & $-$11.25 &  11.0 &   0.23 &   15.56 &   $-$9.36 &  $-$1.15 &   1.77  &  15.55&  8.28&  9.11 &   4.61\\
 NGC 3201&        &   277.23 &   8.64 &   9.0 &   0.21 &   14.24 &   $-$7.49 &  $-$1.58 &   1.30  &  28.45&  8.81&  9.23 &   2.69\\
 NGC 4147&        &   252.85 &  77.19 &  21.3 &   0.02 &   16.48 &   $-$6.16 &  $-$1.83 &   1.80  &   6.31&  7.49&  8.67 &   3.48\\
 NGC 4372&        &   300.99 &  $-$9.88 &   7.1 &   0.39 &   15.01 &   $-$7.77 &  $-$2.09 &   1.30  &  34.82&  8.90&  9.59 &   2.09\\
 NGC 4590& M 68   &   299.63 &  36.05 &  10.1 &   0.05 &   15.19 &   $-$7.35 &  $-$2.06 &   1.64  &  30.34&  8.67&  9.29 &   2.54\\
 NGC 4833&        &   303.61 &  $-$8.01 &   6.9 &   0.33 &   14.92 &   $-$8.01 &  $-$1.79 &   1.25  &  17.85&  8.71&  9.34 &   3.06\\
 NGC 5024& M 53   &   332.96 &  79.76 &  18.8 &   0.02 &   16.38 &   $-$8.77 &  $-$1.99 &   1.78  &  21.75&  8.79&  9.69 &   3.04\\
 NGC 5634&        &   342.21 &  49.26 &  21.9 &   0.05 &   17.22 &   $-$7.75 &  $-$1.82 &   1.60  &   8.36&  8.61&  9.28 &   3.12\\
 NGC 5694&        &   331.06 &  30.36 &  29.1 &   0.09 &   17.98 &   $-$7.81 &  $-$1.86 &   1.84  &   4.29&  7.86&  9.15 &   4.03\\
 IC 4499 &        &   307.35 & $-$20.47 &  15.7 &   0.23 &   17.09 &   $-$7.33 &  $-$1.60 &   1.11  &  12.35&  9.37&  9.66 &   1.49\\
 NGC 5824&        &   332.55 &  22.07 &  25.8 &   0.13 &   17.93 &   $-$8.84 &  $-$1.85 &   2.45  &  15.50&  7.88&  9.33 &   4.66\\
 NGC 5904& M 5    &     3.86 &  46.80 &   6.2 &   0.03 &   14.46 &   $-$8.81 &  $-$1.29 &   1.83  &  28.40&  8.26&  9.53 &   3.91\\
 NGC 5927&        &   326.60 &   4.86 &   4.5 &   0.45 &   15.81 &   $-$7.80 &  $-$0.37 &   1.60  &  16.68&  8.29&  8.98 &   3.87\\
 NGC 5946&        &   327.58 &   4.19 &   7.4 &   0.54 &   17.21 &   $-$7.60 &  $-$1.38 &   2.50c &  24.03&  7.06&  8.95 &   4.50\\
 NGC 5986&        &   337.02 &  13.27 &   4.8 &   0.27 &   15.94 &   $-$8.42 &  $-$1.58 &   1.22  &  10.52&  8.94&  9.23 &   3.30\\
 NGC 6093& M 80   &   352.67 &  19.46 &   3.8 &   0.18 &   15.56 &   $-$8.23 &  $-$1.75 &   1.95  &  13.28&  7.73&  8.86 &   4.76\\
 NGC 6139&        &   342.37 &   6.94 &   3.6 &   0.75 &   17.35 &   $-$8.36 &  $-$1.68 &   1.80  &   8.52&  7.56&  9.04 &   4.66\\
 NGC 6171& M 107  &     3.37 &  23.01 &   3.3 &   0.33 &   15.06 &   $-$7.13 &  $-$1.04 &   1.51  &  17.44&  8.05&  9.31 &   3.13\\
 NGC 6205& M 13   &    59.01 &  40.91 &   8.7 &   0.02 &   14.48 &   $-$8.70 &  $-$1.54 &   1.51  &  25.18&  8.80&  9.30 &   3.33\\
 NGC 6229&        &    73.64 &  40.31 &  30.0 &   0.01 &   17.46 &   $-$8.07 &  $-$1.43 &   1.61  &   5.38&  8.36&  9.19 &   3.40\\
 NGC 6218& M 12   &    15.72 &  26.31 &   4.5 &   0.19 &   14.02 &   $-$7.32 &  $-$1.48 &   1.39  &  17.60&  8.10&  9.02 &   3.23\\
 NGC 6235&        &   358.92 &  13.52 &   2.9 &   0.36 &   16.11 &   $-$6.14 &  $-$1.40 &   1.33  &   7.61&  8.11&  8.67 &   3.11\\
 NGC 6256&        &   347.79 &   3.31 &   2.1 &   1.03 &   17.31 &   $-$6.02 &  $-$0.70 &   2.50c &   7.59&  5.36&  8.40 &   5.70\\
 NGC 6266& M 62   &   353.58 &   7.32 &   1.7 &   0.47 &   15.64 &   $-$9.19 &  $-$1.29 &   1.70c:&   8.97&  7.64&  9.19 &   5.14\\
 NGC 6273& M 19   &   356.87 &   9.38 &   1.6 &   0.37 &   15.85 &   $-$9.08 &  $-$1.68 &   1.53  &  14.50&  8.50&  9.34 &   3.96\\
 NGC 6284&        &   358.35 &   9.94 &   6.9 &   0.28 &   16.70 &   $-$7.87 &  $-$1.32 &   2.50c &  23.08&  7.15&  9.16 &   4.44\\
 NGC 6287&        &     0.13 &  11.02 &   1.7 &   0.60 &   16.51 &   $-$7.16 &  $-$2.05 &   1.60  &  10.51&  7.85&  8.66 &   3.85\\
 NGC 6293&        &   357.62 &   7.83 &   1.4 &   0.41 &   15.99 &   $-$7.77 &  $-$1.92 &   2.50c &  14.23&  6.24&  8.91 &   5.22\\
 NGC 6304&        &   355.83 &   5.38 &   2.1 &   0.52 &   15.54 &   $-$7.32 &  $-$0.59 &   1.80  &  13.25&  7.38&  8.89 &   4.39\\
 NGC 6316&        &   357.18 &   5.76 &   3.2 &   0.51 &   16.78 &   $-$8.35 &  $-$0.55 &   1.55  &   5.93&  7.72&  9.00 &   4.21\\
 NGC 6325&        &     0.97 &   8.00 &   2.0 &   0.89 &   17.68 &   $-$7.35 &  $-$1.17 &   2.50c &   9.49&  5.94&  8.92 &   5.40\\
 NGC 6342&        &     4.90 &   9.73 &   1.7 &   0.46 &   16.10 &   $-$6.44 &  $-$0.65 &   2.50c &  14.86&  6.09&  8.66 &   4.77\\
 NGC 6356&        &     6.72 &  10.22 &   7.6 &   0.28 &   16.77 &   $-$8.52 &  $-$0.50 &   1.54  &   7.97&  8.33&  9.26 &   3.76\\
 NGC 6355&        &   359.58 &   5.43 &   1.0 &   0.75 &   16.62 &   $-$7.48 &  $-$1.50 &   2.50c &  15.18&  5.95&  8.71 &   4.95\\
 IC 1257 &        &    16.53 &  15.14 &  17.9 &   0.73 &   19.25 &   $-$6.15 &  $-$1.70 &         &       &      &       &       \\
 NGC 6362&        &   325.55 & $-$17.57 &   5.3 &   0.08 &   14.79 &   $-$7.06 &  $-$0.95 &   1.10  &  16.67&  9.07&  9.31 &   2.22\\
 NGC 6380& Ton 1  &   350.18 &  $-$3.42 &   3.2 &   1.17 &   18.77 &   $-$7.46 &  $-$0.50 &   1.55c:&  12.06&  8.39&  8.87 &   3.70\\
 NGC 6388&        &   345.56 &  $-$6.74 &   4.4 &   0.40 &   16.54 &   $-$9.82 &  $-$0.60 &   1.70  &   6.21&  7.90&  9.24 &   5.31\\
 NGC 6402& M 14   &    21.32 &  14.81 &   3.9 &   0.60 &   16.61 &   $-$9.02 &  $-$1.39 &   1.60  &  33.24&  9.07&  9.36 &   3.30\\
 NGC 6401&        &     3.45 &   3.98 &   0.8 &   0.85 &   17.07 &   $-$7.62 &  $-$1.12 &   1.69  &  12.10&  7.74&  9.29 &   4.10\\
 NGC 6397&        &   338.17 & $-$11.96 &   6.0 &   0.18 &   12.36 &   $-$6.63 &  $-$1.95 &   2.50c &  15.81&  4.90&  8.46 &   5.68\\
 NGC 6440&        &     7.73 &   3.80 &   1.3 &   1.07 &   17.95 &   $-$8.75 &  $-$0.34 &   1.70  &   6.31&  7.54&  8.76 &   5.28\\
 NGC 6441&        &   353.53 &  $-$5.01 &   3.5 &   0.44 &   16.62 &   $-$9.47 &  $-$0.53 &   1.85  &   8.00&  7.72&  9.13 &   5.23\\
 NGC 6453&        &   355.72 &  $-$3.87 &   3.3 &   0.61 &   17.13 &   $-$7.05 &  $-$1.53 &   2.50c &  21.50&  6.87&  8.36 &   4.72\\
 NGC 6517&        &    19.23 &   6.76 &   4.3 &   1.08 &   18.51 &   $-$8.28 &  $-$1.37 &   1.82  &   4.10&  6.90&  8.88 &   5.20\\
 NGC 6522&        &     1.02 &  $-$3.93 &   0.6 &   0.48 &   15.94 &   $-$7.67 &  $-$1.44 &   2.50c &  16.44&  6.32&  8.90 &   5.31\\
 NGC 6539&        &    20.80 &   6.78 &   3.1 &   0.97 &   17.63 &   $-$8.30 &  $-$0.66 &   1.60  &  21.46&  8.60&  9.37 &   3.62\\
 NGC 6540& Djorg 3&     3.29 &  $-$3.31 &   4.4 &   0.60 &   14.68 &   $-$5.38 &  $-$1.20 &   2.50c &   9.49&  5.01&  7.08 &   5.92\\
 NGC 6544&        &     5.84 &  $-$2.20 &   5.4 &   0.73 &   14.33 &   $-$6.56 &  $-$1.56 &   1.63c:&   2.05&  5.05&  8.35 &   5.75\\
 NGC 6569&        &     0.48 &  $-$6.68 &   1.2 &   0.56 &   16.43 &   $-$7.88 &  $-$0.86 &   1.27  &   6.95&  8.25&  9.17 &   3.76\\
\hline 
\end{tabular}
\caption{Main Parameters of the Clusters: Cols.\ 1 and 2 give the cluster
identification number and other commonly used cluster name, Cols.\ 3 and
4 the Galactic longitude and latitude (degrees), Col.\ 5 gives the
distance from the Galactic center (kpc), assuming $R_0=8.0$ kpc, Col.\ 6
the foreground reddening, Col.\ 7 the apparent visual distance modulus,
Col.\ 8 the absolute visual magnitude, Col.\ 9 the metallicity [Fe/H],
Col.\ 10 the central concentration ($c = \log(r_t/r_c)$, a ``c'' denotes
a core-collapsed cluster), Col.\ 11 the logarithm of core relaxation
time ($\log$ (years)), Col.\ 12 the logarithm of relaxation time at the
half-mass radius ($\log$ (years)), Col.\ 13 the logarithm of central
luminosity density ($L_\odot$pc$^{-3}$ ). \label{par1}} 
\end{table*}

\begin{table*}
\begin{tabular}{llrrrcccclrccc}
\hline  \multicolumn{1}{l}{~~~~ID} & Name & $l$~~ & $b$~~ & $R_{\rm GC}$
 & $E(B-V)$ & $(m-M)_V$ & $M_{Vt}$ & [Fe/H] & ~$c$ & $r_t$~~  &$\log
 t_c$&$\log t_h$& $\log \rho_0$ \\ \hline
 NGC 6584&        &   342.14 & $-$16.41 &   7.0 &   0.10 &   15.95 &   $-$7.68 &  $-$1.49 &   1.20  &   9.37&  9.01&  9.09 &   2.92\\
 NGC 6624&        &     2.79 &  $-$7.91 &   1.2 &   0.28 &   15.37 &   $-$7.50 &  $-$0.42 &   2.50c &  20.55&  6.62&  8.74 &   5.25\\
 NGC 6638&        &     7.90 &  $-$7.15 &   1.6 &   0.40 &   15.85 &   $-$6.83 &  $-$0.99 &   1.40  &   6.63&  7.93&  8.51 &   4.05\\
 NGC 6637& M 69   &     1.72 & $-$10.27 &   1.6 &   0.16 &   15.16 &   $-$7.52 &  $-$0.71 &   1.39  &   8.35&  8.15&  8.79 &   3.81\\
 NGC 6642&        &     9.81 &  $-$6.44 &   1.6 &   0.41 &   15.70 &   $-$6.57 &  $-$1.35 &   1.99  &  10.07&  6.94&  8.49 &   4.72\\
 NGC 6652&        &     1.53 & $-$11.38 &   2.4 &   0.09 &   15.19 &   $-$6.57 &  $-$0.96 &   1.80  &   4.48&  6.66&  8.55 &   4.54\\
 NGC 6681& M 70   &     2.85 & $-$12.51 &   2.1 &   0.07 &   14.98 &   $-$7.11 &  $-$1.51 &   2.50c &   7.91&  5.62&  8.83 &   5.41\\
 NGC 6712&        &    25.35 &  $-$4.32 &   3.5 &   0.45 &   15.60 &   $-$7.50 &  $-$1.01 &   0.90  &   7.44&  8.86&  8.98 &   3.14\\
 NGC 6717& Pal 9  &    12.88 & $-$10.90 &   2.3 &   0.20 &   14.95 &   $-$5.67 &  $-$1.29 &   2.07c:&   9.87&  6.61&  8.26 &   4.65\\
 NGC 6723&        &     0.07 & $-$17.30 &   2.6 &   0.05 &   14.87 &   $-$7.86 &  $-$1.12 &   1.05  &  10.51&  8.99&  9.30 &   2.81\\
 NGC 6760&        &    36.11 &  $-$3.92 &   4.8 &   0.77 &   16.74 &   $-$7.86 &  $-$0.52 &   1.59  &  12.96&  7.94&  9.39 &   3.84\\
 NGC 6838& M 71   &    56.74 &  $-$4.56 &   6.7 &   0.25 &   13.75 &   $-$5.56 &  $-$0.73 &   1.15  &   8.96&  7.64&  8.41 &   3.05\\
 NGC 6864& M 75   &    20.30 & $-$25.75 &  12.8 &   0.16 &   16.87 &   $-$8.35 &  $-$1.32 &   1.88  &   7.28&  7.85&  9.08 &   4.51\\
 NGC 6934&        &    52.10 & $-$18.89 &  14.3 &   0.09 &   16.48 &   $-$7.65 &  $-$1.54 &   1.53  &   8.37&  8.43&  9.07 &   3.37\\
 NGC 6981& M 72   &    35.16 & $-$32.68 &  12.9 &   0.05 &   16.31 &   $-$7.04 &  $-$1.40 &   1.23  &   9.15&  8.93&  9.20 &   2.35\\
 NGC 7078& M 15   &    65.01 & $-$27.31 &  10.4 &   0.10 &   15.37 &   $-$9.17 &  $-$2.25 &   2.50c &  21.50&  7.02&  9.35 &   5.38\\
 NGC 7089& M 2    &    53.38 & $-$35.78 &  10.4 &   0.06 &   15.49 &   $-$9.02 &  $-$1.62 &   1.80  &  21.45&  8.54&  9.32 &   3.90\\
 NGC 7099& M 30   &    27.18 & $-$46.83 &   7.1 &   0.03 &   14.62 &   $-$7.43 &  -2.12 &   2.50c &  18.34&  6.38&  8.95 &   5.04\\
\hline 
\end{tabular}
\caption{Main Parameters of the Clusters (continued). \label{par2}}
% K: No need to repeat the caption.
%: Col.\ 1 and 2 give the cluster identification number and other
%commonly used cluster name, Col.\ 3 and 4 the Galactic longitude and
%latitude (degrees), Col.\ 5 gives the distance from Galactic center
%(kpc), assuming $R_0=8.0$ kpc, Col.\ 6 the foreground reddening, Col.\
%7 the apparent visual distance modulus, Col.\ 8 the absolute visual
%magnitude, Col.\ 9 the metallicity [Fe/H], Col.\ 10 the central
%concentration ($c = \log(r_t/r_c)$, a ``c'' denotes a core-collapsed
%cluster), Col.\ 11 the logarithm of core relaxation time ($\log
%(years)$), Col.\ 12 the logarithm of relaxation time at the half-mass
%radius ($\log (years)$), Col.\ 13 the logarithm of central luminosity
%density (Solar luminosities per cubic parsec). \label{par2}} 
\end{table*}

\subsection{Instrumental magnitudes}

The photometric reduction was carried out using the
\noun{daophot~II/allframe} package (Stetson \cite{stetson87},
\cite{stetson94}).  Preliminary photometry was carried out in order to
construct an approximate list of stars for each single frame.  This list
was used to match the different frames accurately.  With the correct
coordinate transformations among the frames, we obtained a single image,
combining all the frames, regardless of the filter.  In this way we
could eliminate all the cosmic rays and obtain the highest signal/noise
image for star finding.  We ran the \noun{daophot/find} routine on the
stacked image and performed PSF-fitting photometry in order to obtain
the deepest list of stellar objects free from spurious detections.  The
subtracted image was searched again for objects missed in the first
pass, and the new list was appended to the existing one.  Finally, the
entire star list was given as input to \noun{allframe}, for the
simultaneous PSF-fitting photometry of all the individual frames.  In
some cases we could not construct a PSF from our images, since there
were not enough sufficiently isolated stars in any of the four chips.
In those instances, the PSFs used were the high-S/N PSFs extracted by
P.\ B.\ Stetson (private communication) from a large set of uncrowded
and unsaturated WFPC2 images.  
%GP Added this sentence
The clusters for which Stetson PSFs have been used are marked with
an asterisk in Tables~\ref{obs1},\ref{obs2}.
For each of the WFPC2 chips, the
(typically two) F555W and (typically three) F439W magnitude lists were
combined to create a raw color-magnitude diagram (CMD).  First a catalog
of mean magnitudes was created for each of the two filters, and then
they were combined to obtain the F439W$-$F555W colors. In this process,
we used the programs \noun{daomatch/daomaster} (kindly provided by P.\
B.\ Stetson), which yield magnitudes in the instrumental photometric
system of the two F555W and F439W frames that were chosen as references.

\subsection{The correction for CTE}

In order to calibrate our photometry, we followed the procedure
outlined in Dolphin (\cite{dolphin-cte}; D00).  This accounts for both
the charge transfer (in)efficiency (CTE) and the variation of the
effective pixel area across the WFPC2 field of view, and it yields
final calibrated magnitudes in either the Johnson photometric system
or the HST one.  As a first step, 
% K: You need to be more specific about what you mean by the following.
% (Or does it need to say ``as follows''?)
a softened background was calculated 
from the actual output of \noun{allframe} (after multiplying by the
gain value \( G=7 \) or 15, so counts are expressed in electrons)
as follows.
Negative values of the background were set to zero, and then the
background counts ($B$) were replaced by \(\sqrt{(1+B^{2})}\).  The
counts in electrons \( D=G\times 10^{-0.4\, (m-25)} \) were then
computed for each star magnitude. (Note that DAOPHOT sets a star
magnitude to $m=25$ for stars with a flux corresponding to one count
above the sky background.)  The \( B \) and \( D \) values were then
used to find the CTE (and pixel area) correction to magnitude \( m \),
which is computed as \( m=m-C \), where \( C=Y+X \) and
\vspace{-0.5pt}
{\small
\begin{align}
X&=\frac{x}{800}(0.024+0.002\cdot yr)\cdot \notag \\
&\cdot \exp{[-0.196(\ln D-7)-0.126(\ln B-1)]}\notag \\
\notag \\
Y&=\frac{y}{800}\big\{0.018+(0.097+0.041\cdot yr)\cdot \notag \\
& \cdot (0.088+ \exp{[-0.507(\ln D-7)]})\cdot \notag \\
& \cdot\exp{[-0.035(\ln B-1)-0.042\cdot B]}\big\}. \notag
% K: All of the preceding would look better if you used \exp instead of
% e^. 
%GP Francesca, puoi farlo tu? Non vorrei fare poi errori sulla formula
\end{align}
}
\vspace{-0.3pt}
The CTE correction depends on the epoch of the observations ($yr$),
which is expressed, in the D00 equations, relative to the reference epoch
1996.3. The coefficients of the equations were taken from Table~1 of
D00, which is appropriate for observations made after April 23, 1994,
when the camera was cooled from $-76^{\circ }$C to $-88^{\circ }$C.
The corrected magnitudes were then calibrated to both the Johnson and
HST flight systems, going through the steps described in the following
Section. 

\subsection{Transformation to the standard photometric systems}

The first step was to find the aperture corrections from
\noun{allframe} magnitudes to the reference aperture of \( 0.5'' \) used
by Holtzman et al. (\cite{holtzman}; H95).  A set of bright isolated
objects was selected, all their neighbors were subtracted, and
aperture photometry was performed within the chosen set of radii.  The
aperture corrections were then defined as \( AC=m_{\rm PSF}-m_{0.5''}
\), and median values were computed. Generally the agreement between
the zero points of the four chips is good (\( \Delta m<0.01 \)
magnitudes), but in some cases the procedure gave poor aperture
corrections.  This normally happened for the more crowded PC chip.  In
such cases, the aperture corrections were changed by a few hundredths
of a magnitude in order to bring the PC photometry into agreement with
the WF zero points.  This procedure will of course erase any true
magnitude offset introduced by a patchy reddening on arcmin scales, so
we warn the readers that these data are not suitable for mapping the
reddening within the sky area covered by the WFPC2.

Once the aperture corrections had been applied, the following procedure
was followed.  First, the aperture corrections \( AC_{555} \) and \(
AC_{439} \), and the absorptions \( A_{555} \) and \( A_{439} \) were
subtracted from the instrumental magnitues.  Since the absorptions
depend on the true colors, which are not known at the beginning, we
started with null values for \( A_{555} \) and \( A_{439} \).  From the
corrected instrumental magnitudes \( m_{555} \) and \( m_{439} \) the
counts were computed in the usual way, i.e., \(
D_{555}=10^{-0.4(m_{555}-25)} \) and \( D_{439}=10^{-0.4(m_{439}-25)}
\). The \( D \) values were used to compute the provisional flight
magnitudes, as given by the D00 equations:
\[
\begin{array}{l}
F555W=-2.5\log (D_{555}/t_{555})+21.734+\Delta Z_{\rm CG}\\
\\
F439W=-2.5\log (D_{439}/t_{439})+20.086+\Delta Z_{\rm CG}.\\
\end{array}
\]
In these equations, \( t_{555} \) and \( t_{439} \) are the exposure
times in seconds, the filter- and temperature-dependent zero-points were
taken from Table~6 of D00 (\emph{Cold \( Z_{\rm FG} \)} column), and the
chip to chip zero-point differences \( \Delta Z_{\rm CG} \) were in turn
taken from Table~5. Explicitly, for gain \( 7 \) the values are \( 0.701
\), \( 0.761 \), \( 0.749 \), and \( 0.722 \) for the PC, WF2, WF3 and
WF4 chips, respectively. For gain \( 15 \) the values are, in the same
order of chips, \( -0.044 \), \( 0.007 \), \( -0.007 \), and \( -0.005
\).

%PC1  & \(-0.044\pm0.001\) & \( 0.701\pm0.001\) \nl
%WFC2 & \( 0.007\pm0.000\) & \( 0.761\pm0.000\) \nl
%WFC3 & \(-0.007\pm0.000\) & \( 0.749\pm0.000\) \nl
%WFC4 & \(-0.005\pm0.000\) & \( 0.722\pm0.000\) \nl

The determination of the Johnson \( B \) and \( V \) magnitudes is complicated
by the fact that they depend on the true color of the star, so we followed an
iterative procedure. Assuming \( B-V=1 \), we computed the provisional magnitudes
as prescribed by D00:
\[
\begin{array}{l}
V=F555W-0.060\times (B-V)+0.033\times (B-V)^{2}\\
\\
B=F439W+0.003\times (B-V)-0.088\times (B-V)^{2}.\\
\end{array}
\]
We then computed an updated value of \( B-V \) and iterated until the
difference between successive values of the magnitudes dropped below \(
0.001 \).

At this stage we are still ignoring absorption. However, we now have
an estimate of the Johnson magnitudes, so we can compute a first provisional
value of \( A_{555} \) and \( A_{439} \). They can be subtracted, together
with the aperture corrections, from the instrumental magnitudes, and the previous
cycle can be repeated until new values for the absorptions are obtained. Indeed,
the cycle was repeated until again the differences between successive values
of \( V \) and \( B \) were smaller than 0.001.

The absorptions were computed in the following manner. The absorptions
for two stars of spectral type K5 and O6 are given in H95 for
the different HST filters, as a function of the reddening \( E_{B-V}
\). In order to compute the absorption for a star of any spectral type,
two fiducial \( B-V \) colors were assigned to the two reference types,
\( B-V=1.15 \) and \( B-V=-0.32 \), respectively. The absorptions for
stars of intermediate colors were computed as linear interpolations
between the values at the two color extremes. Since the range in color
of our stellar populations is not extreme, the same linear
interpolation was used also to compute the absorptions for stars falling
outside the preferred color range. The average reddening of each
globular cluster was taken from the Harris (\cite{harris-catalog})
catalog.

%The final photometric catalog then contains the true Johnson magnitudes \( V_{0} \)
%and \( B_{0} \), ST magnitudes \( F555W_{0} \) and \( F439W_{0} \), and also
%the provisional magnitudes obtained in the first cycle, in order to allow the
%possibility of a recalibration of the whole dataset.

% K: The photometric columns should be given to only 2 decimal digits.
%GP Ha ragione. Francesca, puoi correggere. Ovviamente le tabelle in
%   rete restano come sono!
% K2: I think the dots should be removed from the table headings.  They
% confused me.
\begin{table*}
\begin{center}
\begin{tabular}{rrrrrrrrc}
\hline  ID & $x$~~~ & $y$~~~ &\( V \)~~~~ &\( B \)~~~~ & F555W & F439W & $\sigma_{F555W}$ & $\sigma_{F439W}$ \\ \hline
 4   &506.598 &59.980 &17.97 &18.74 &18.00 &18.79 &0.10 &0.10 \\
 7   &200.220 &60.927 &18.94 &19.28 &18.96 &19.29 &0.24 &0.08 \\
 8   &235.644 &61.244 &16.49 &17.15 &16.52 &17.19 &0.15 &0.07 \\
 9   &545.239 &61.390 &14.68 &15.62 &14.70 &15.70 &0.08 &0.04 \\
 5   &688.846 &61.697 &21.49 &21.90 &21.51 &21.91 &0.26 &0.43 \\
 15  &736.285 &61.745 &19.66 &20.30 &19.68 &20.34 &0.15 &0.10 \\
 5603&537.317 &62.062 &18.96 &19.48 &18.99 &19.50 &0.16 &0.08 \\
 12  &445.775 &62.208 &18.65 &19.37 &18.67 &19.42 &0.09 &0.06 \\
 14  &636.699 &62.301 &18.00 &18.65 &18.02 &18.69 &0.08 &0.06 \\
 11  &119.039 &62.725 &17.59 &18.03 &17.61 &18.04 &0.09 &0.06 \\
%... &...&...&...&...&...&...&...&...\\
\hline 
\end{tabular} 
\\
\addvspace{10pt}
\begin{tabular}{ccccrrc}
% IRK fixed 4 headers in this line.
\hline $V_{\rm nr}$~~~ & $B_{\rm nr}$~~~ & F555W$_{\rm nr}$ & F439W$_{\rm nr}$ & $\chi$~~~~ & sharp~ & chip \\ \hline
18.13 &18.94 &18.16 &19.00& 2.206&  0.046& 1\\
19.09 &19.49 &19.11 &19.50& 1.821& $-$0.032& 1\\
16.65 &17.35 &16.67 &17.39& 6.213&  0.157& 1\\
14.83 &15.82 &14.86 &15.90& 9.128&  0.040& 1\\
21.64 &22.10 &21.66 &22.12& 1.015& $-$0.282& 1\\
19.81 &20.51 &19.84 &20.55& 1.421& $-$0.041& 1\\
19.12 &19.68 &19.14 &19.71& 2.239&  0.162& 1\\
18.80 &19.57 &18.83 &19.62& 1.677&  0.079& 1\\
18.15 &18.85 &18.18 &18.90& 2.230&  0.108& 1\\
17.75 &18.23 &17.77 &18.25& 1.887& $-$0.032& 1\\
%&... &...&...&...&...&...&...\\
\hline
\end{tabular}
\caption{Example of a photometry file (NGC 104, 47 Tuc): Col.\ 1 gives a
star identification number, Cols.\ 2 and 3 give the position on the chip,
Cols.\  4 and 5 the \( V \) and \( B \) standard magnitudes (reddening
corrected), Cols.\ 6 and 7 the F555W and F439W magnitudes in the HST
flight system (reddening corrected), Cols.\ 8 and 9 the photometric
errors given by \noun{allframe}, Cols.\ 10 and 11 the \( V \) and \( B \)
standard magnitudes (before the reddening correction), Cols.\ 12 and 13
the F555W and F439W flight magnitudes (before the reddening correction),
Cols.\ 14 and 15 the $\chi$ and sharp parameters as given by
\noun{allframe}, and Col.\ 16 the chip number (1 for PC, and 2, 3, 4 for
WF2, WF3, WF4 respectively). \label{cal}}
\end{center}
\end{table*}

\begin{figure*}
%\resizebox{0.5\columnwidth}{0.4\height}{\includegraphics{crowding.ps}}
\begin{center}
\resizebox{1.5\columnwidth}{!}{\includegraphics{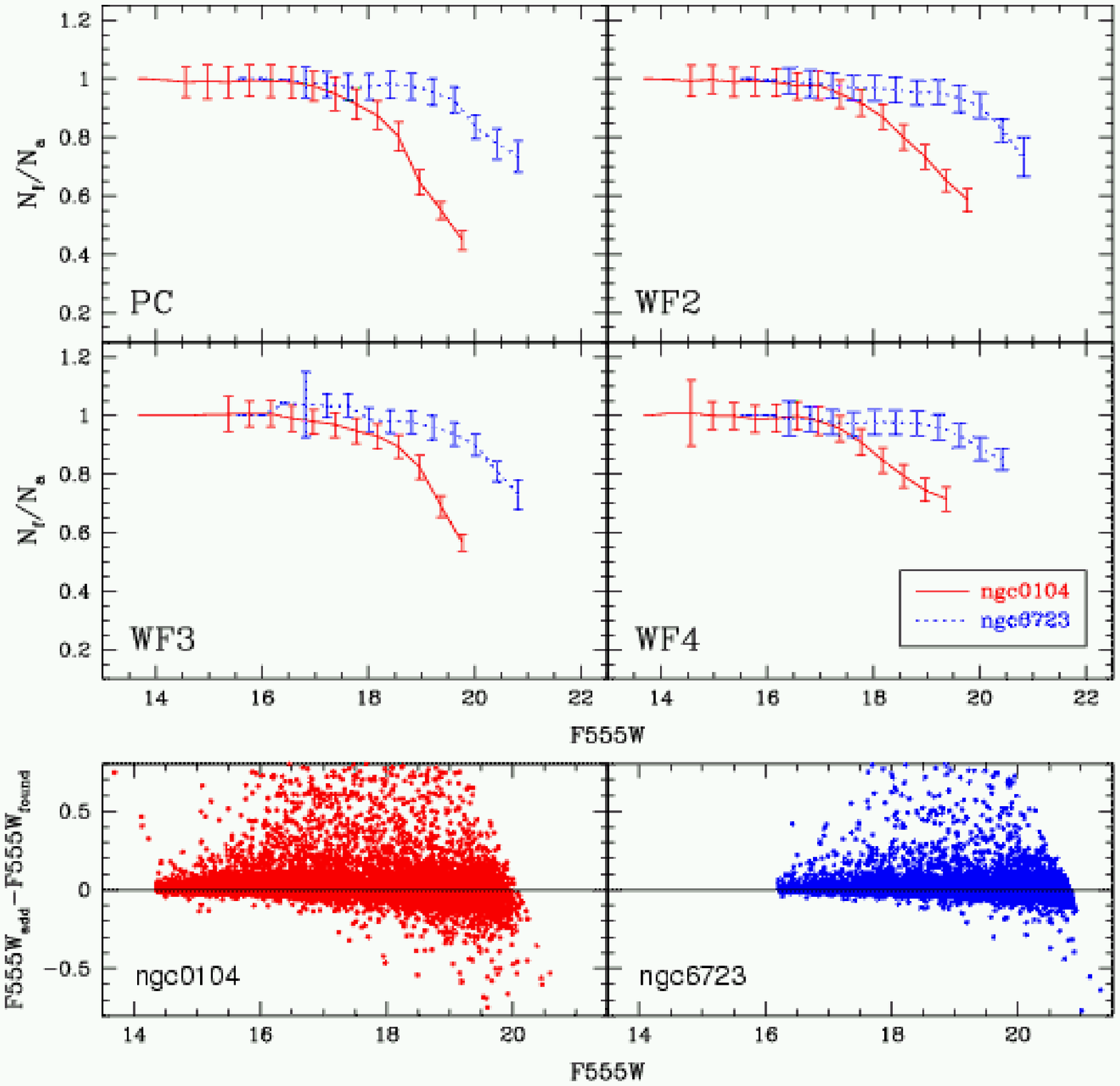}}
\caption{Completeness functions and internal photometric
errors from the artificial star experiments for the case of a high
central density cluster (NGC~104), and of a low density object
(NGC~6723).
}
\end{center}
\label{artificial}
\end{figure*}

\begin{figure}
%\resizebox{0.4\columnwidth}{0.2\height}{\includegraphics{figure_hb.ps}}
\resizebox{1\columnwidth}{!}{\includegraphics{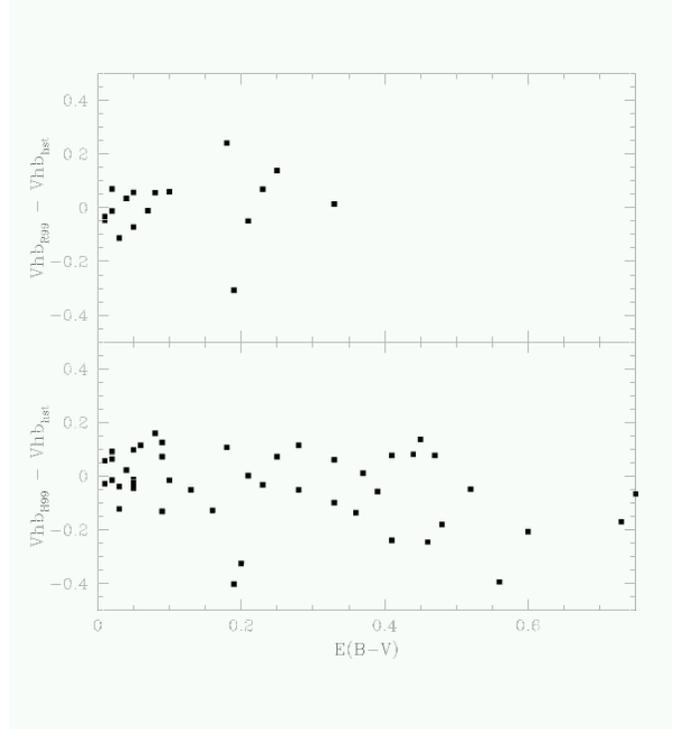}}
\caption{({\it Upper panel:} differences between the \( V_{\rm ZAHB} \) of
Rosenberg et al. (1999) and the \( V_{\rm ZAHB} \) for a subsample of
our clusters (De Angeli 2001, Piotto et al. 2002) as a function of
reddening. ({\it Lower panel:} differences between the \( V_{\rm ZAHB} \) of
Harris (\cite{harris-catalog}) and the \( V_{\rm ZAHB} \) for the same
subsample of our clusters as in the {\it upper panel}.}
\label{comp_hb}
\end{figure}

%GP Aggiunto da qui a fine subsection.
In practice, it is impossible to directly compare the photometry of
the data base published in this paper with any photometric catalog
from groundbased data. Our HST images are on the central, very crowded
regions which are not the usual targets of groundbased
investigations. An indirect check of the photometric calibration
is shown in Fig.~\ref{comp_hb}, where the HB magnitude levels of
the HST CMDs and of the CMDs from two groundbased photometric datasets
are compared.
The ({\it upper panel})shows the differences between the \( V
\) magnitudes of the zero age horizontal branch (\( V_{\rm ZAHB} \)) of
Rosenberg et al. (1999) and the \( V_{\rm ZAHB} \) for a subsample of
our clusters (De Angeli 2001, Piotto et al. 2002), as a function of the
reddening.  In the {\it lower panel}, our \( V_{\rm ZAHB} \) are
compared with the corresponding values tabulated by Harris
(\cite{harris-catalog}).  The \( V_{ZAHB} \) for the HST data have
been derived as in Zoccali et al. (1999). There is a good agreement
between the HST and groundbased values. 

\subsection{Artificial star experiments and completeness}

In order to correct the empirical star counts for completeness, we
performed standard artificial star experiments for each GGC.  In order
to optimize the cpu time, in our experiments we tried to add the
largest possible number of artificial stars in a single test, without
artificially increasing the crowding of the original field, i.e.,
avoiding the overlap of two or more artificial-star profiles.  To this
purpose, as described in Piotto \& Zoccali (1999), the artificial
stars were added in a spatial grid such that the separation of
the centers in each star pair was two PSF radii plus one pixel.  The
relative position of each star was fixed within the grid. However, the
grid was randomly moved on the frame for each different experiment.

For each artificial star test, the frame-to-frame coordinate
transformations (as calculated from the original photometry) were
used to ensure that the artificial stars were added exactly in the
same position in each frame.  We started by adding stars in one $V$
frame at random magnitudes; the corresponding \( B \) magnitude for each
star was chosen according to the fiducial points representing the
instrumental CMD.  The frames obtained in this way were
processed following the same procedure used for the reduction of the
original images.

The completeness fraction, typically in 0.4 magnitude intervals, was
then computed as the ratio between the number of the added artificial
stars and the number of artificial stars found in the same magnitude
range.

We performed separate experiments for the HB, the subgiant and red giant
branches, and the blue stragglers and main sequence. In each of the
three CMD branches, we ran 8 independent experiments for each of the
4 WFPC2 chips, 
%GP Added the following row:
adding in each experiment 600 stars in the PC camera and 700 stars in the WF camera,
for a total of more than 7100 experiments, with more
than 5 million artificial stars added.

A comparison between the added magnitudes and the measured magnitudes
allows us also a realistic estimate of the internal photometric error,
defined as the standard deviation of the differences between the
magnitudes added and those found, as a function of magnitude.

%GP Aggiunto da qui a fine subsection.
An example of the completeness functions, and of the internal photometric
errors obtained from the artificial star experiments is shown in 
Fig.~\ref{artificial}. We have selected two typical situations:
(a) the case of a high central density cluster (NGC~104), and (b)
the case of a low density object (NGC~6723).

\begin{figure*}
\begin{tabular}{cc}
\resizebox*{0.9\columnwidth}{0.36\height}{\includegraphics{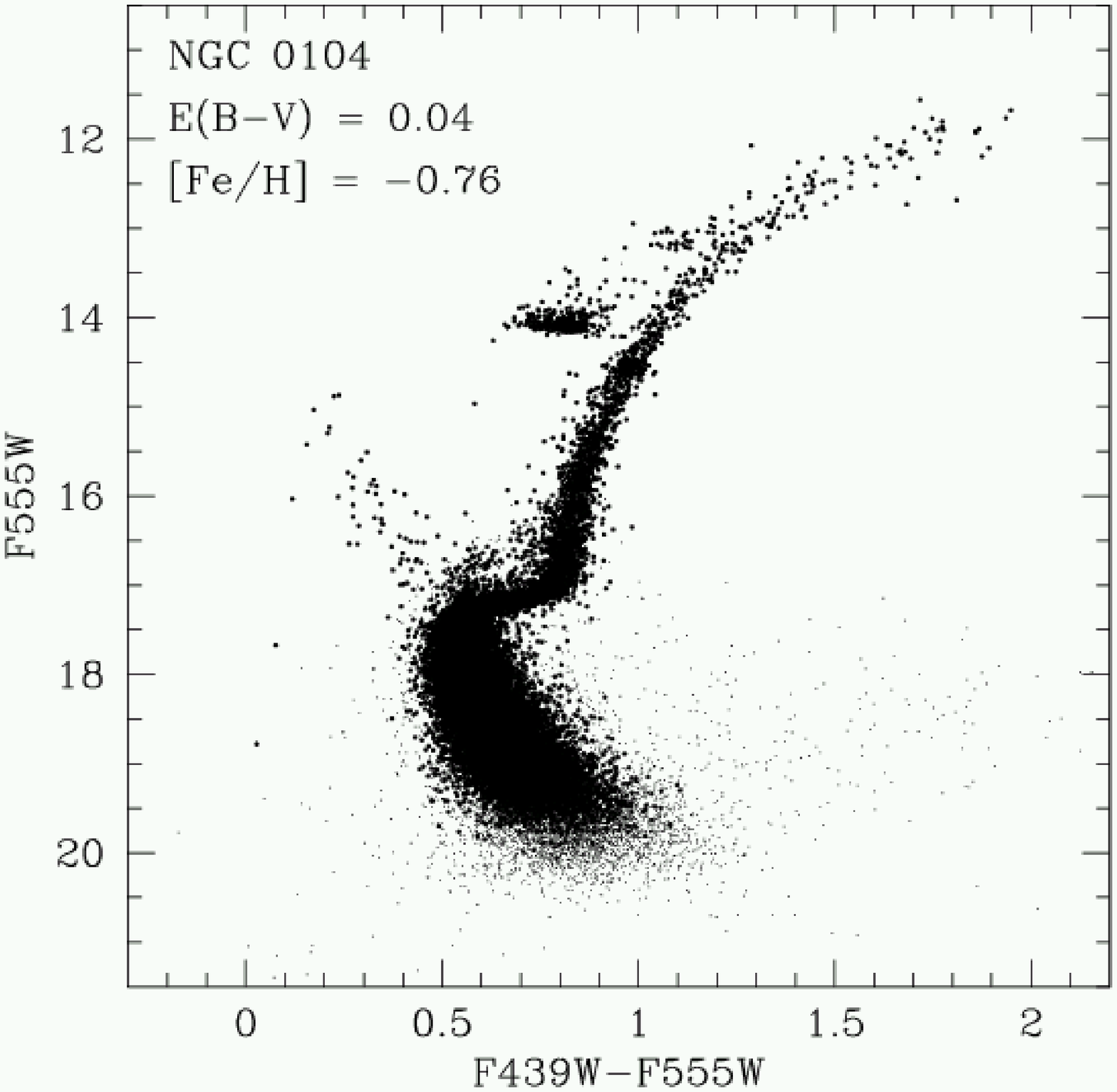}} &
\resizebox*{0.9\columnwidth}{0.36\height}{\includegraphics{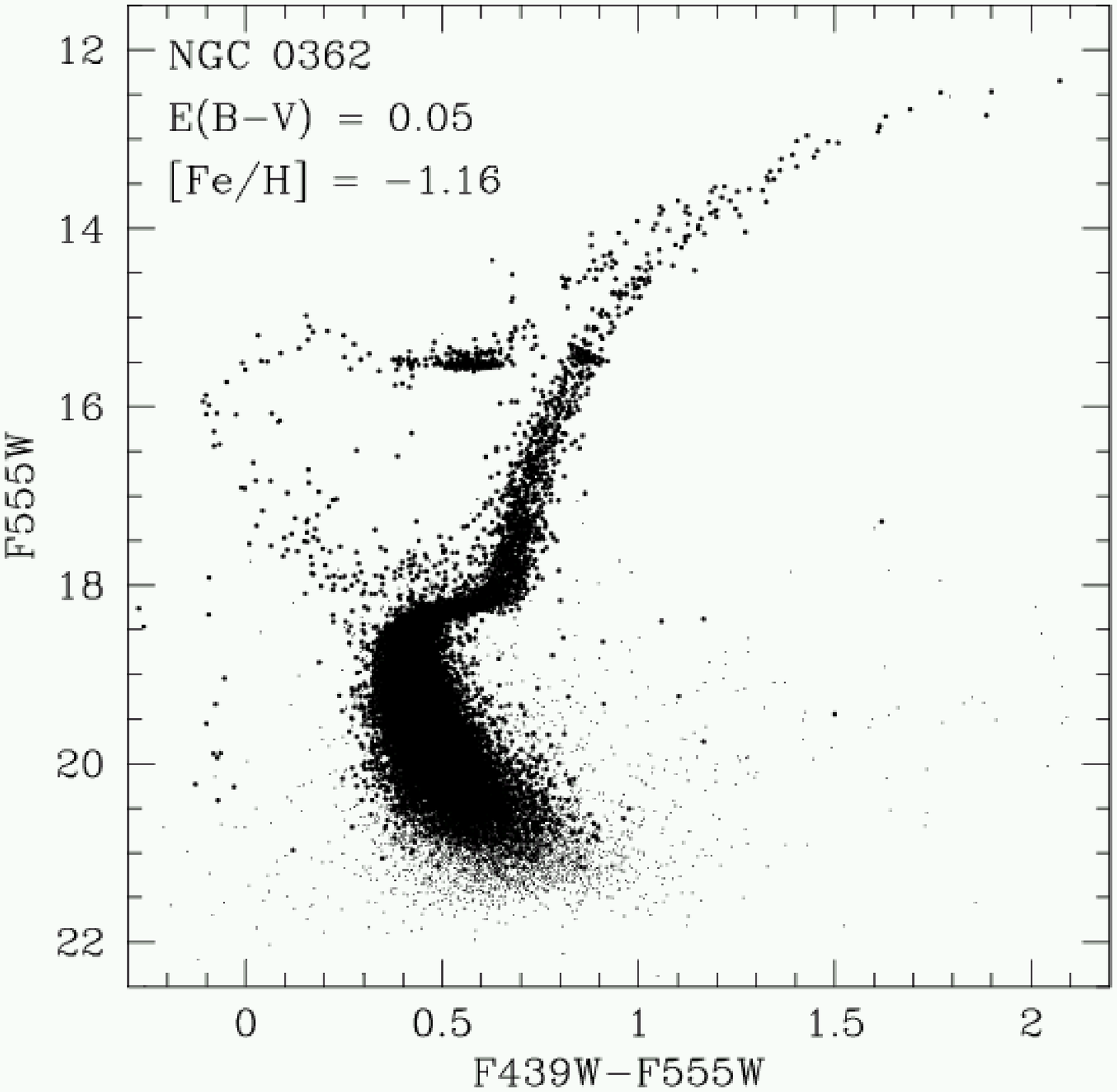}} \\
\resizebox*{0.9\columnwidth}{0.36\height}{\includegraphics{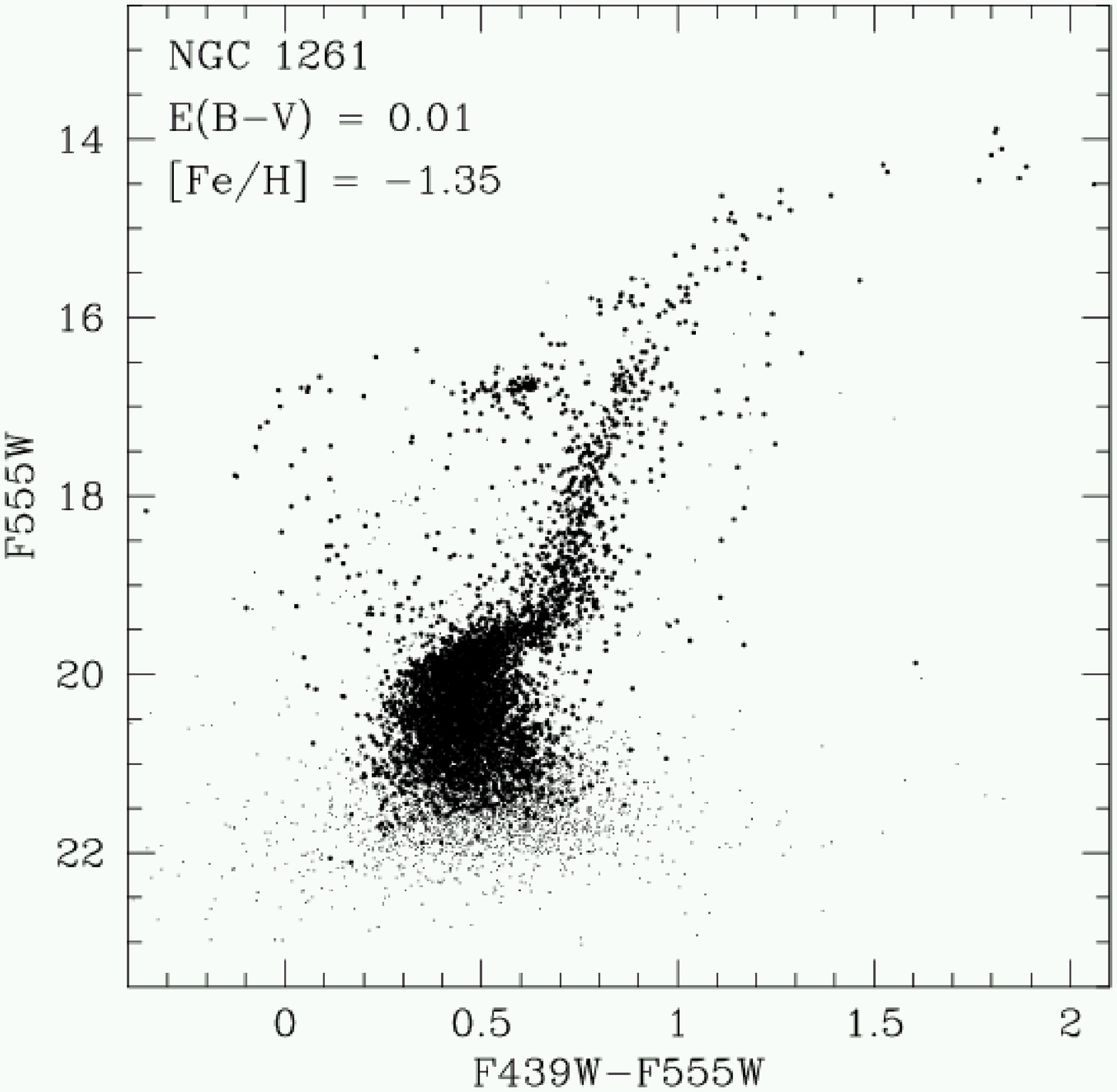}} &
\resizebox*{0.9\columnwidth}{0.36\height}{\includegraphics{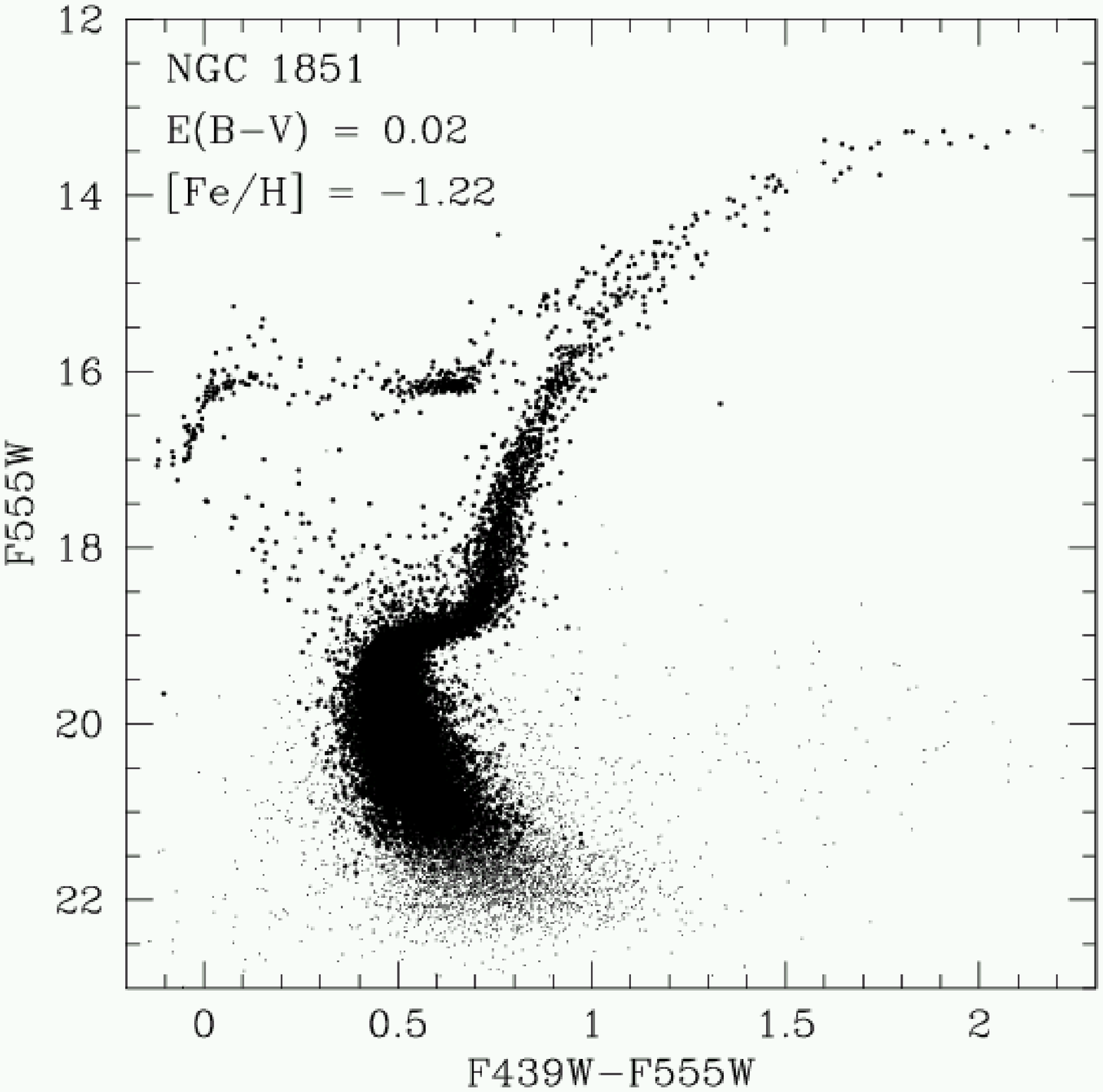}} \\
\resizebox*{0.9\columnwidth}{0.36\height}{\includegraphics{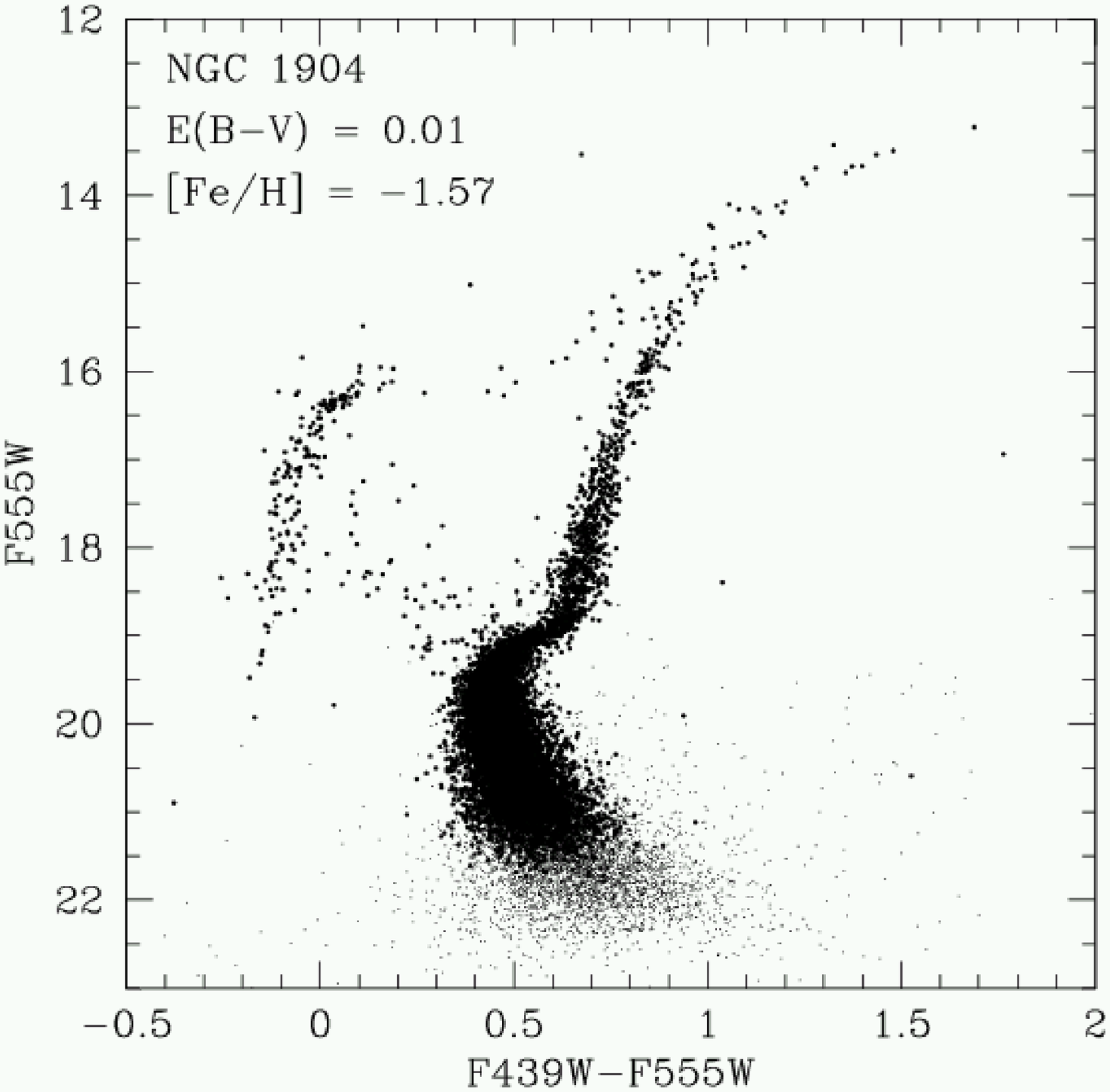}} &
\resizebox*{0.9\columnwidth}{0.36\height}{\includegraphics{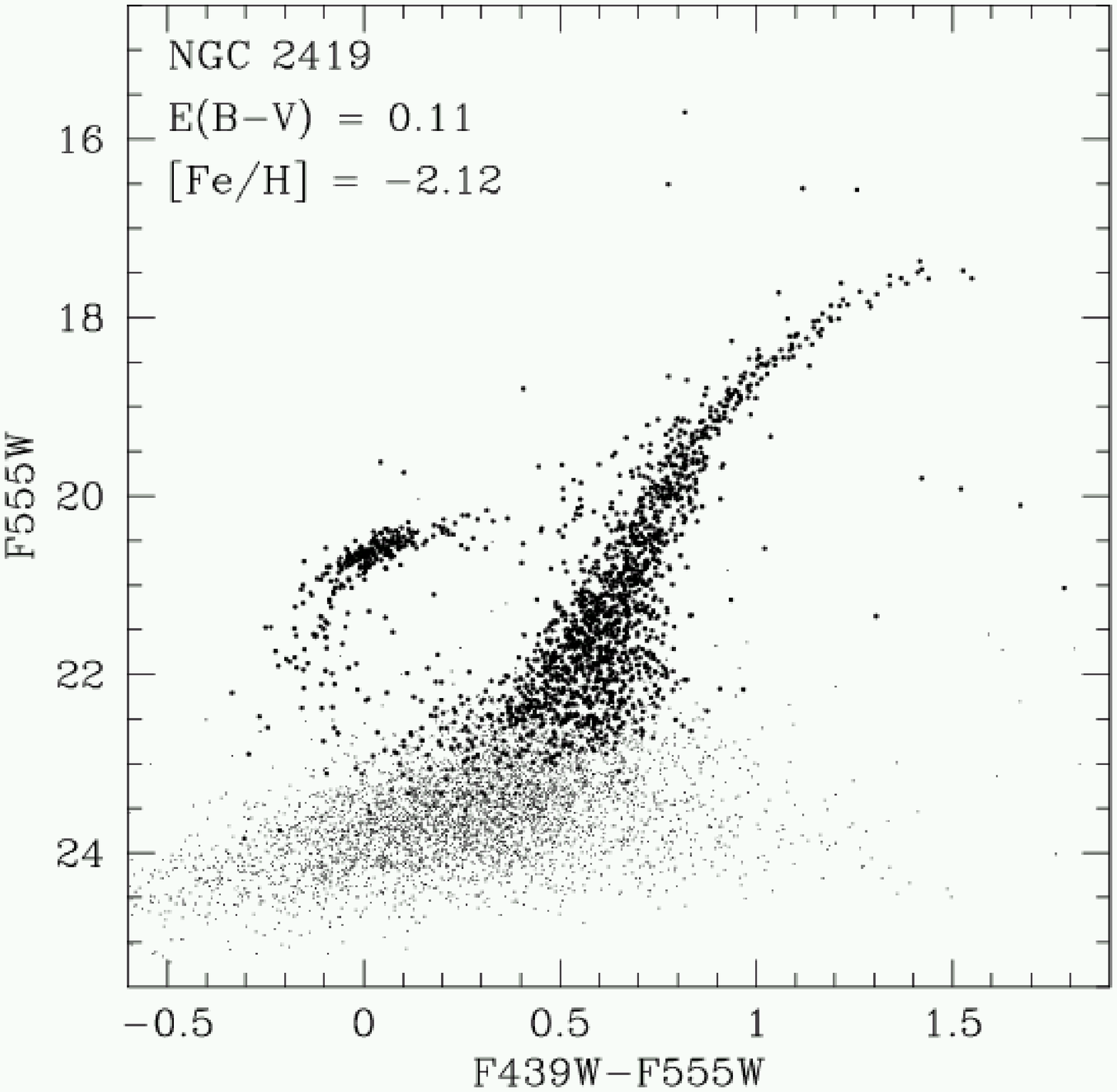}} \\
\end{tabular}
\caption{The F555W vs. F439W$-$F555W (flight system) 
color magnitude diagrams from the combination of the 4 WFPC2 cameras
of 2 clusters of the database.  Note that the magnitude and color
ranges covered by each figure are always of the same size (though
magnitude and color intervals start at different values).  Heavier
dots correspond to stars with an internal total error less than 0.1
magnitudes.}
\end{figure*}

\begin{figure*}
\begin{tabular}{cc}
\resizebox*{0.9\columnwidth}{0.36\height}{\includegraphics{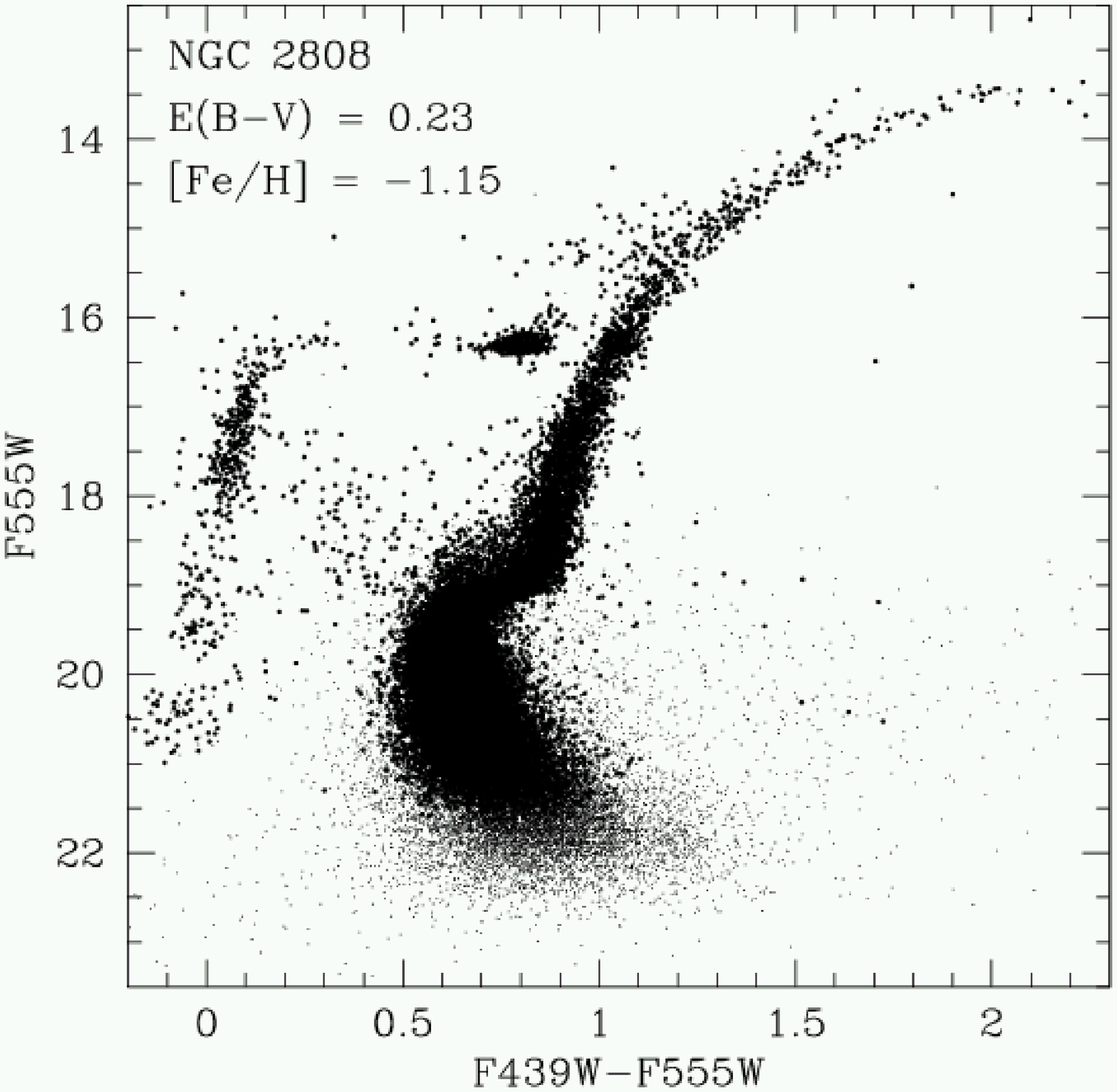}} &
\resizebox*{0.9\columnwidth}{0.36\height}{\includegraphics{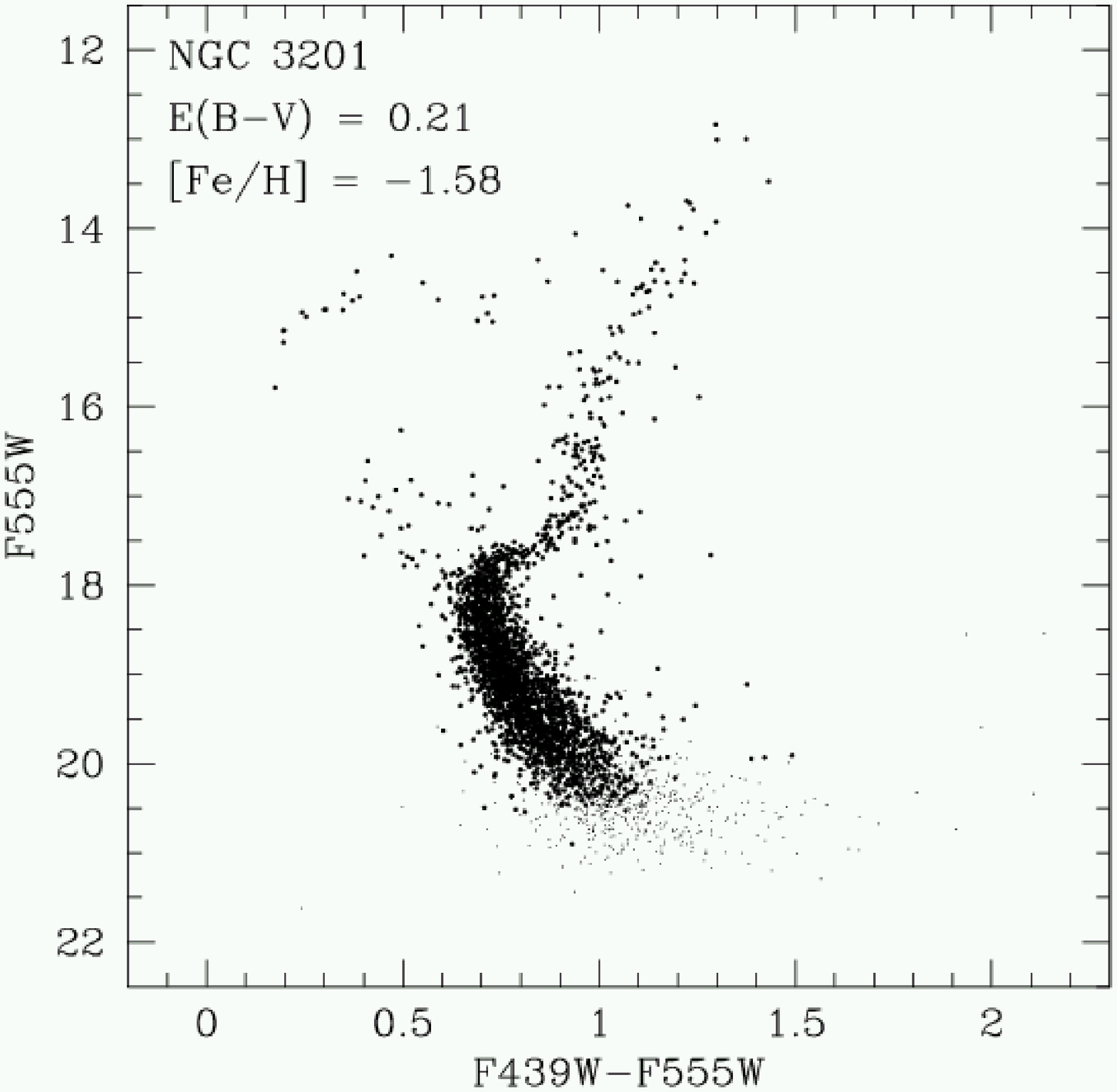}} \\
\resizebox*{0.9\columnwidth}{0.36\height}{\includegraphics{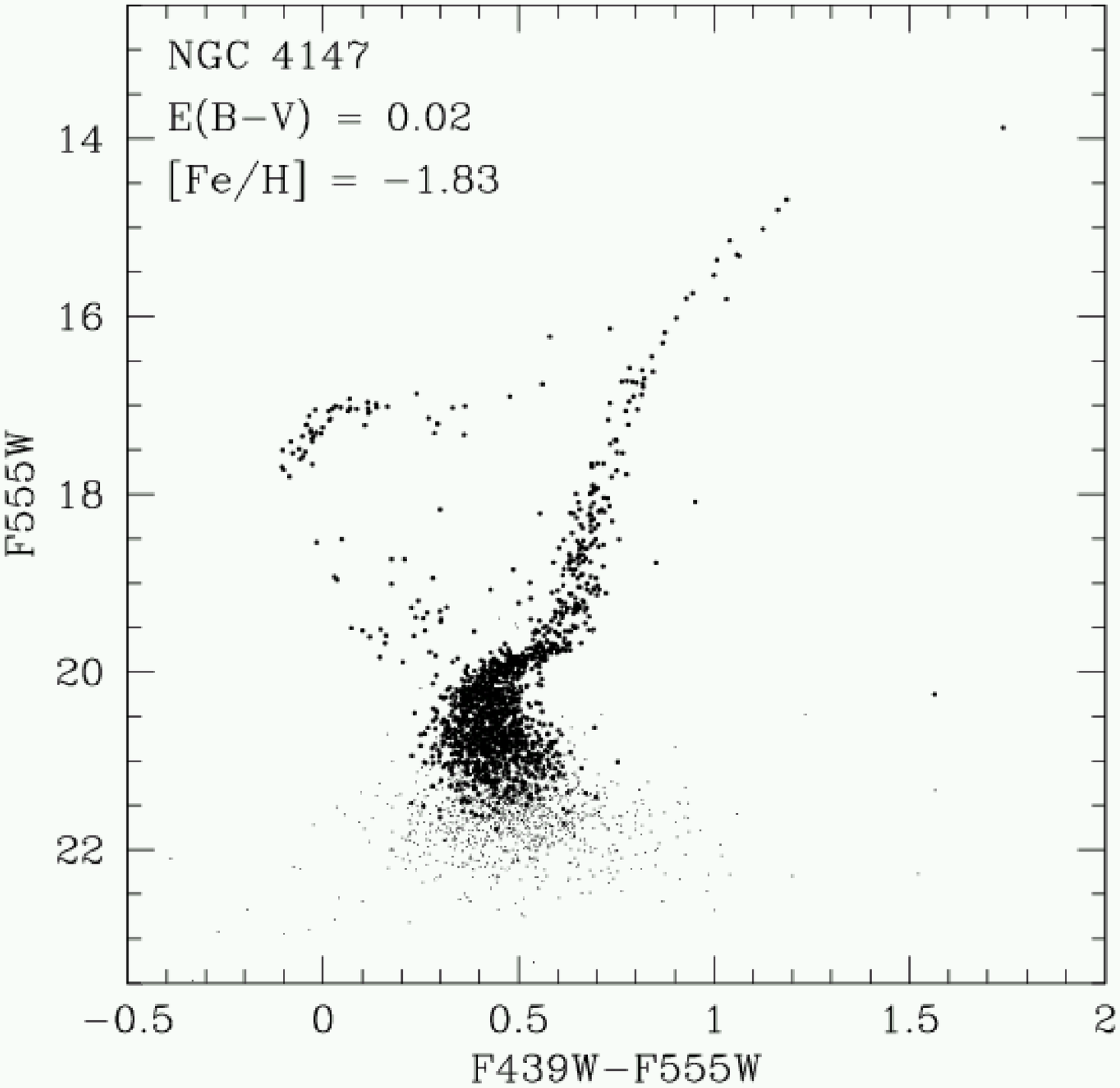}} &
\resizebox*{0.9\columnwidth}{0.36\height}{\includegraphics{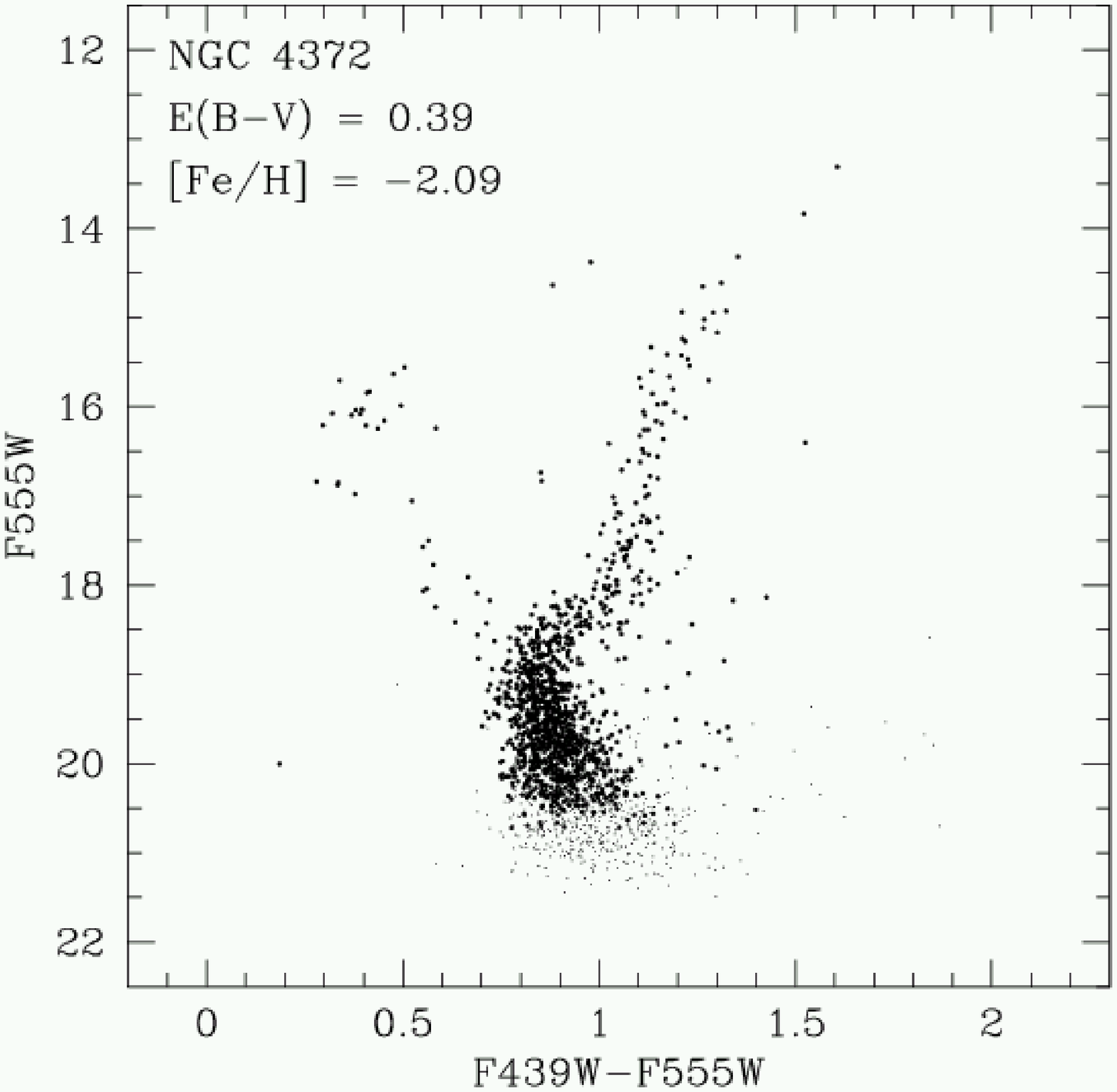}} \\
\resizebox*{0.9\columnwidth}{0.36\height}{\includegraphics{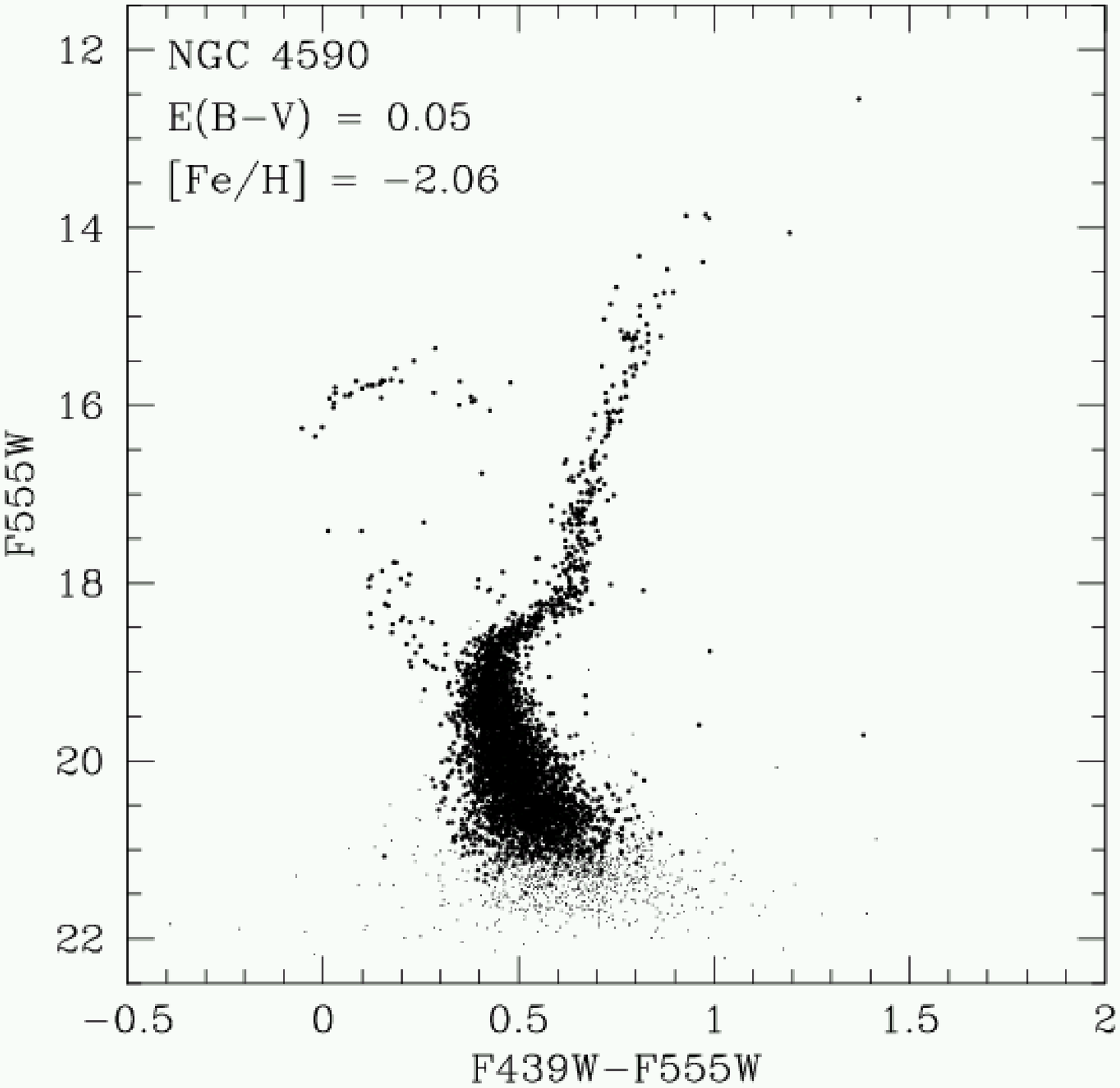}} &
\resizebox*{0.9\columnwidth}{0.36\height}{\includegraphics{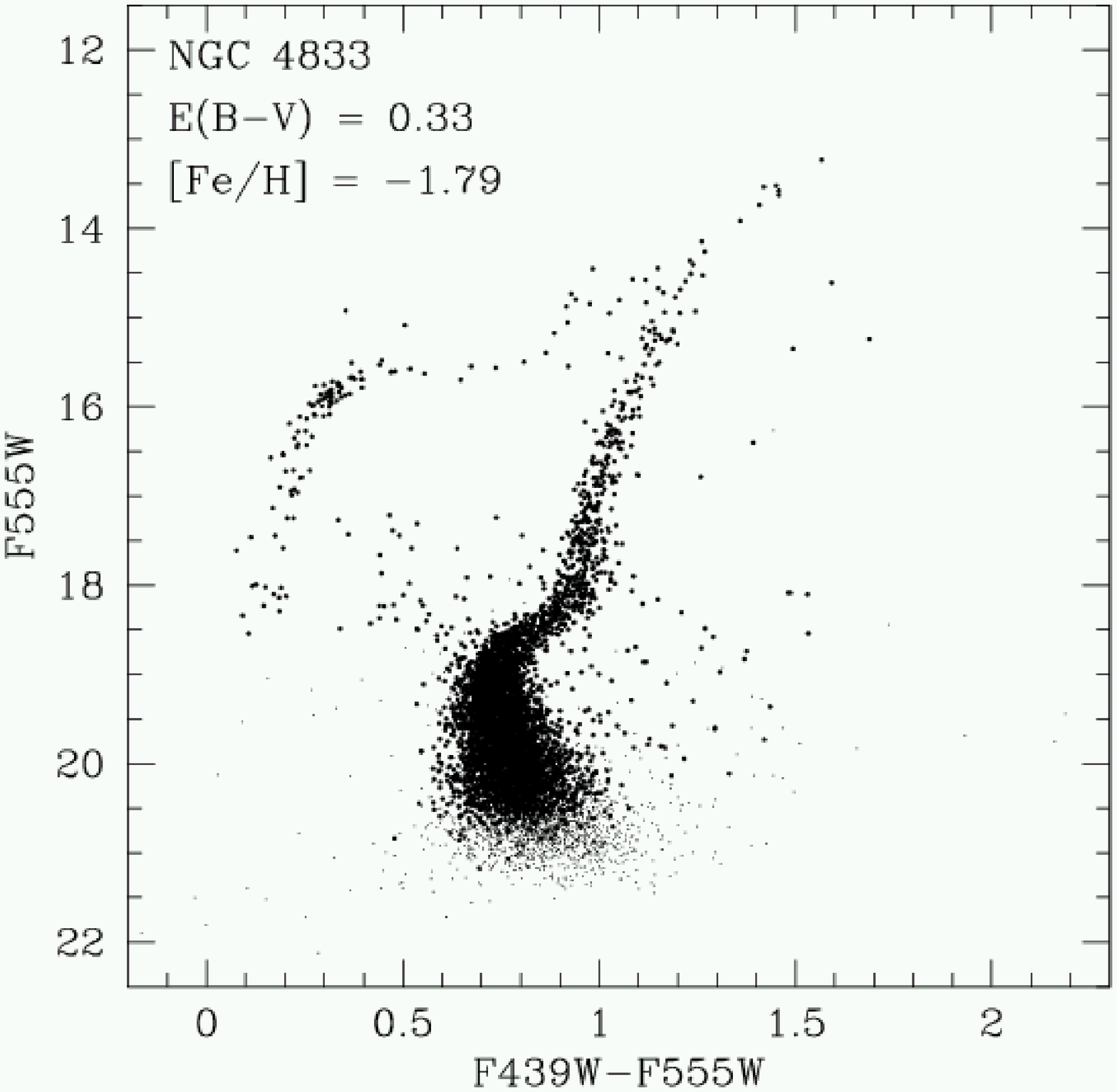}} \\
\end{tabular}
\caption{The color magnitude diagrams (cont.).}
\end{figure*}

\begin{figure*}
\begin{tabular}{cc}
\resizebox*{0.9\columnwidth}{0.36\height}{\includegraphics{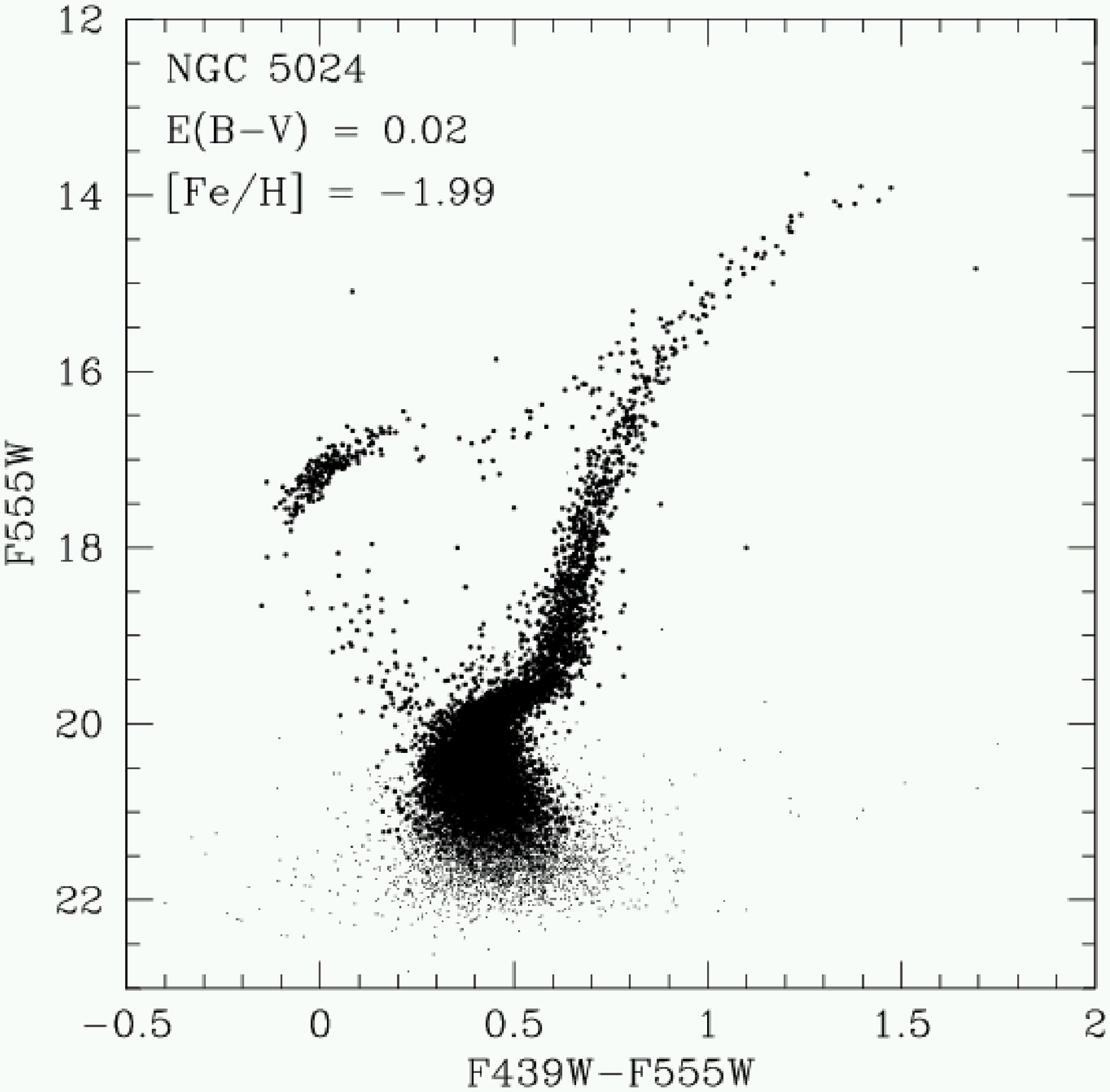}} &
\resizebox*{0.9\columnwidth}{0.36\height}{\includegraphics{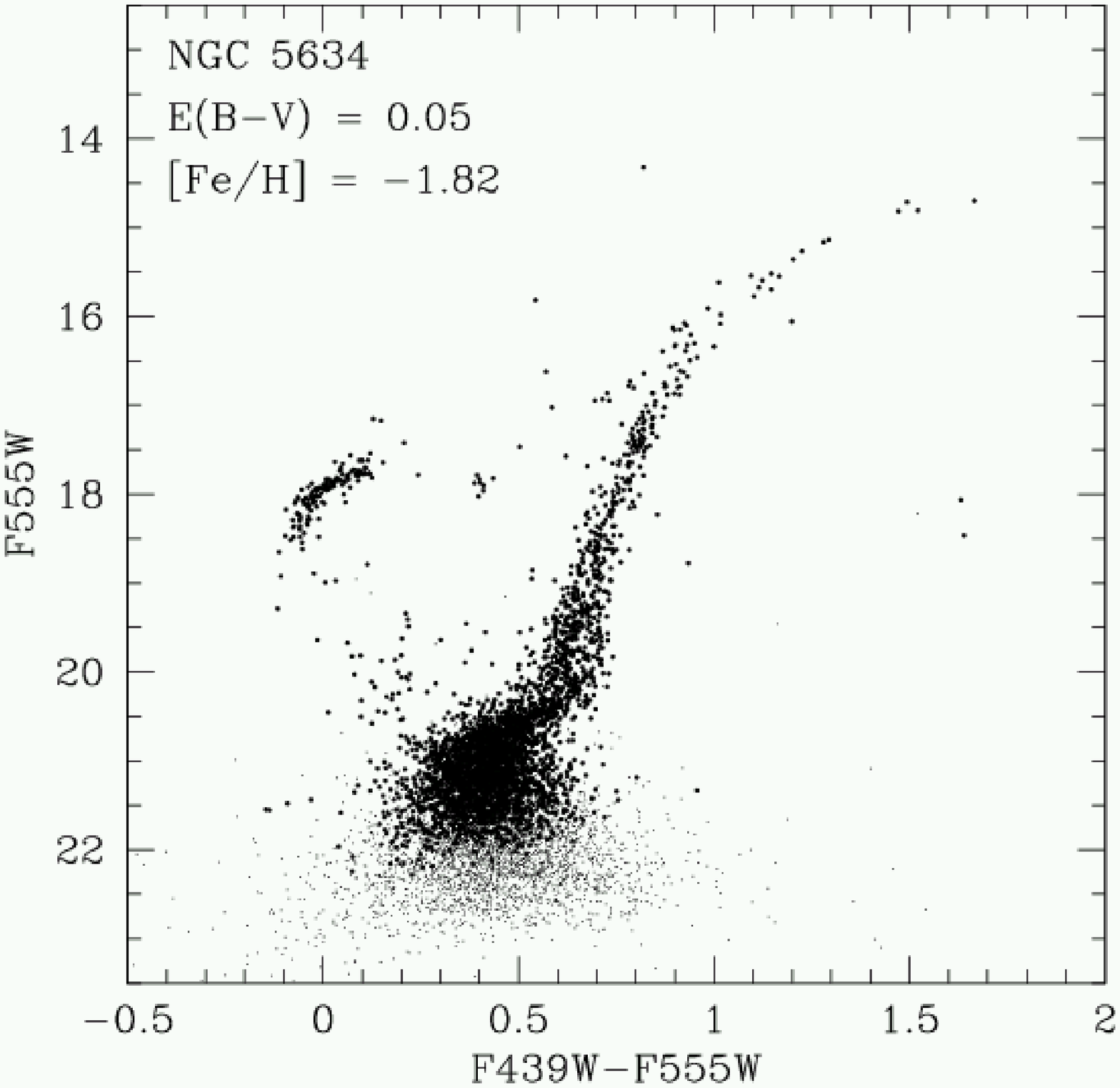}} \\
\resizebox*{0.9\columnwidth}{0.36\height}{\includegraphics{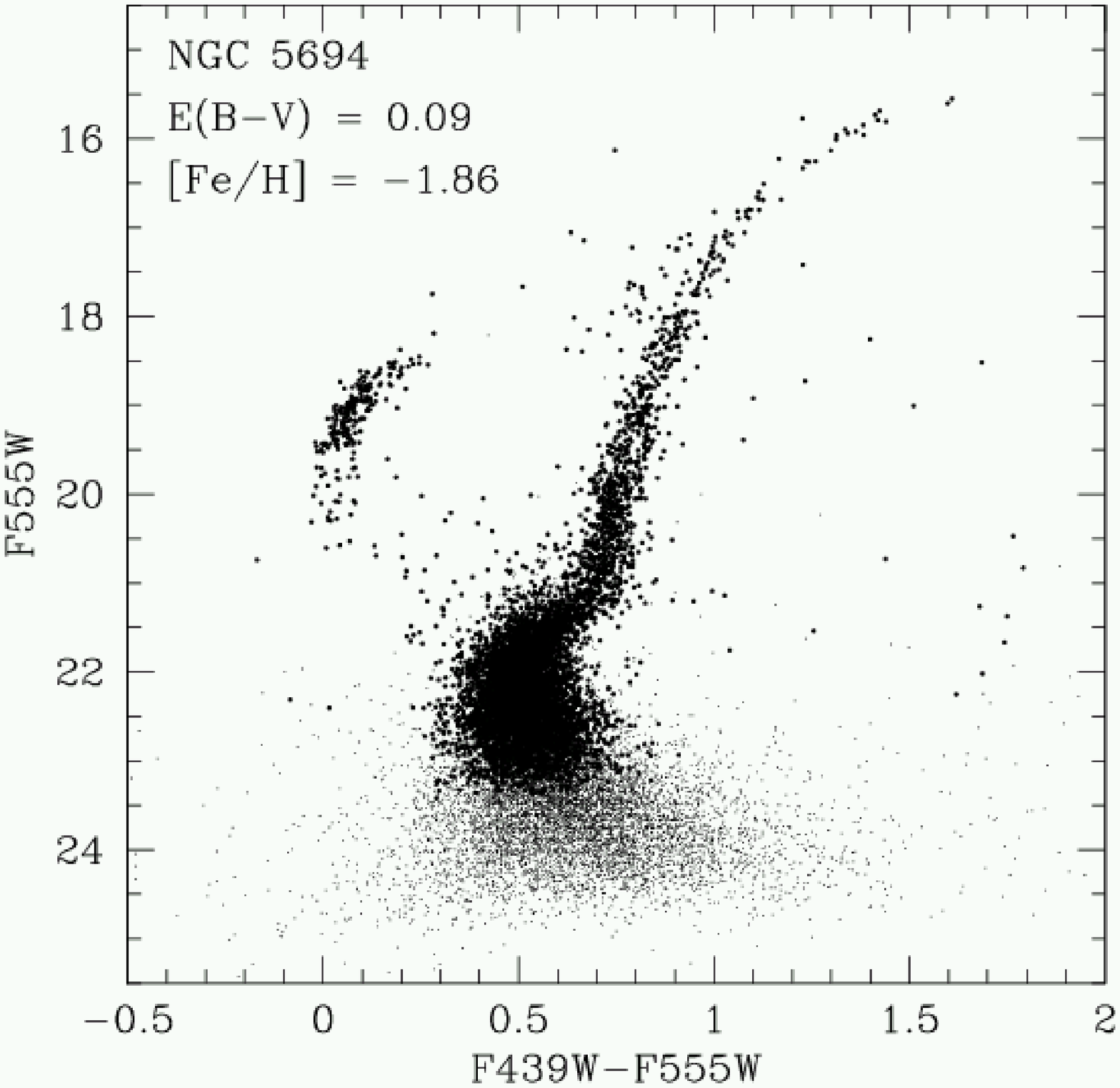}} &
\resizebox*{0.9\columnwidth}{0.36\height}{\includegraphics{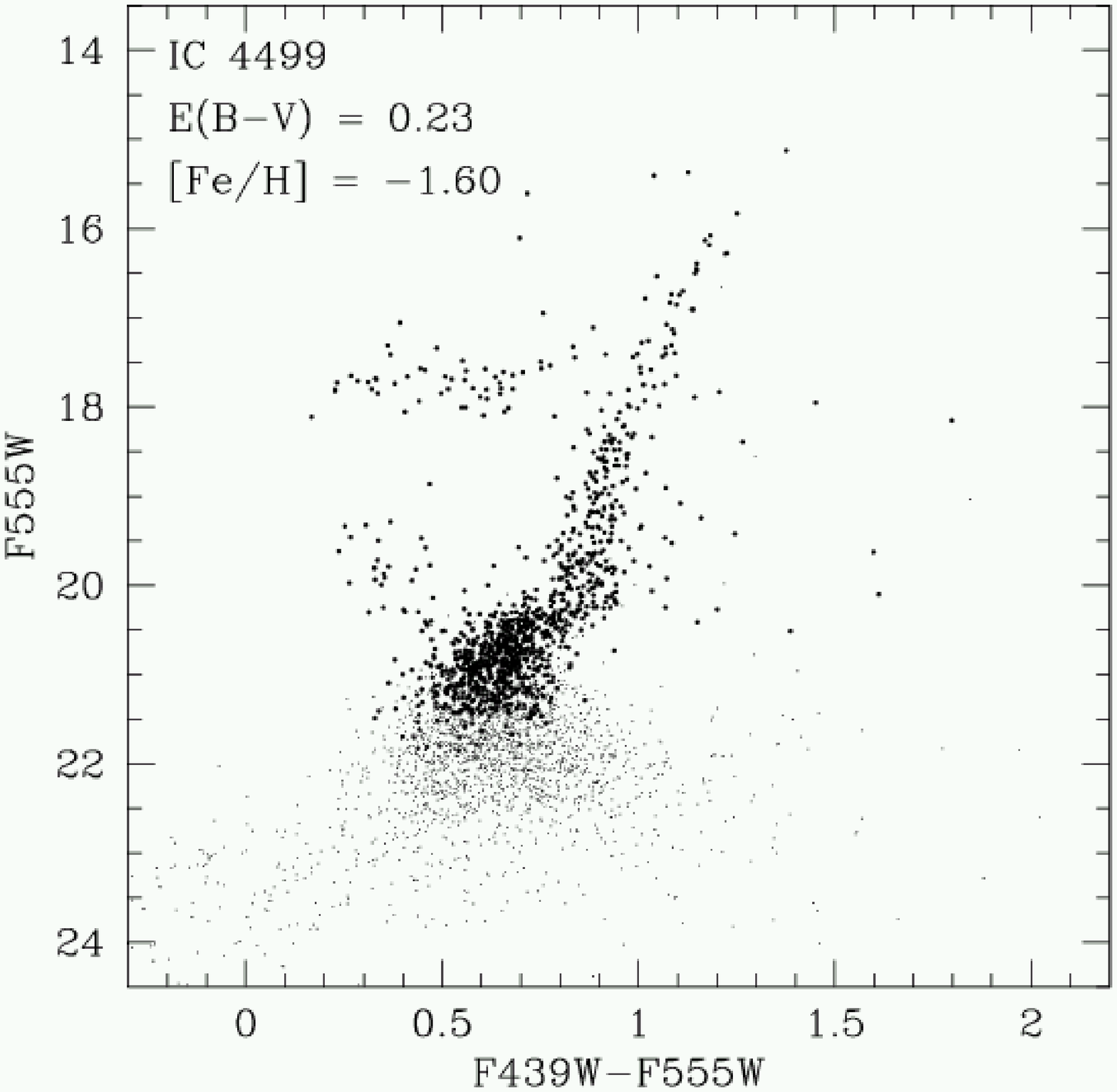}} \\
\resizebox*{0.9\columnwidth}{0.36\height}{\includegraphics{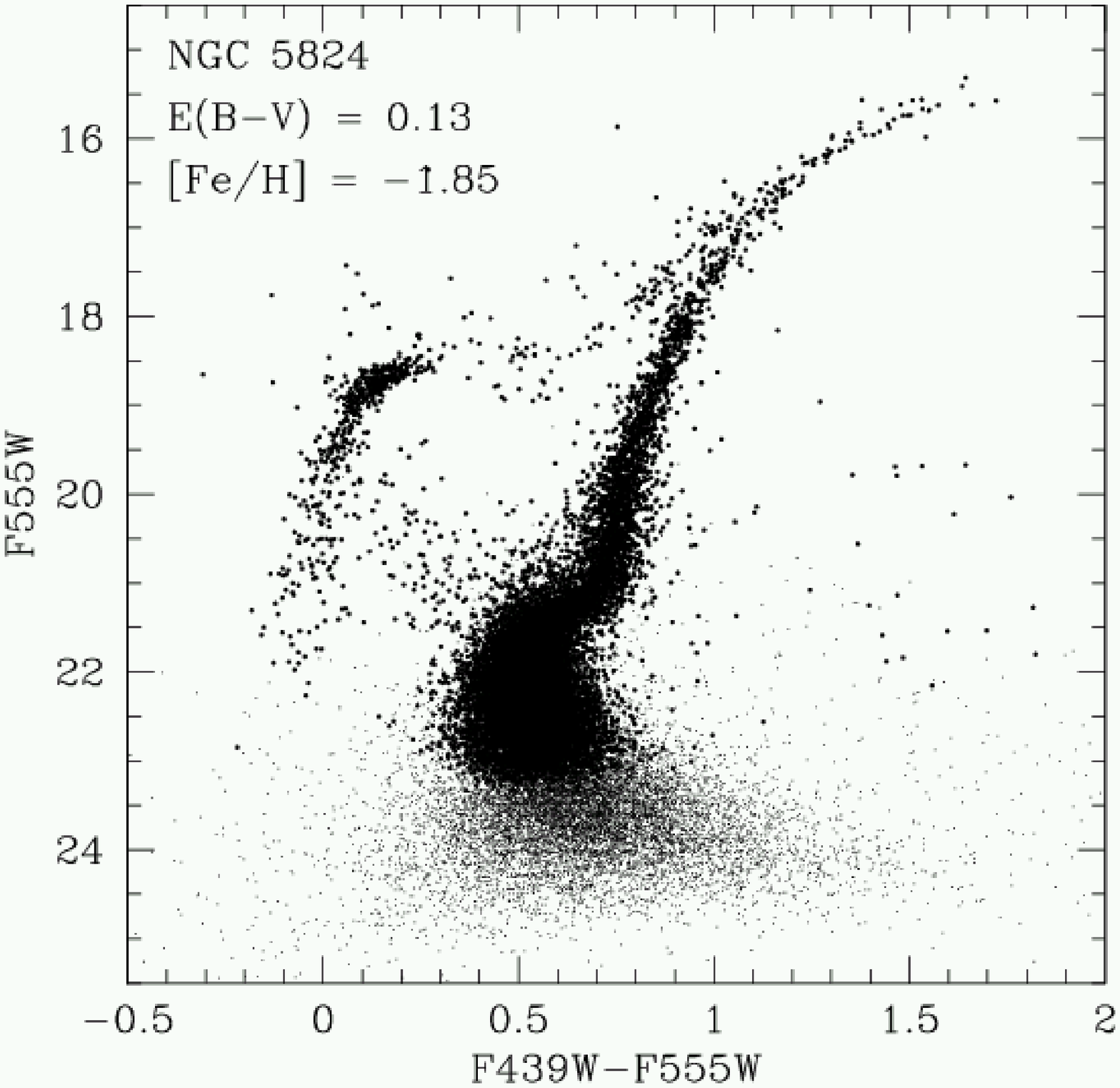}} &
\resizebox*{0.9\columnwidth}{0.36\height}{\includegraphics{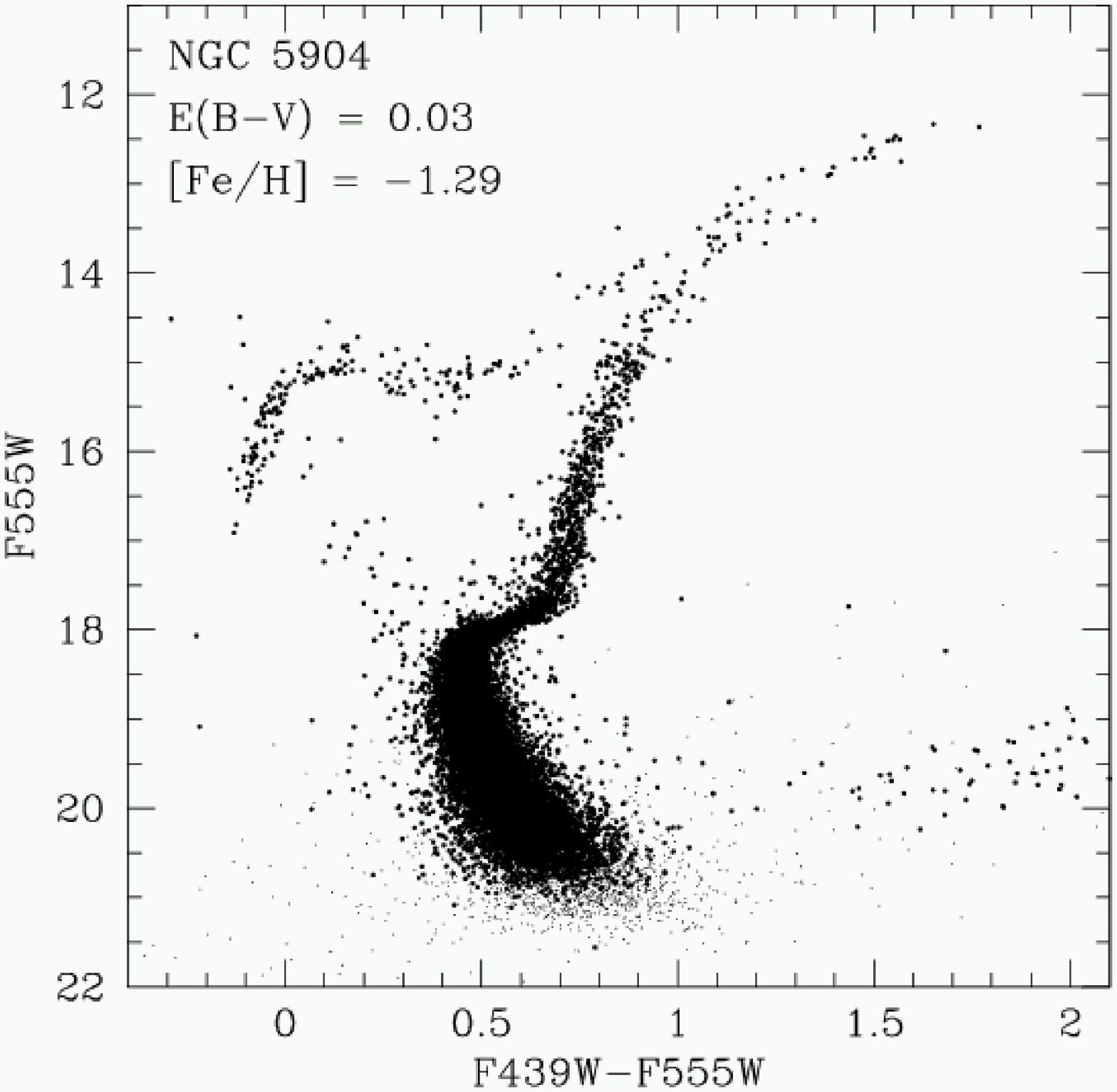}} \\
\end{tabular}
\caption{The color magnitude diagrams (cont.).}
\end{figure*}

\begin{figure*}
\begin{tabular}{cc}
\resizebox*{0.9\columnwidth}{0.36\height}{\includegraphics{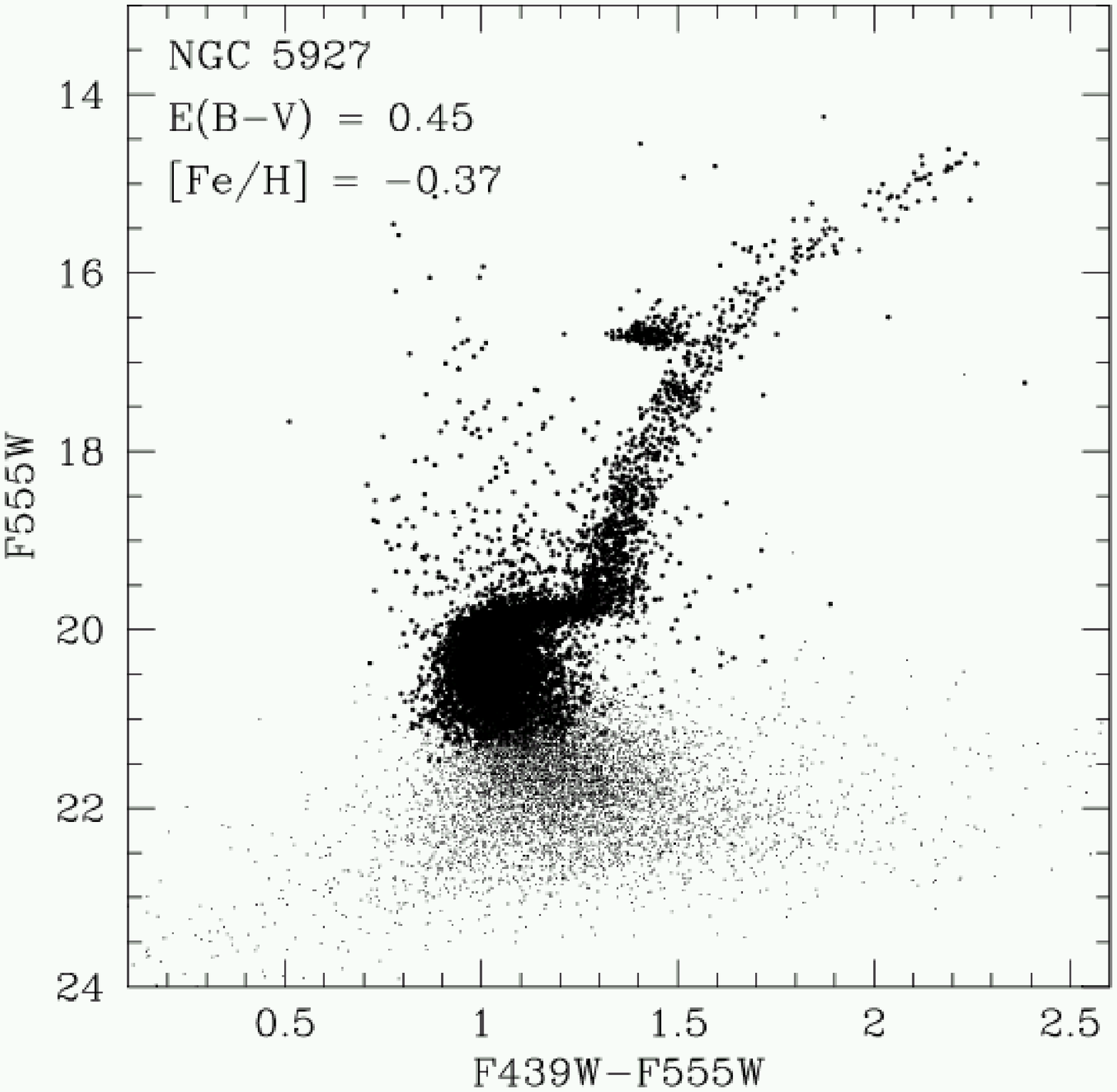}} &
\resizebox*{0.9\columnwidth}{0.36\height}{\includegraphics{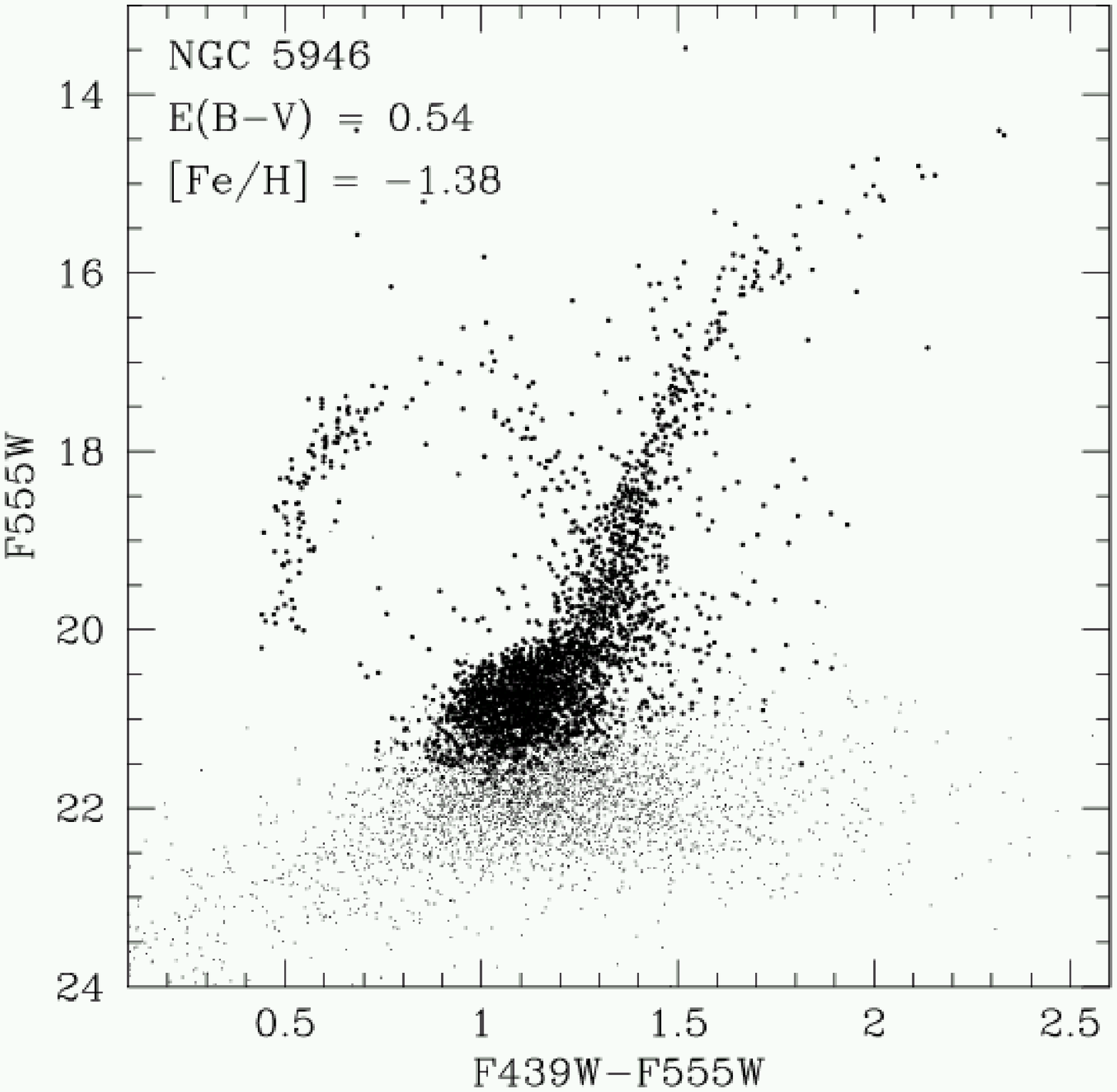}} \\
\resizebox*{0.9\columnwidth}{0.36\height}{\includegraphics{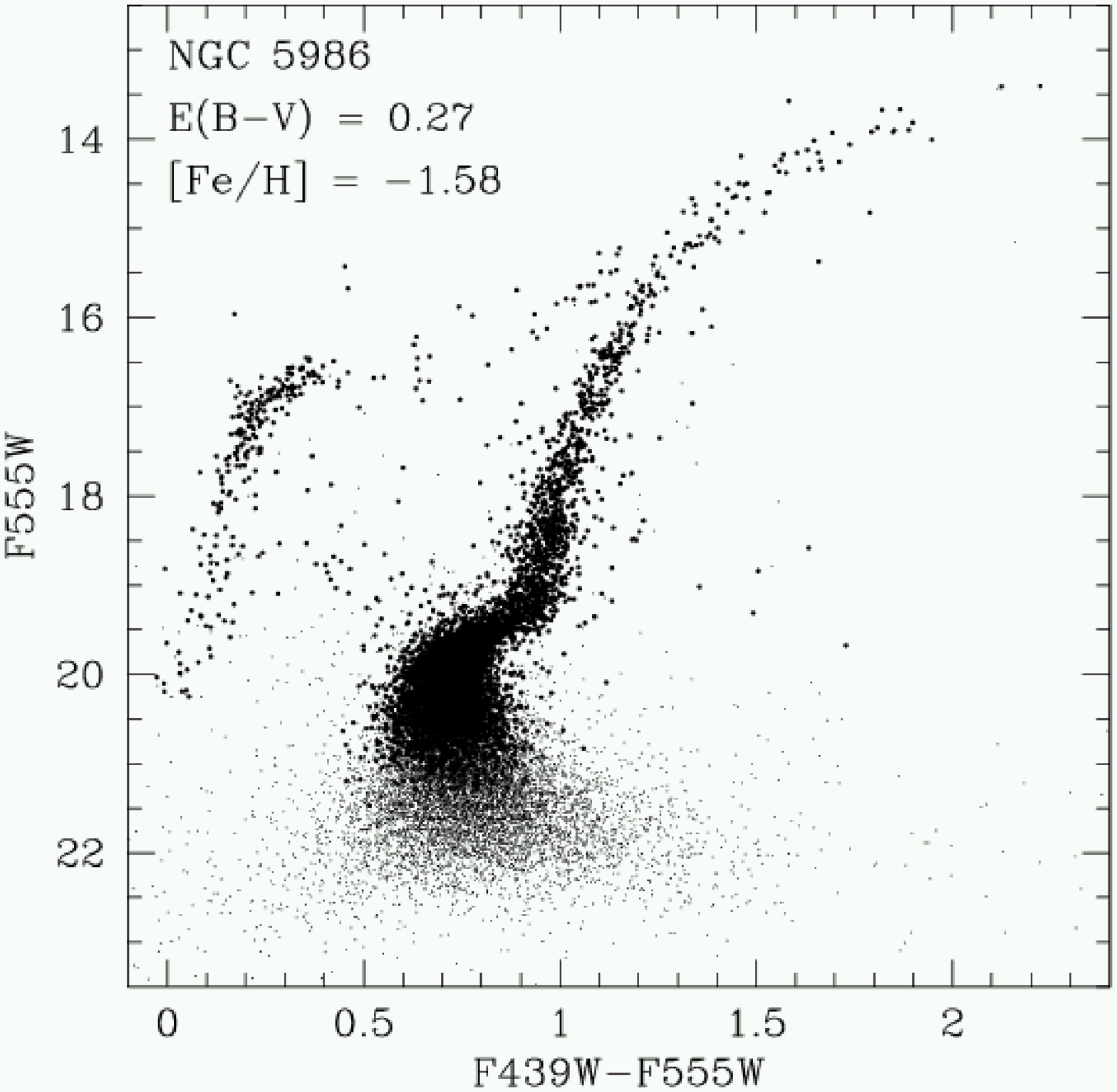}} &
\resizebox*{0.9\columnwidth}{0.36\height}{\includegraphics{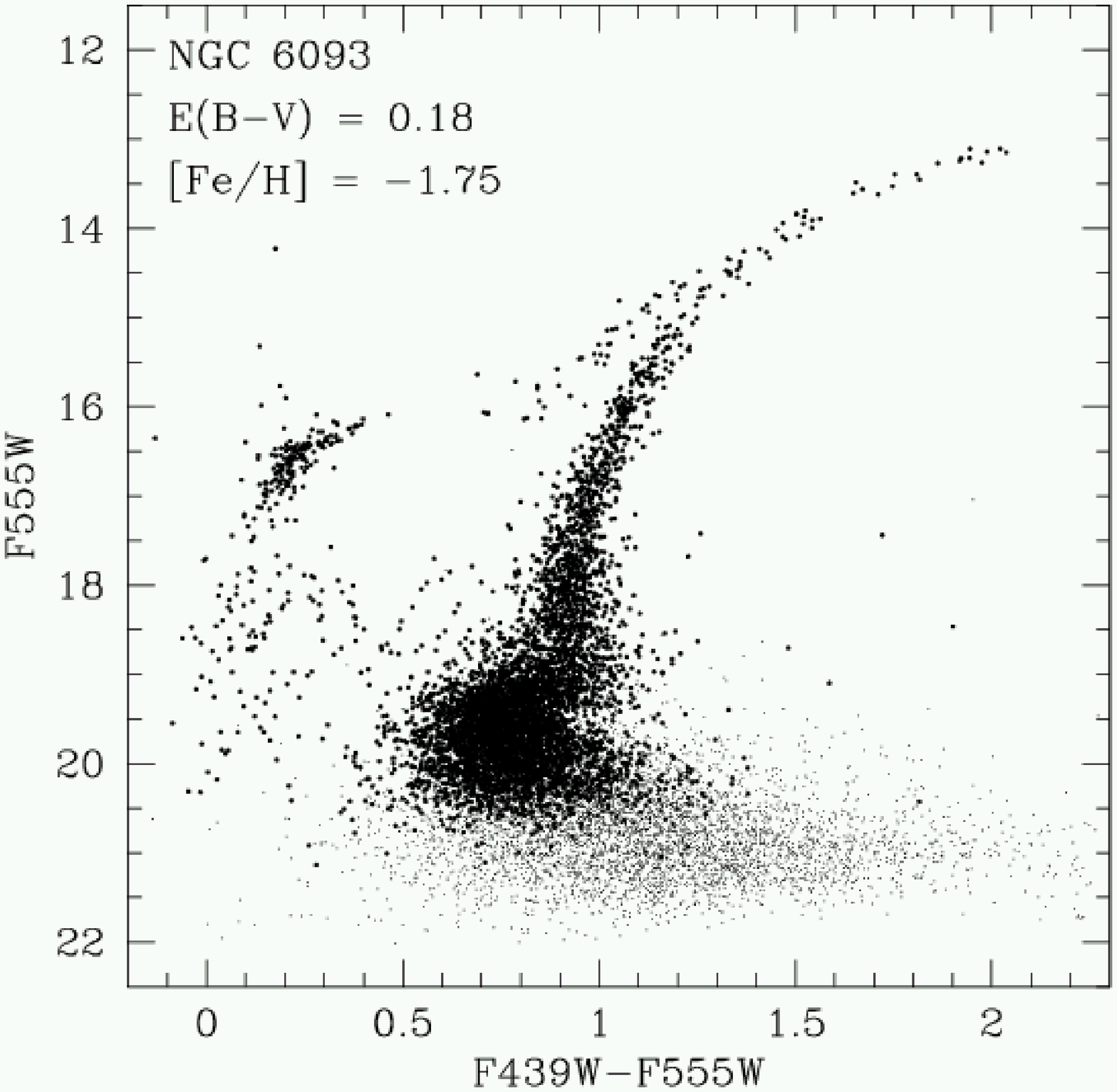}} \\
\resizebox*{0.9\columnwidth}{0.36\height}{\includegraphics{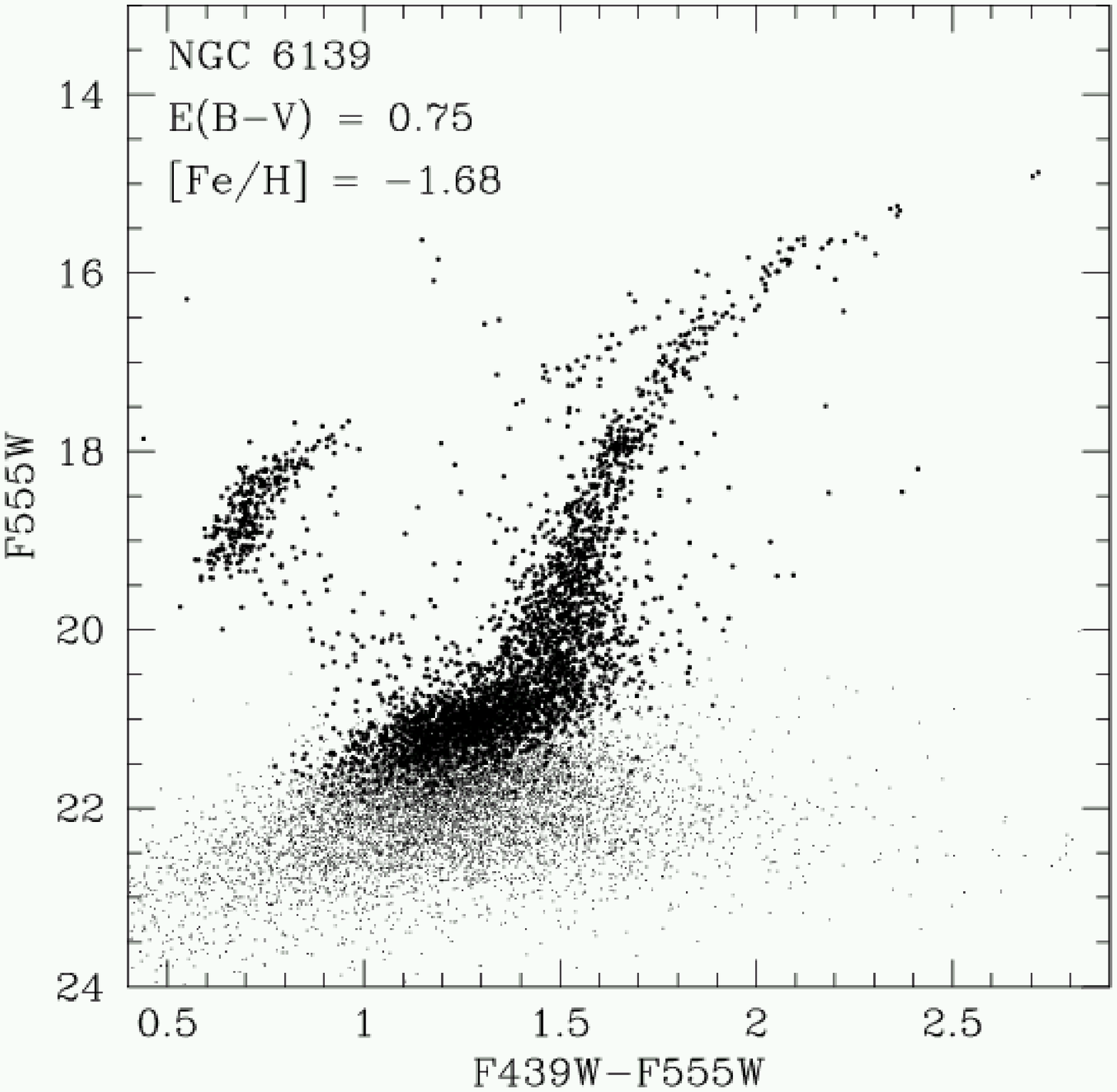}} &
\resizebox*{0.9\columnwidth}{0.36\height}{\includegraphics{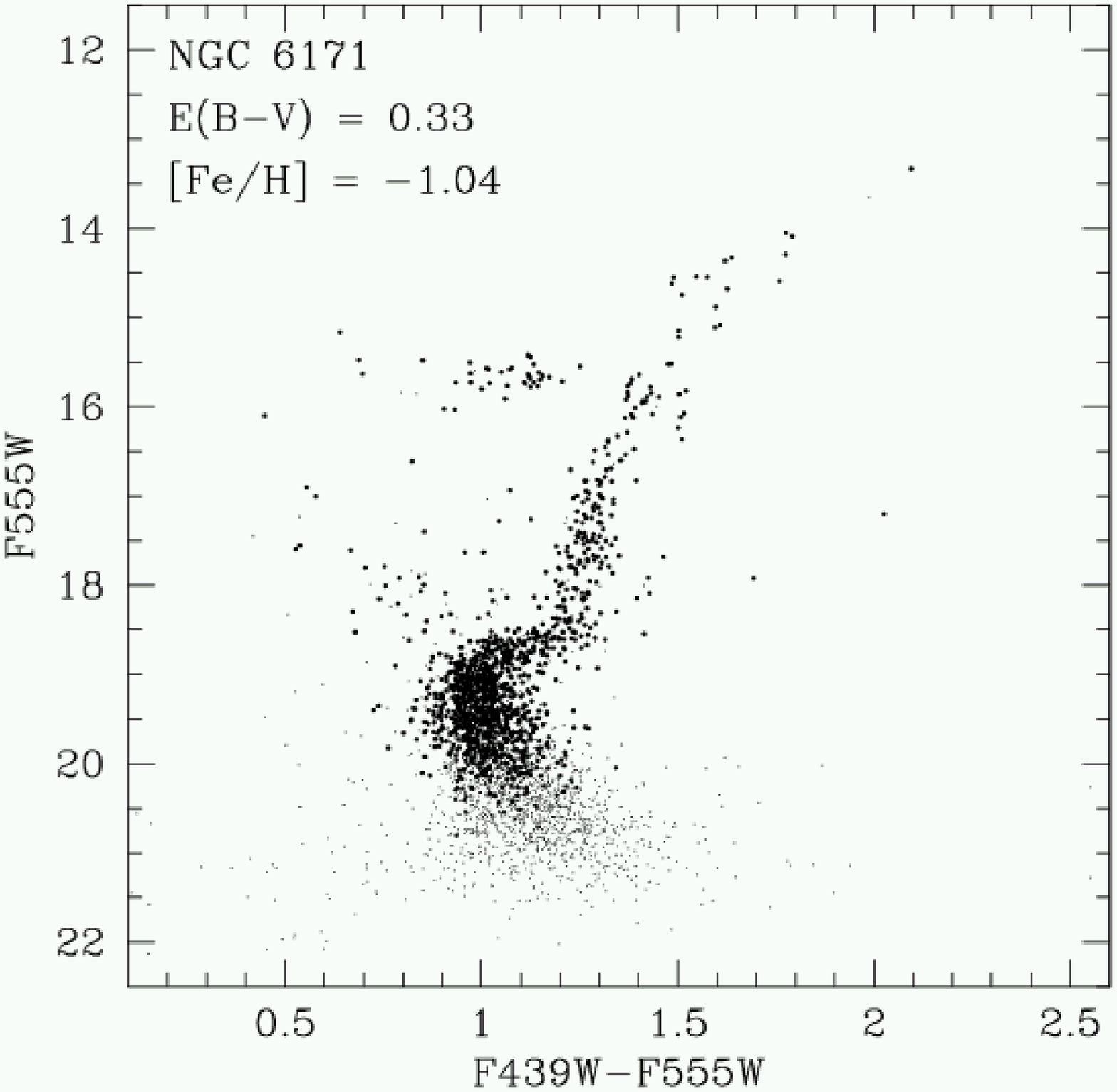}} \\
\end{tabular}
\caption{The color magnitude diagrams (cont.).}
\end{figure*}

\begin{figure*}
\begin{tabular}{cc}
\resizebox*{0.9\columnwidth}{0.36\height}{\includegraphics{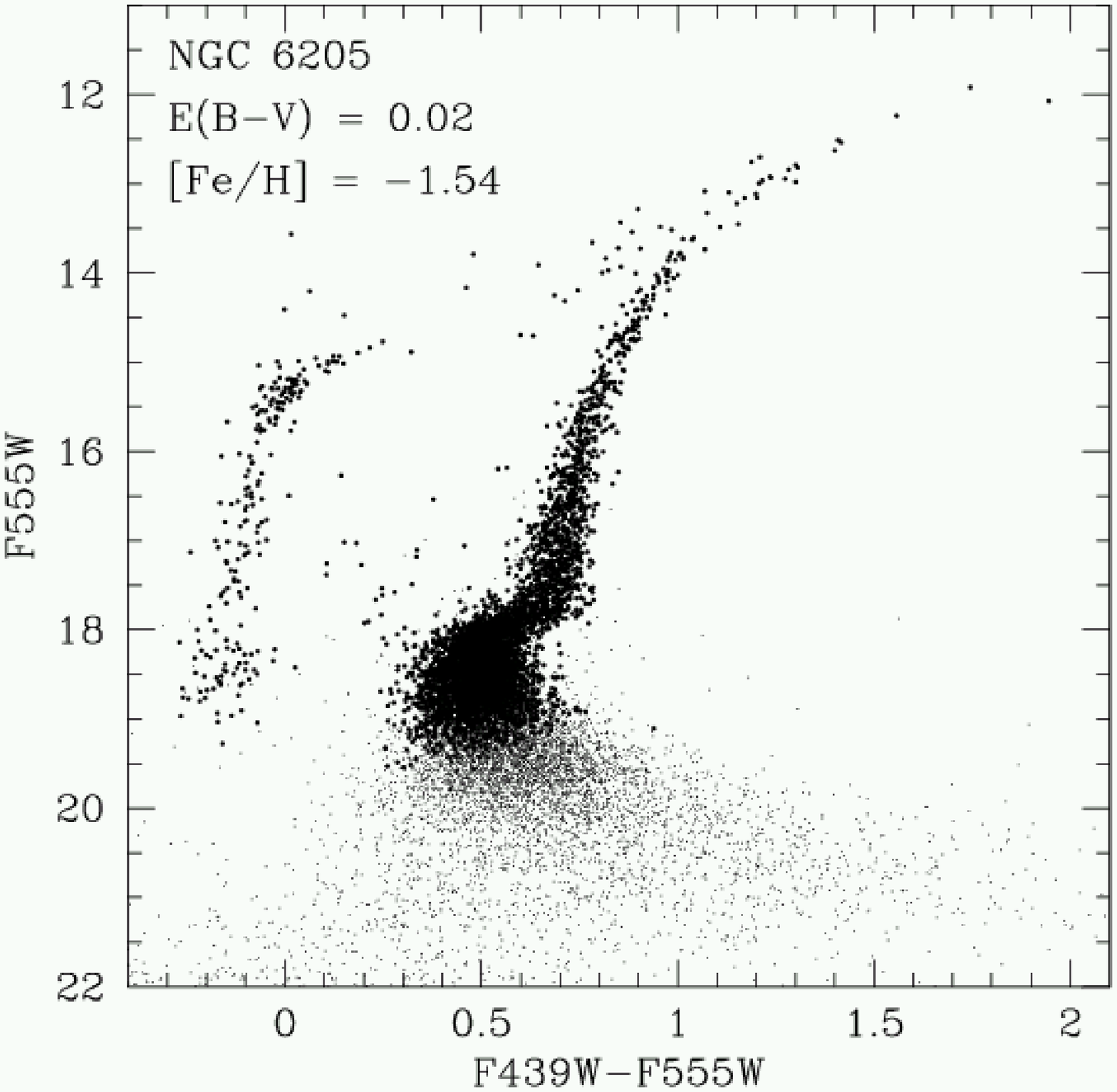}} &
\resizebox*{0.9\columnwidth}{0.36\height}{\includegraphics{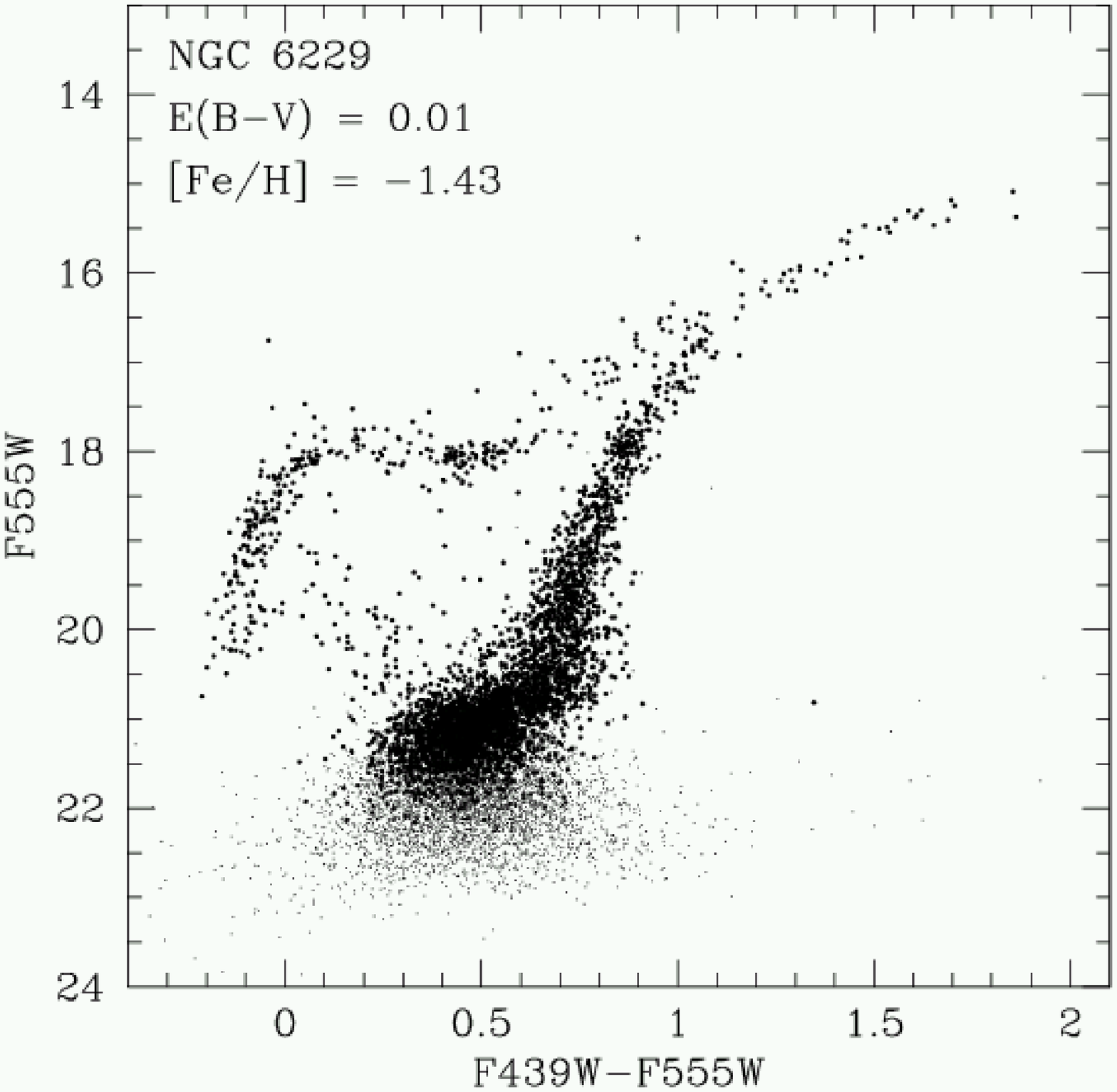}} \\
\resizebox*{0.9\columnwidth}{0.36\height}{\includegraphics{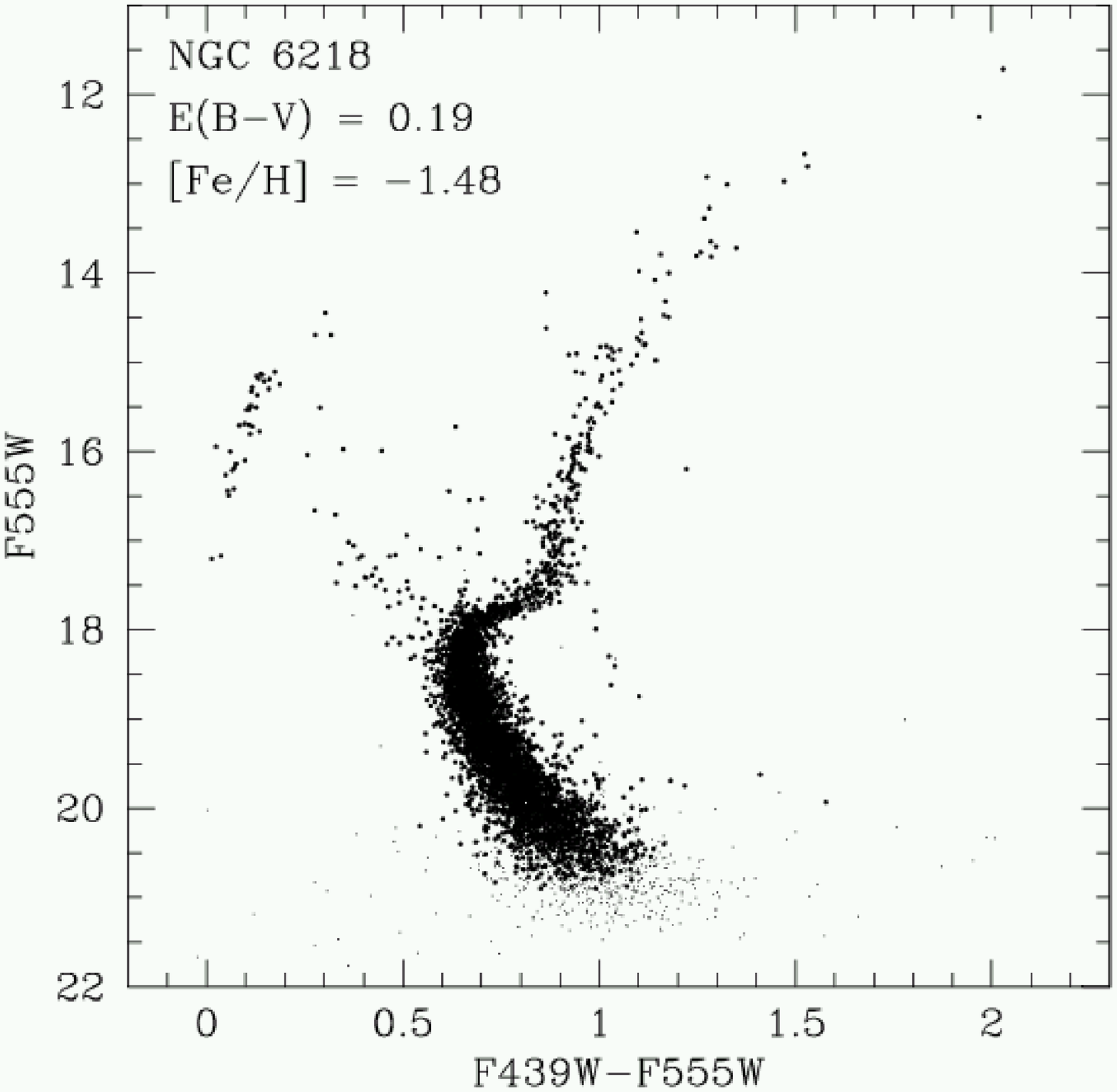}} &
\resizebox*{0.9\columnwidth}{0.36\height}{\includegraphics{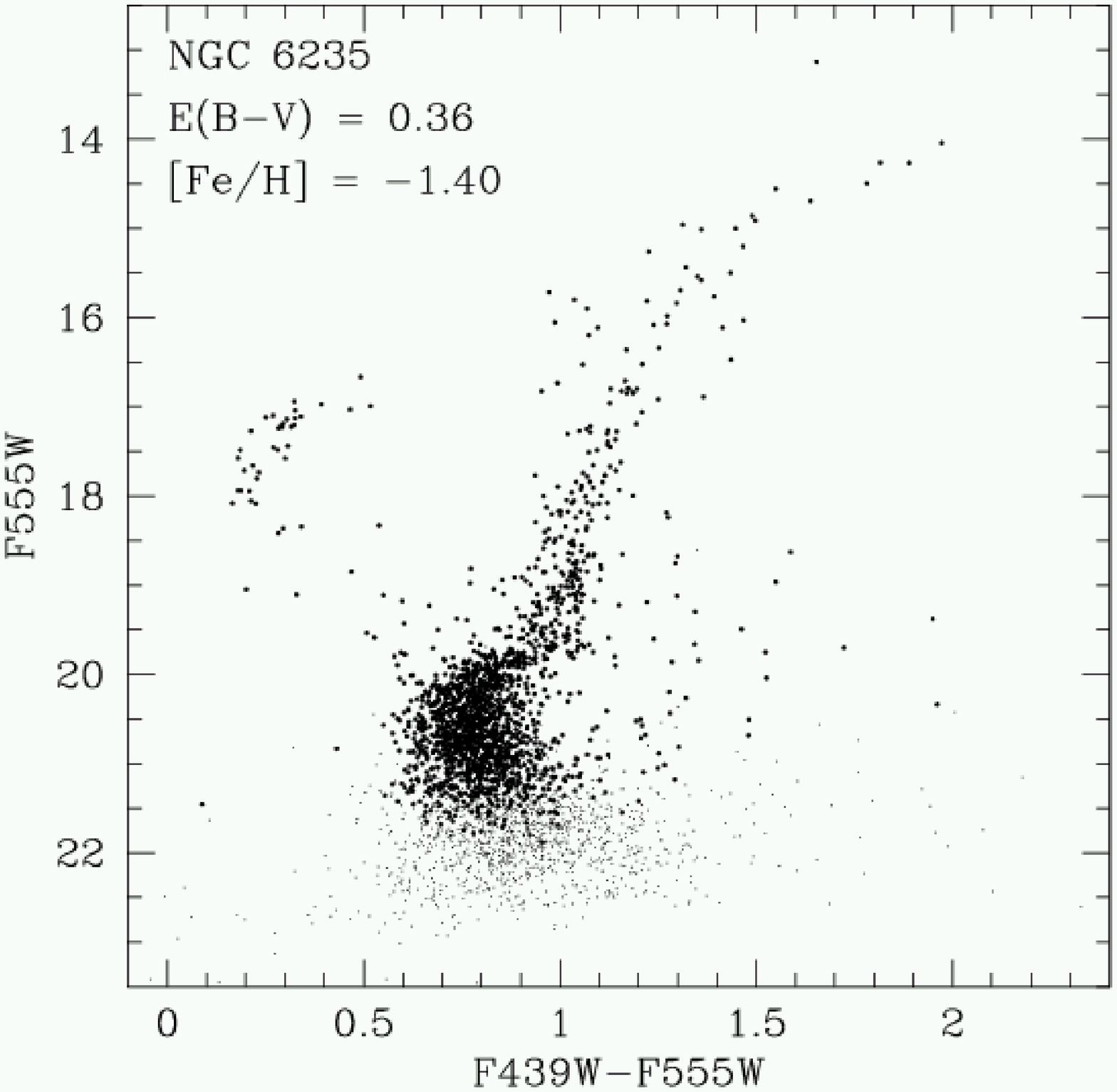}} \\
\resizebox*{0.9\columnwidth}{0.36\height}{\includegraphics{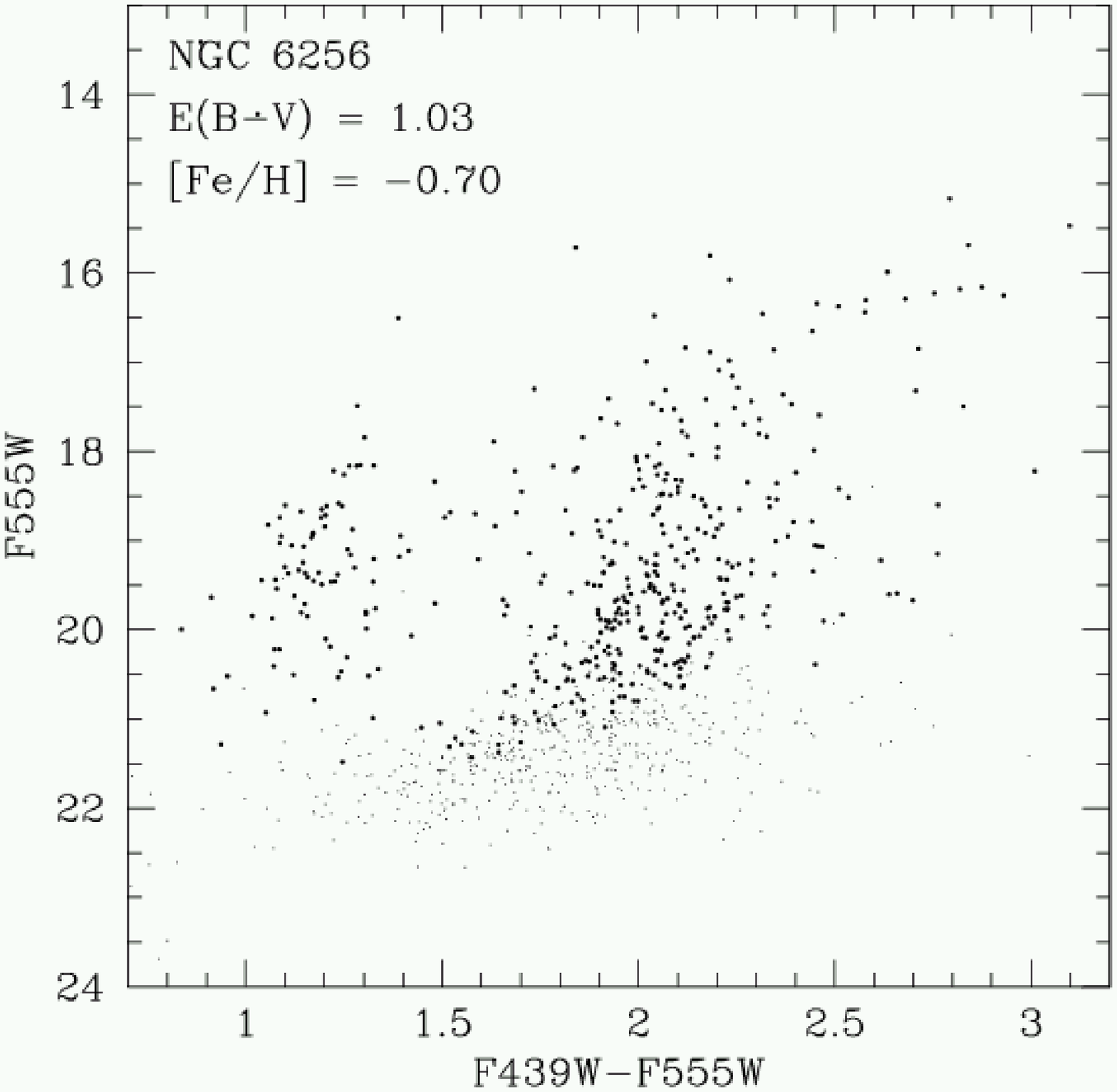}} &
\resizebox*{0.9\columnwidth}{0.36\height}{\includegraphics{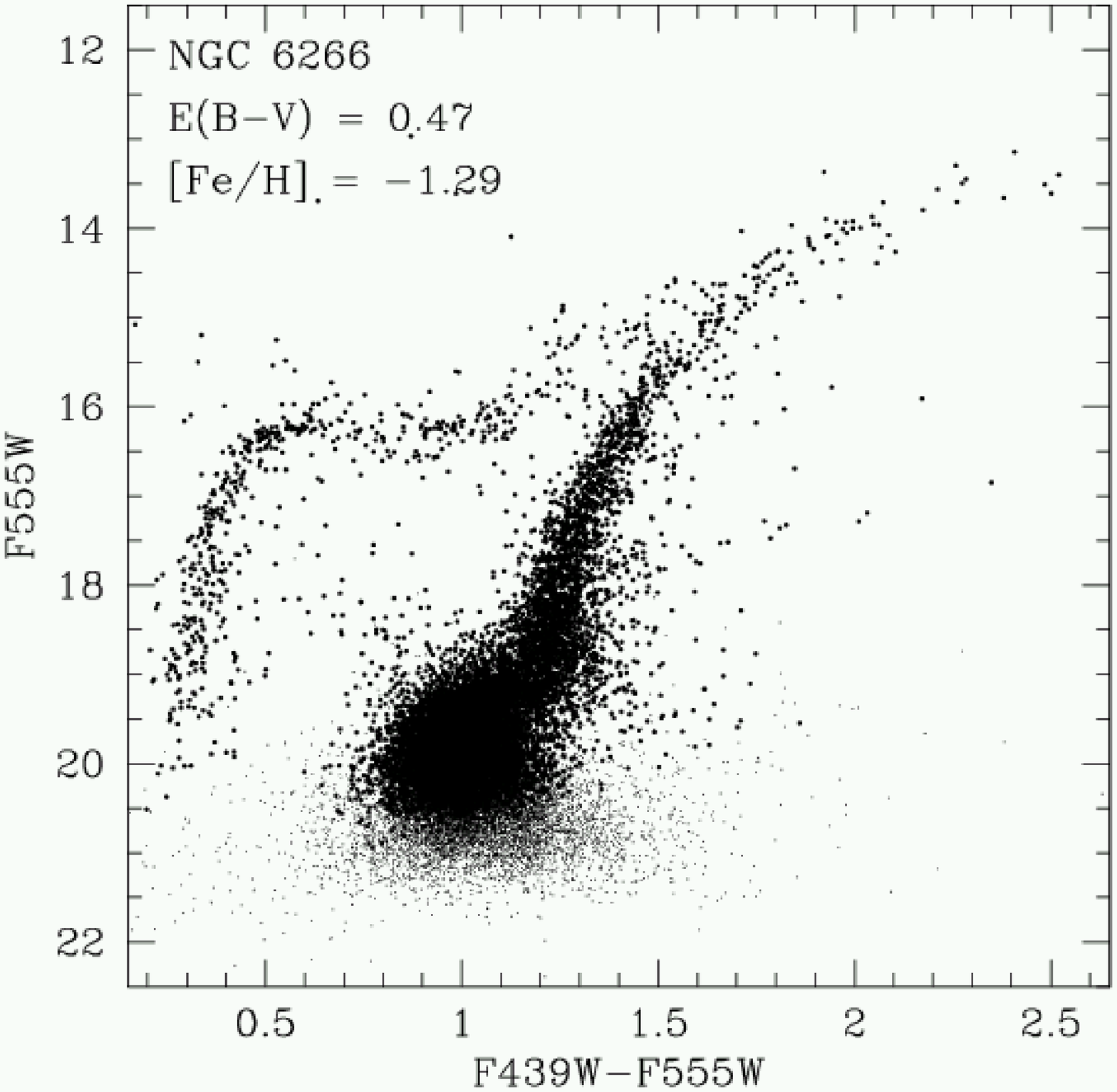}} \\
\end{tabular}
\caption{The color magnitude diagrams (cont.)}
\end{figure*}

\begin{figure*}
\begin{tabular}{cc}
\resizebox*{0.9\columnwidth}{0.36\height}{\includegraphics{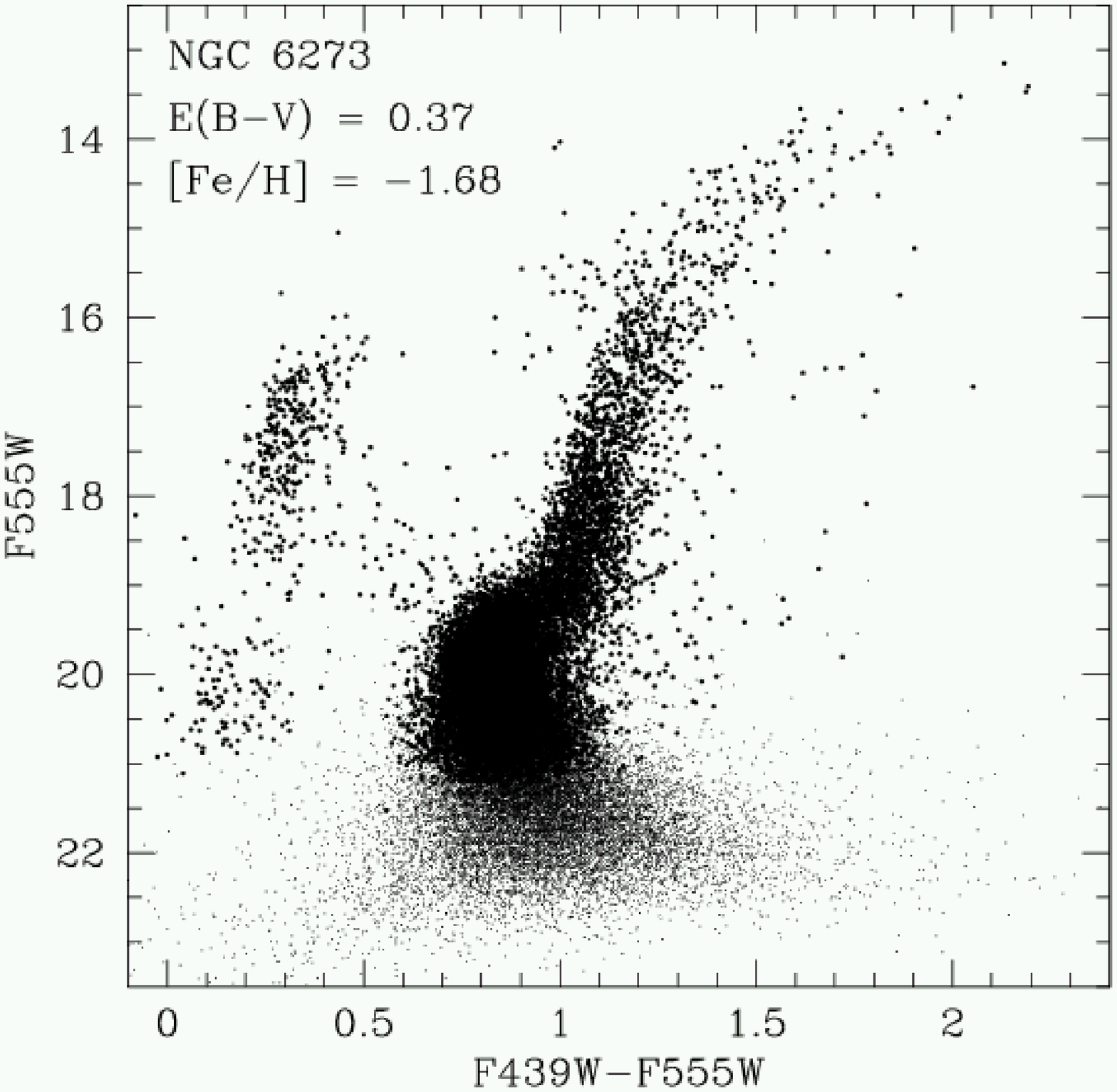}} &
\resizebox*{0.9\columnwidth}{0.36\height}{\includegraphics{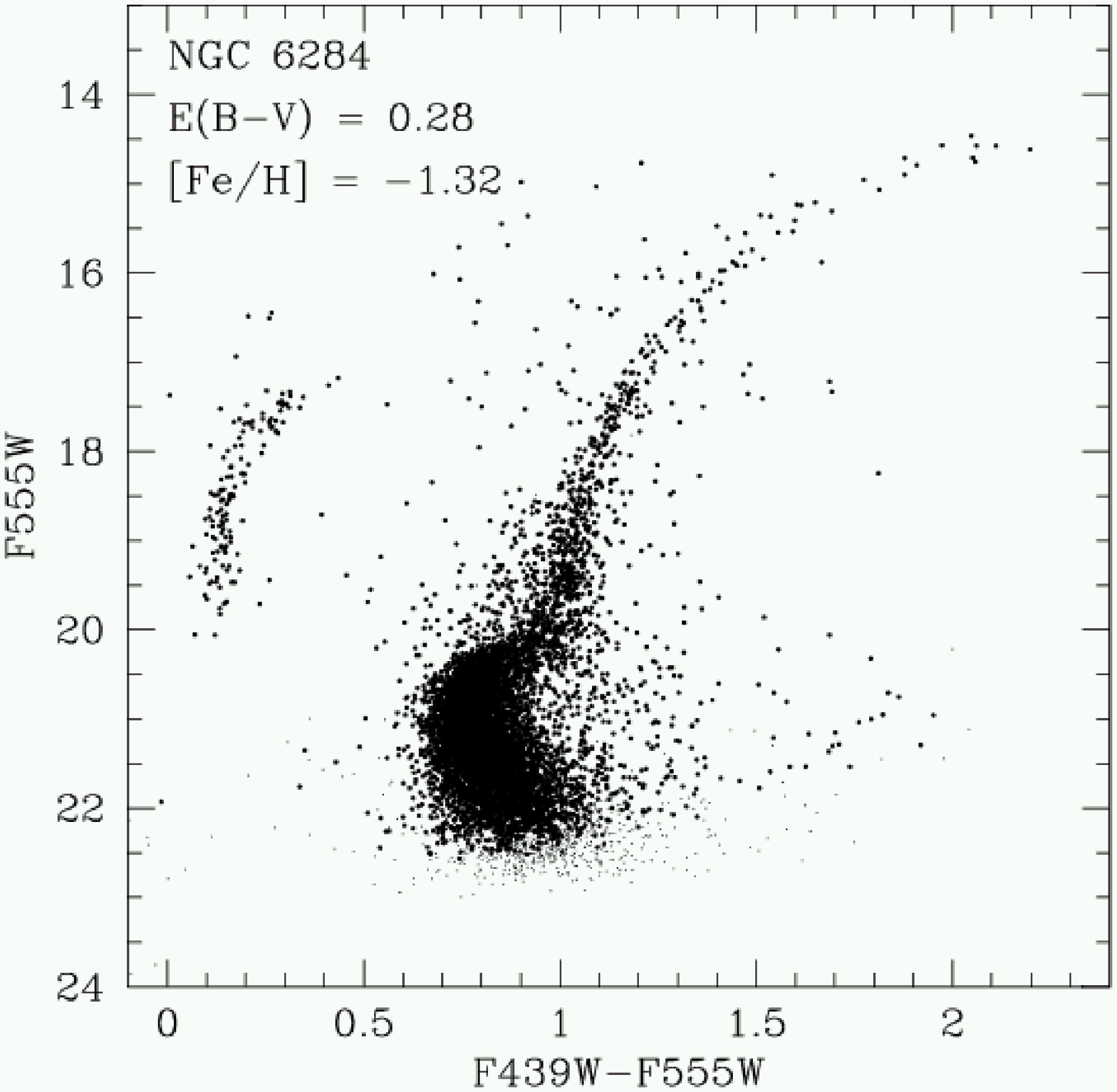}} \\
\resizebox*{0.9\columnwidth}{0.36\height}{\includegraphics{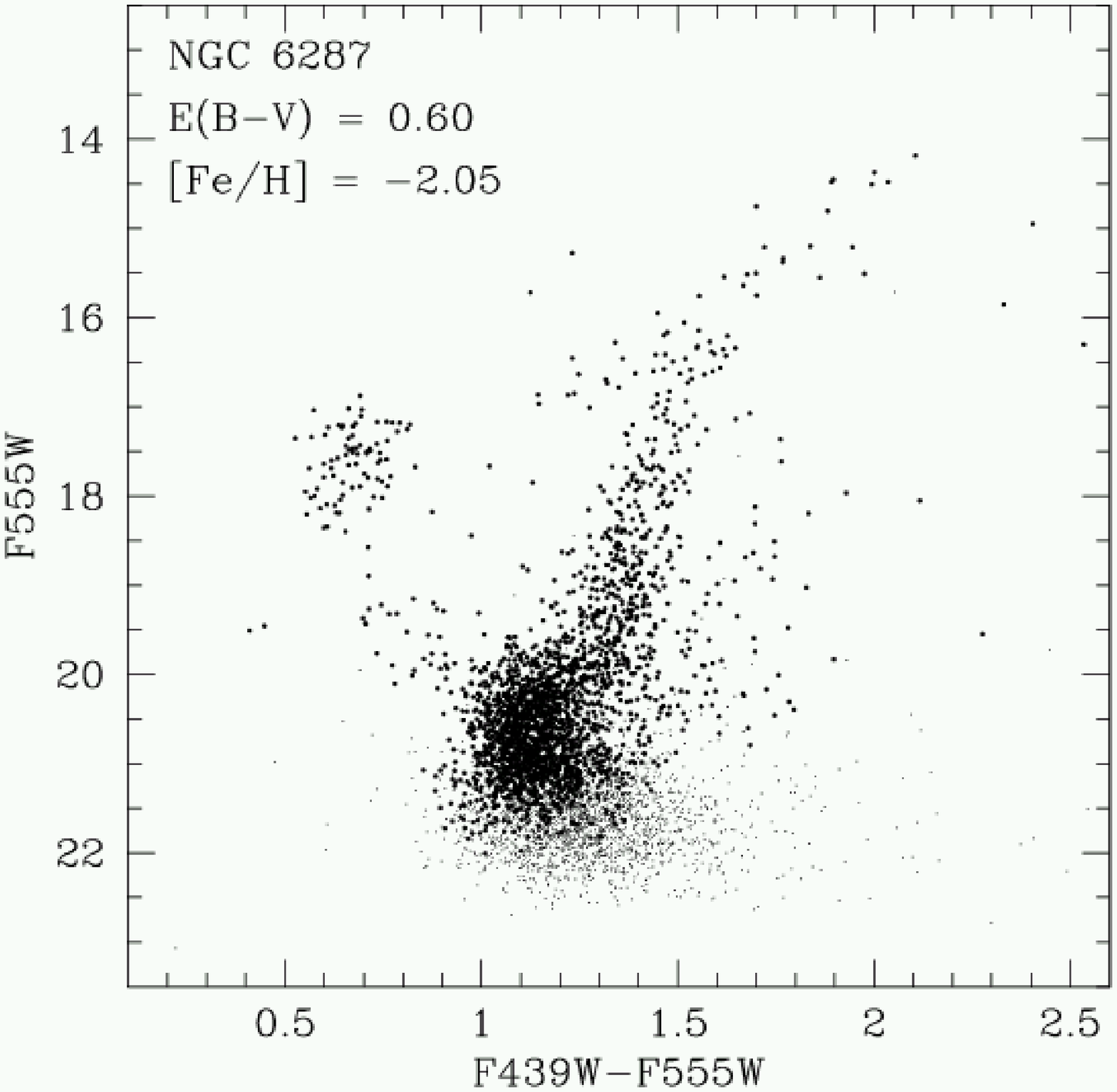}} &
\resizebox*{0.9\columnwidth}{0.36\height}{\includegraphics{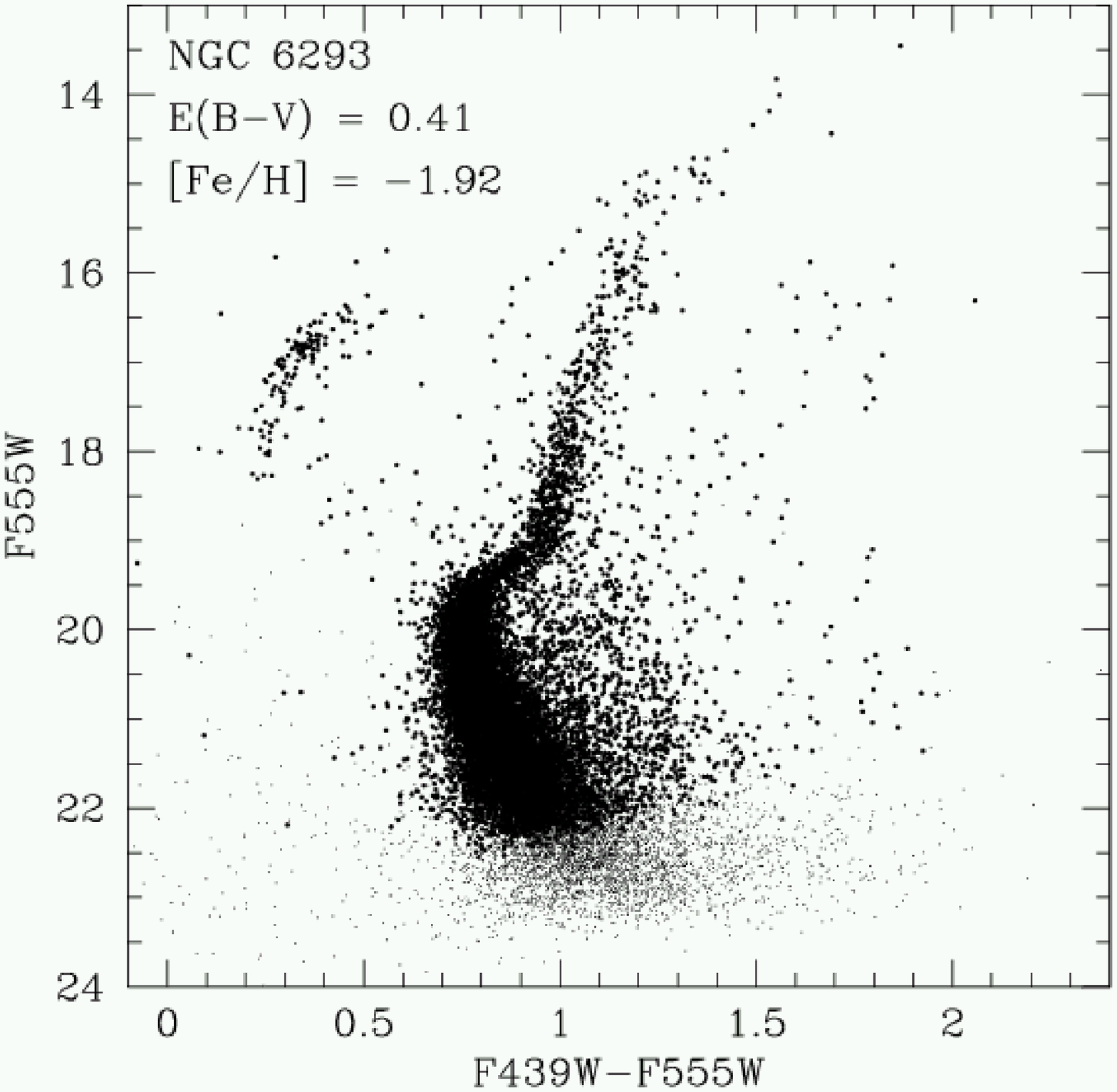}} \\
\resizebox*{0.9\columnwidth}{0.36\height}{\includegraphics{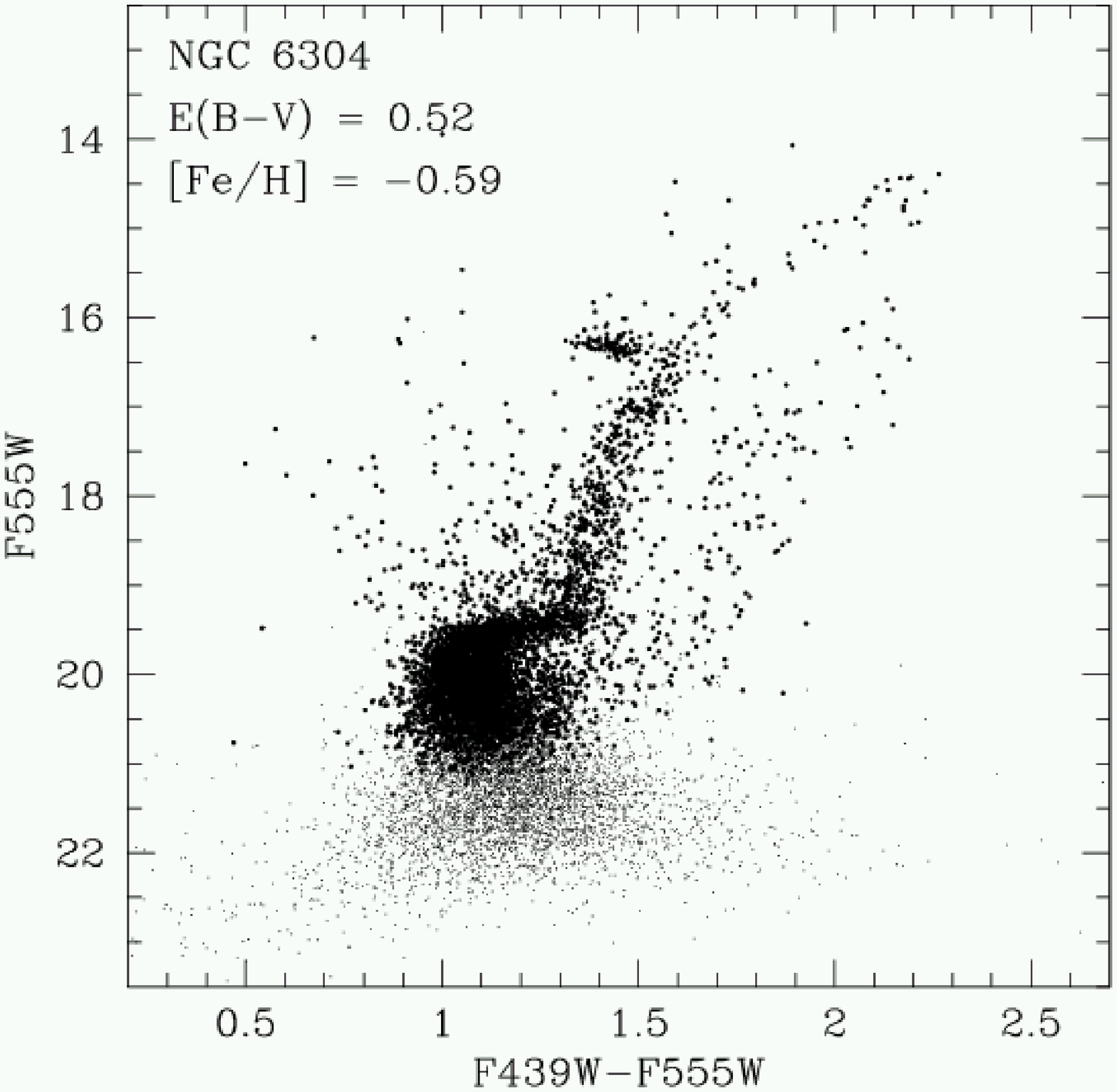}} &
\resizebox*{0.9\columnwidth}{0.36\height}{\includegraphics{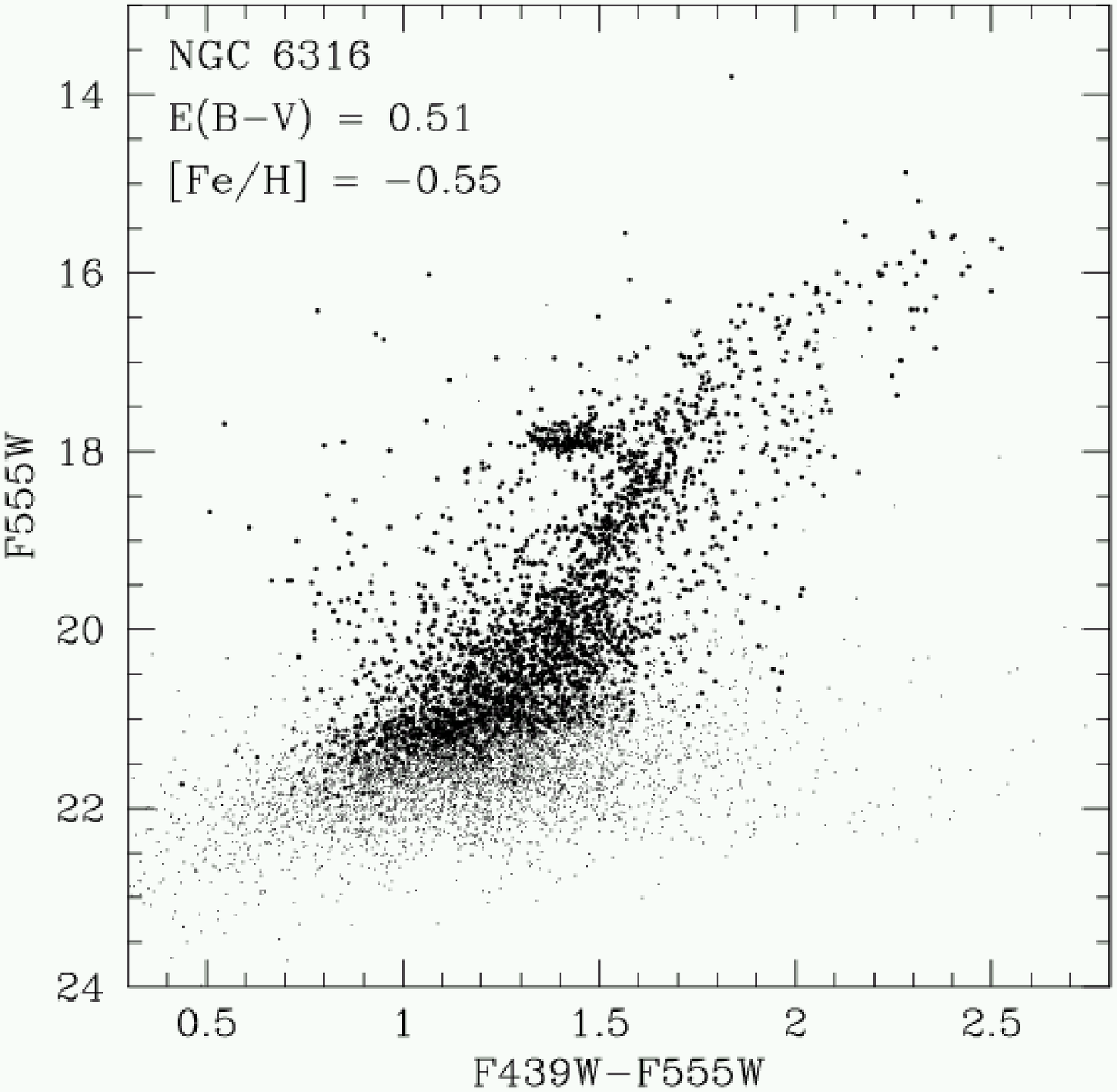}} \\
\end{tabular}
\caption{The color magnitude diagrams (cont.).}
\end{figure*}

\begin{figure*}
\begin{tabular}{cc}
\resizebox*{0.9\columnwidth}{0.36\height}{\includegraphics{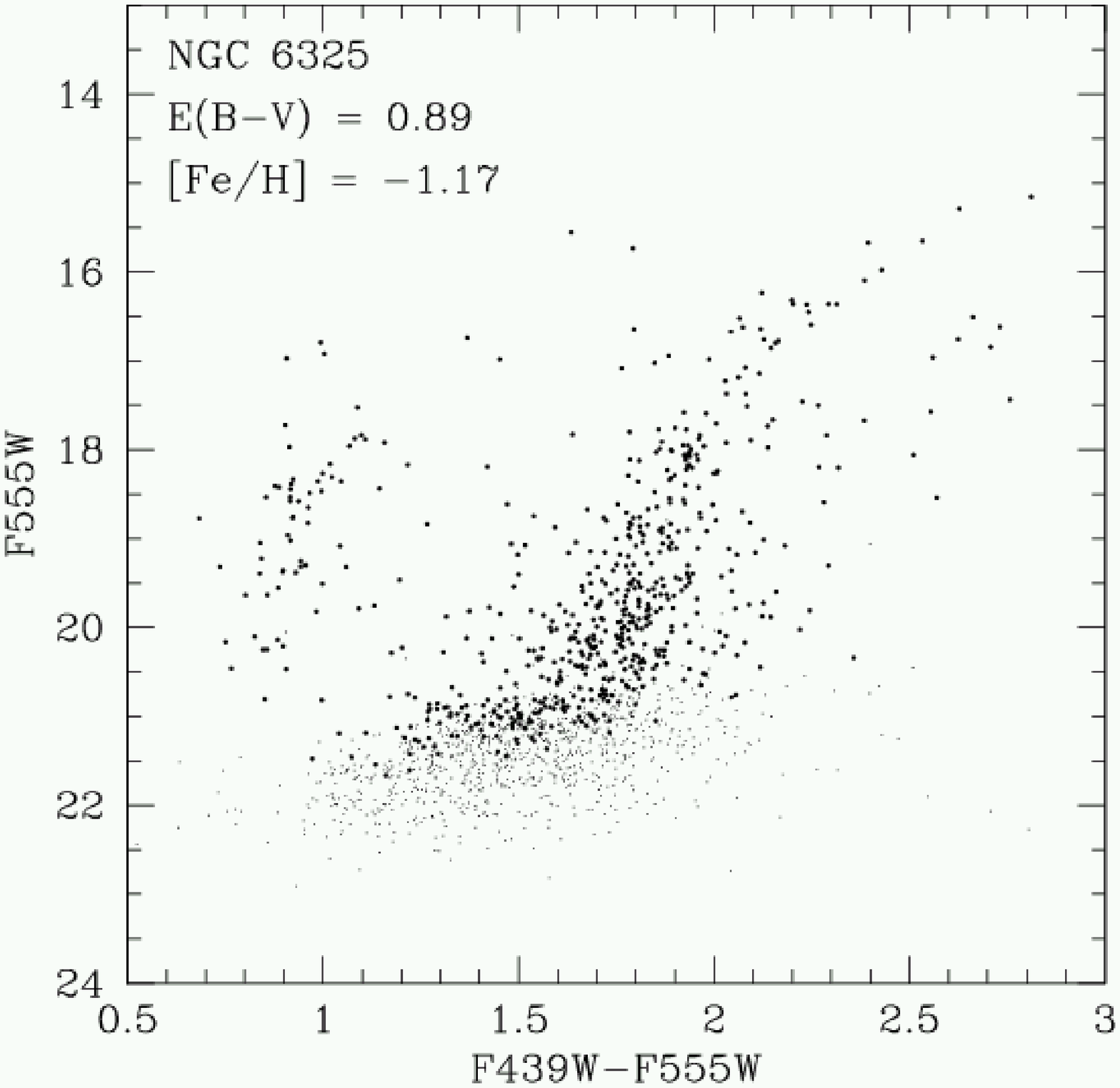}} &
\resizebox*{0.9\columnwidth}{0.36\height}{\includegraphics{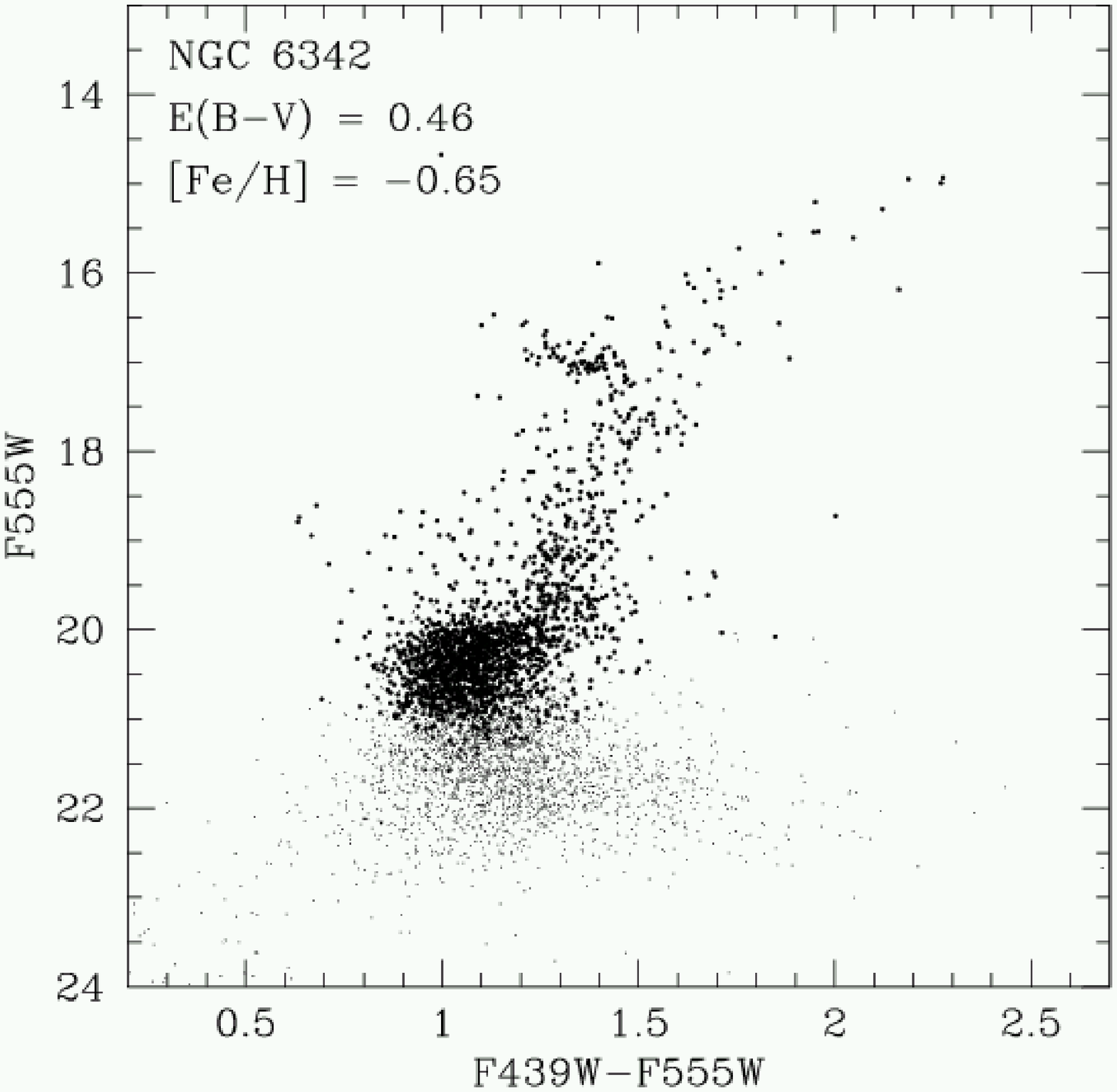}} \\
\resizebox*{0.9\columnwidth}{0.36\height}{\includegraphics{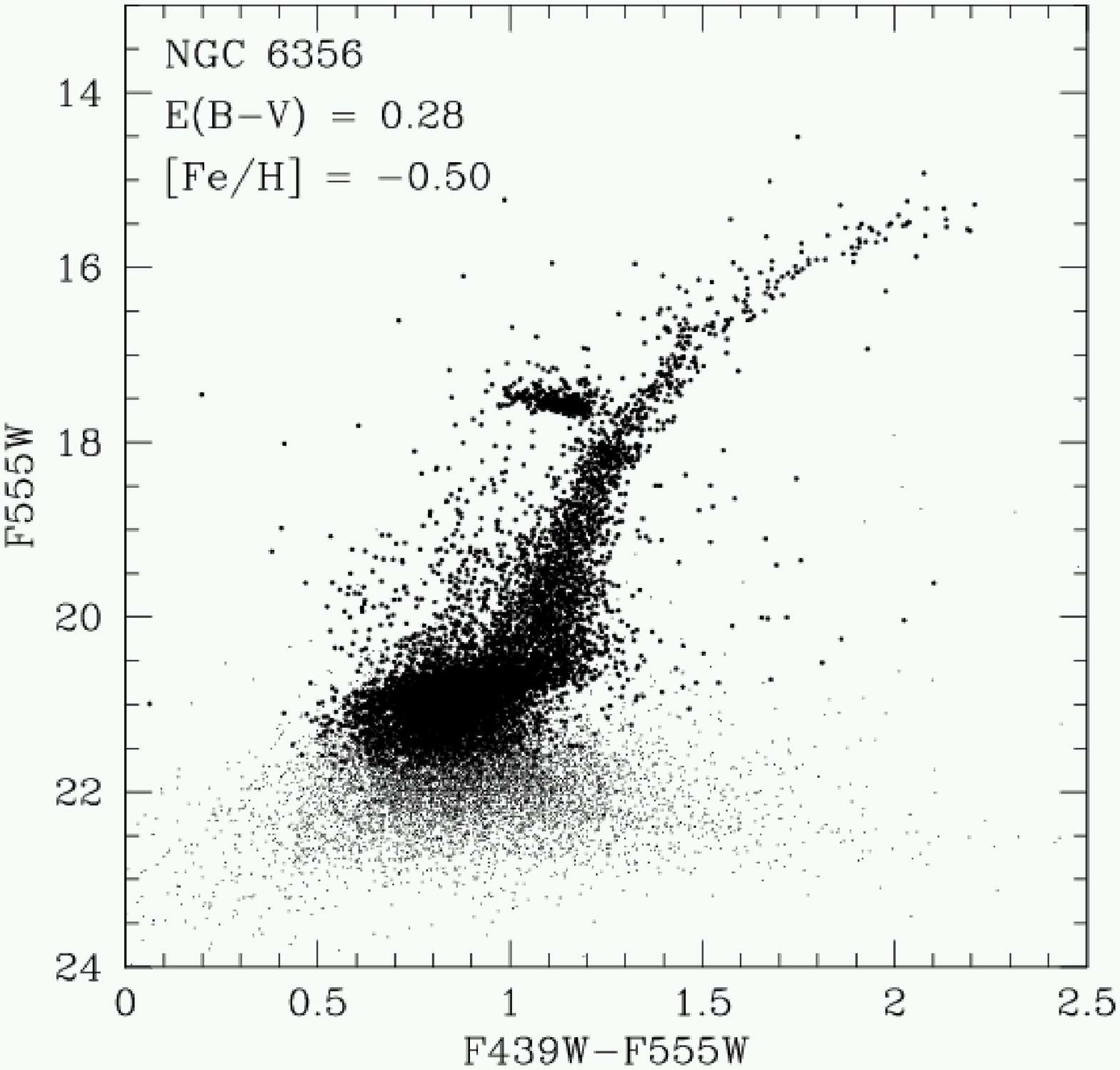}} &
\resizebox*{0.9\columnwidth}{0.36\height}{\includegraphics{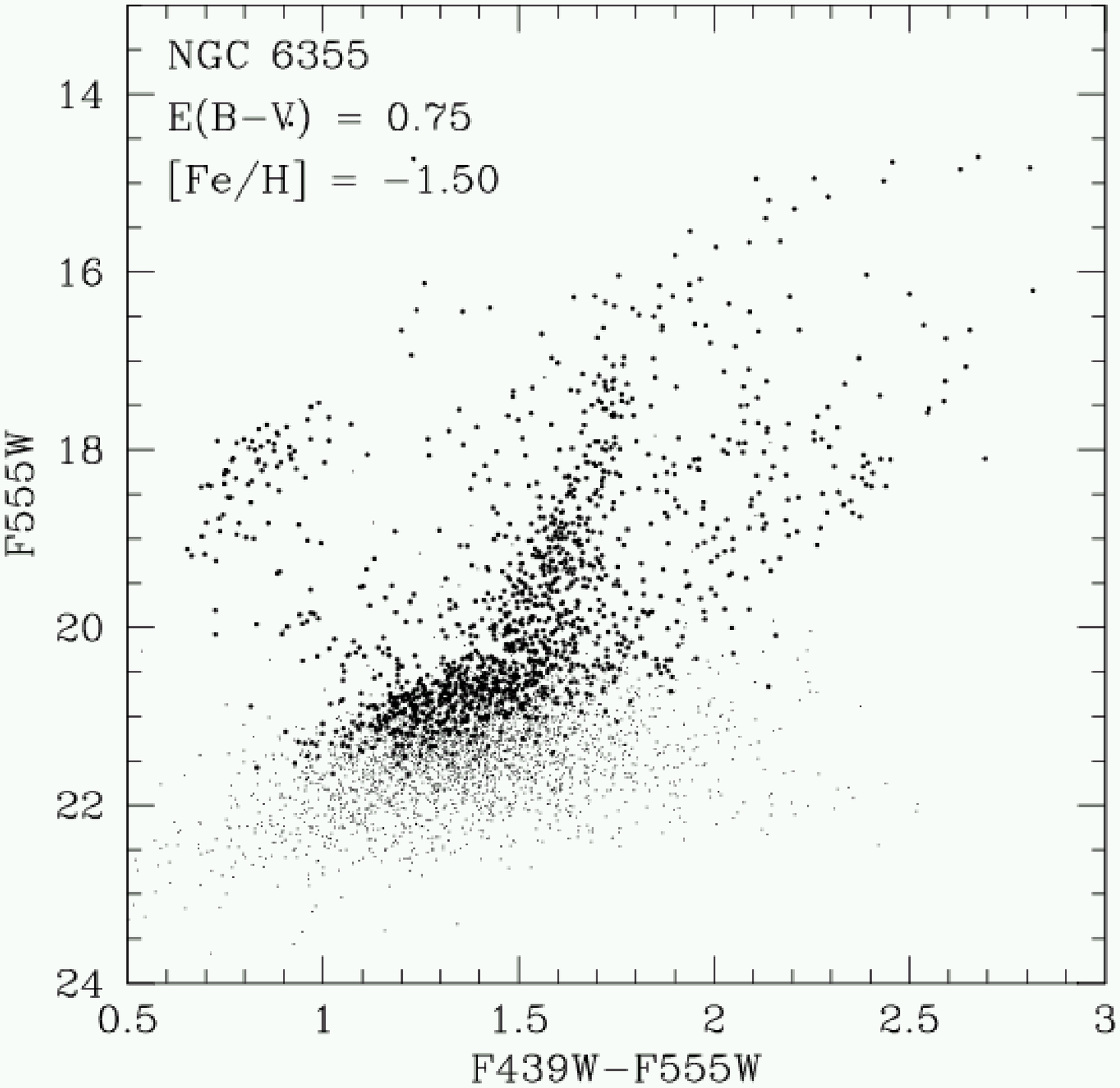}} \\
\resizebox*{0.9\columnwidth}{0.36\height}{\includegraphics{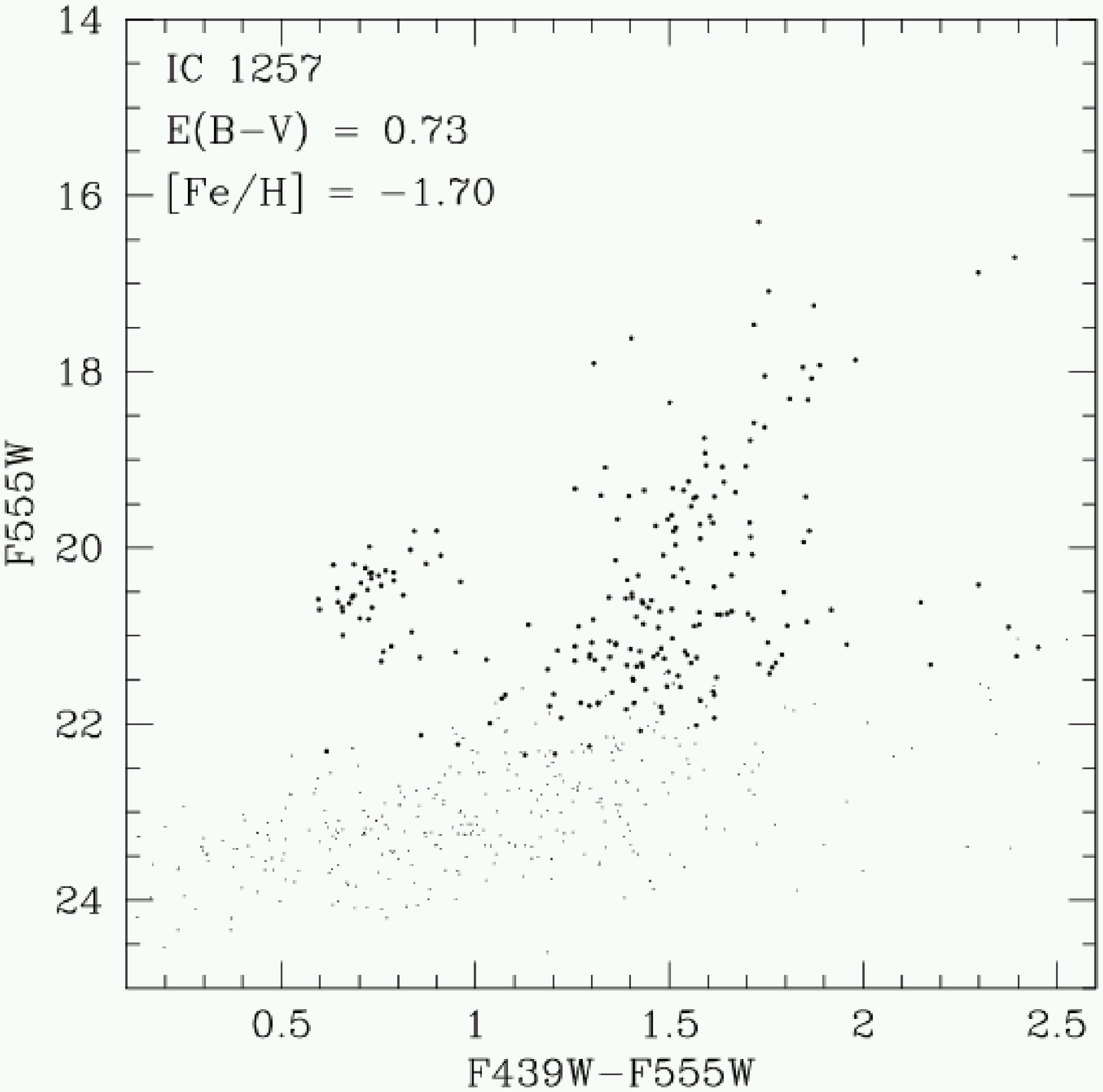}} &
\resizebox*{0.9\columnwidth}{0.36\height}{\includegraphics{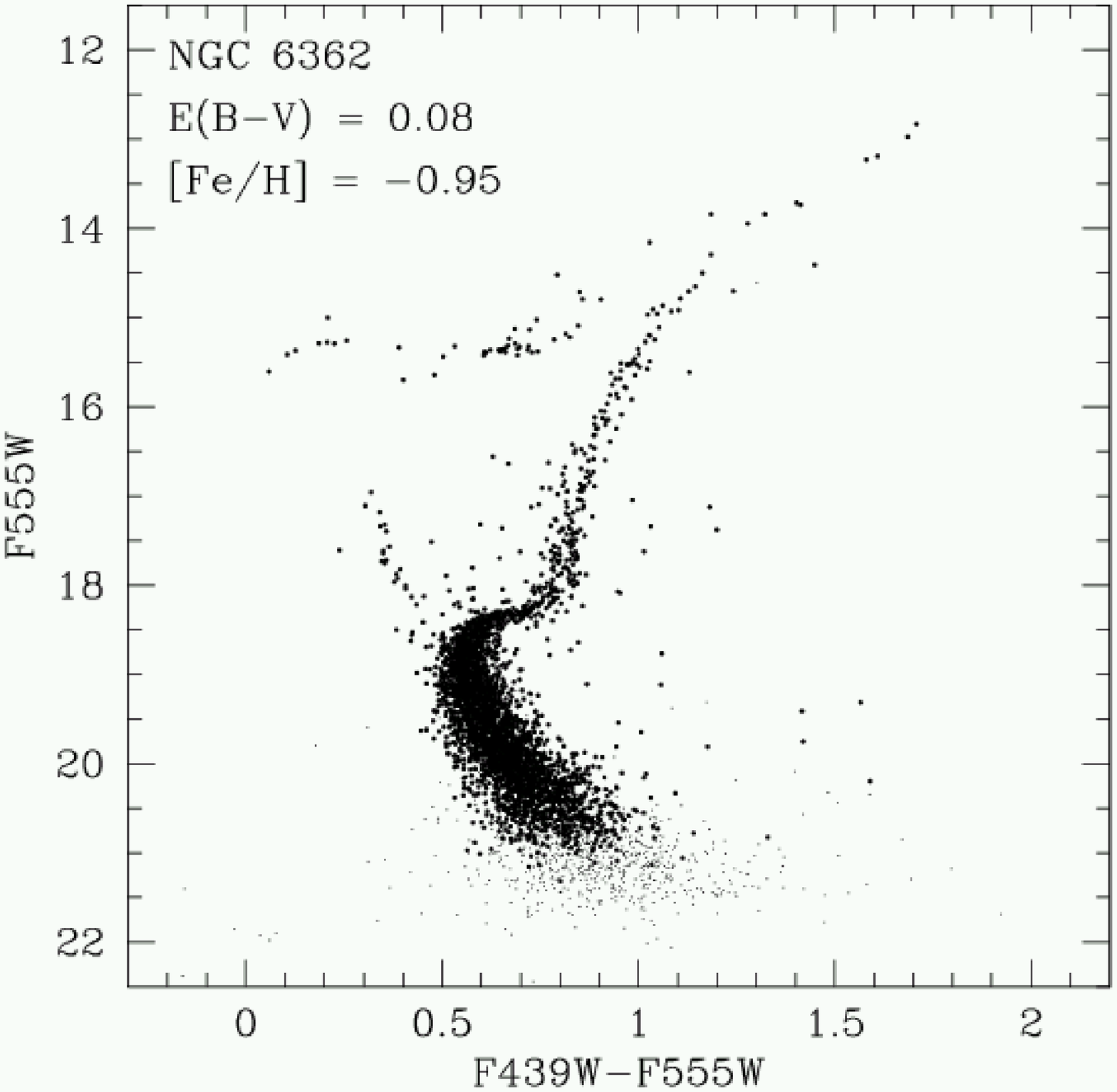}} \\
\end{tabular}
\caption{The color magnitude diagrams (cont.)}
\end{figure*}

\begin{figure*}
\begin{tabular}{cc}
\resizebox*{0.9\columnwidth}{0.36\height}{\includegraphics{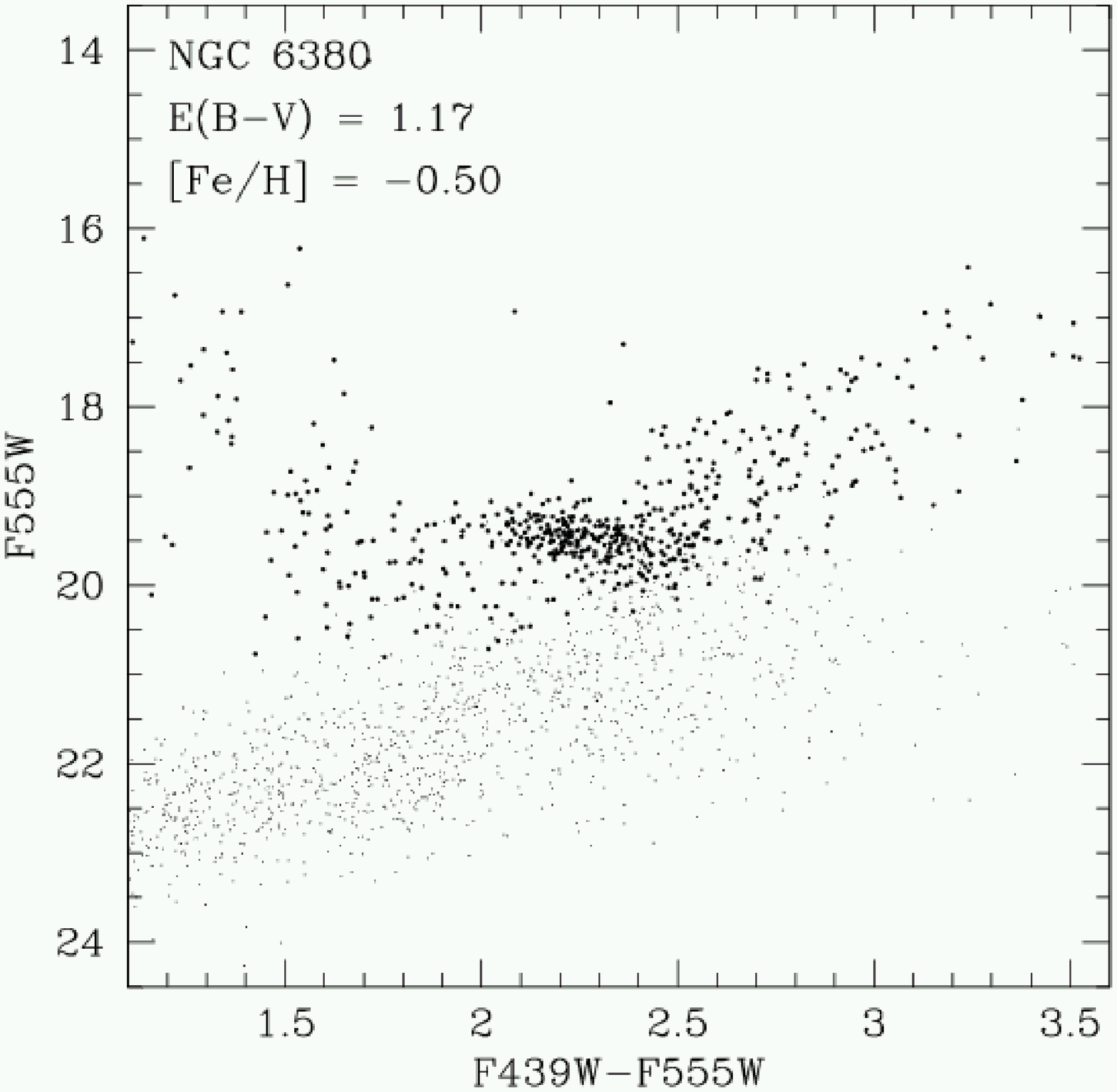}} &
\resizebox*{0.9\columnwidth}{0.36\height}{\includegraphics{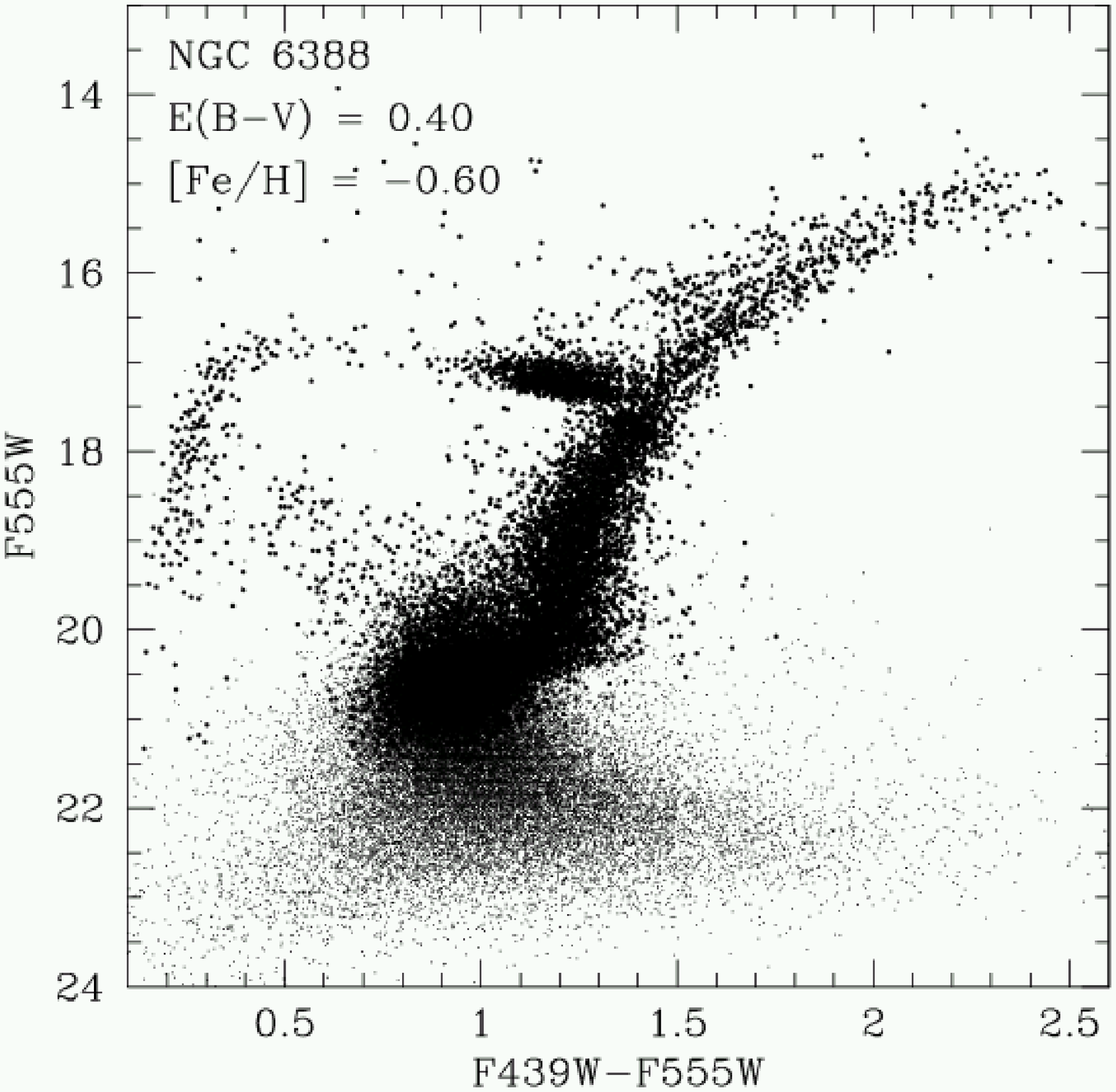}} \\
\resizebox*{0.9\columnwidth}{0.36\height}{\includegraphics{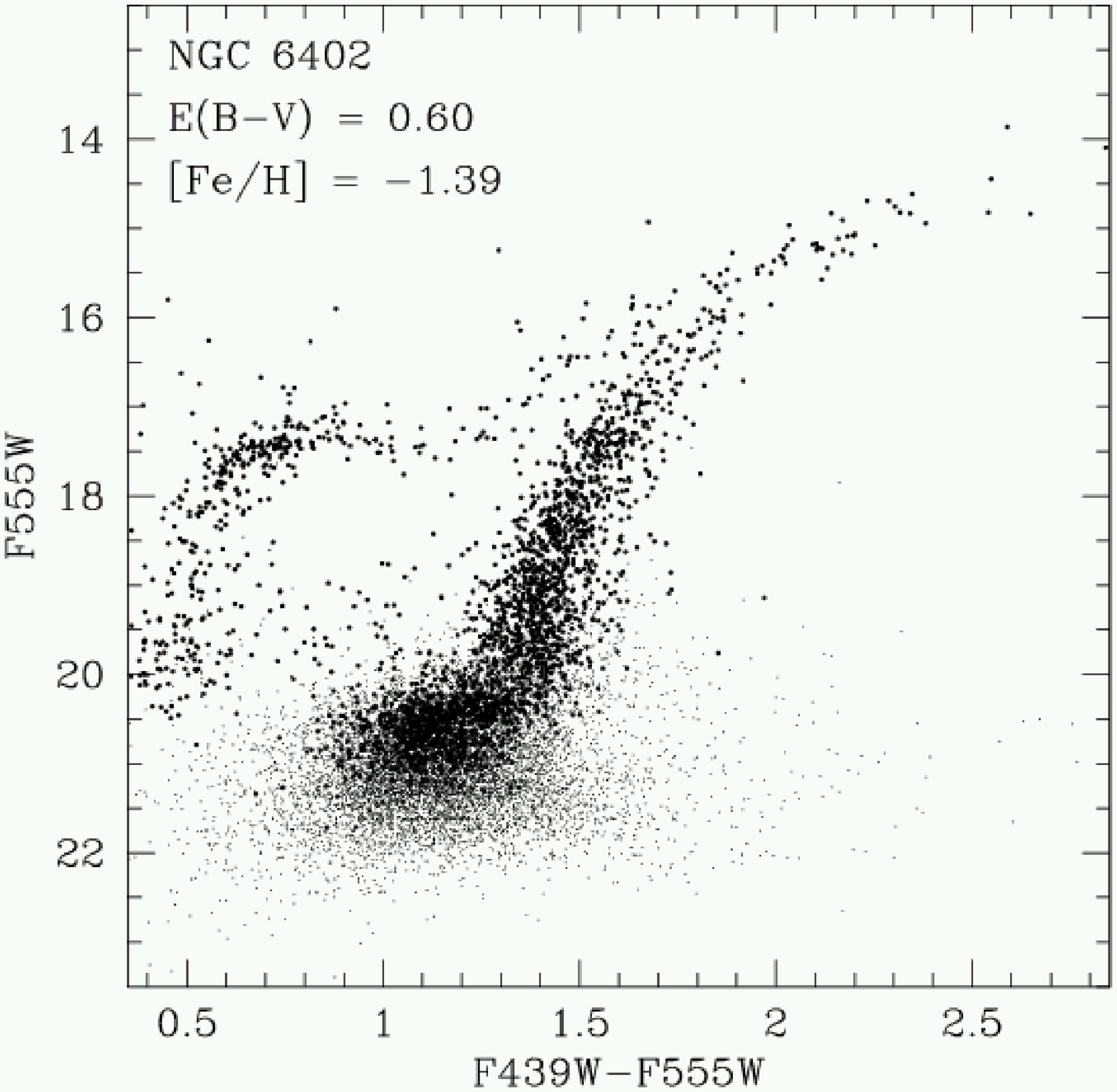}} &
\resizebox*{0.9\columnwidth}{0.36\height}{\includegraphics{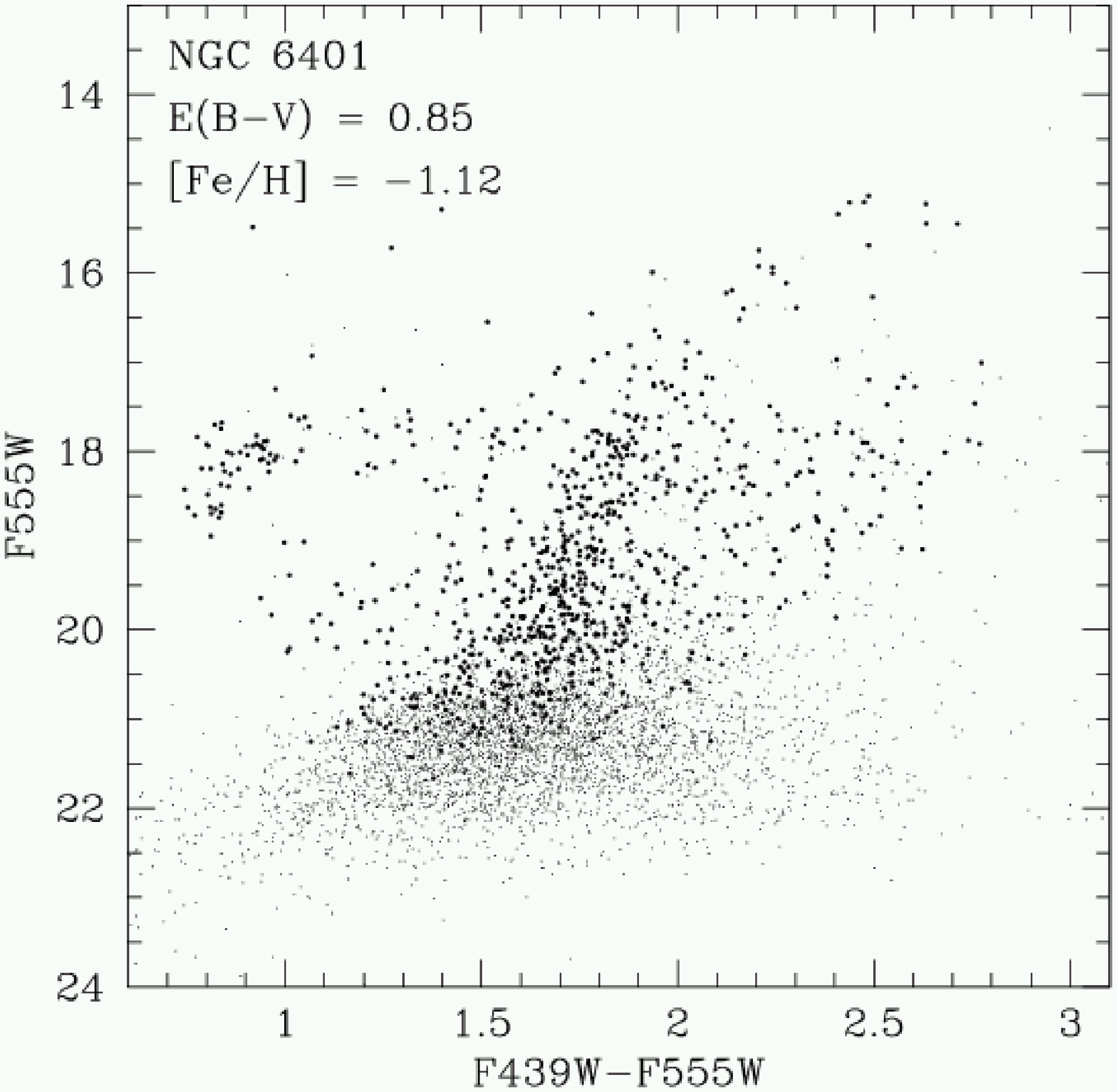}} \\
\resizebox*{0.9\columnwidth}{0.36\height}{\includegraphics{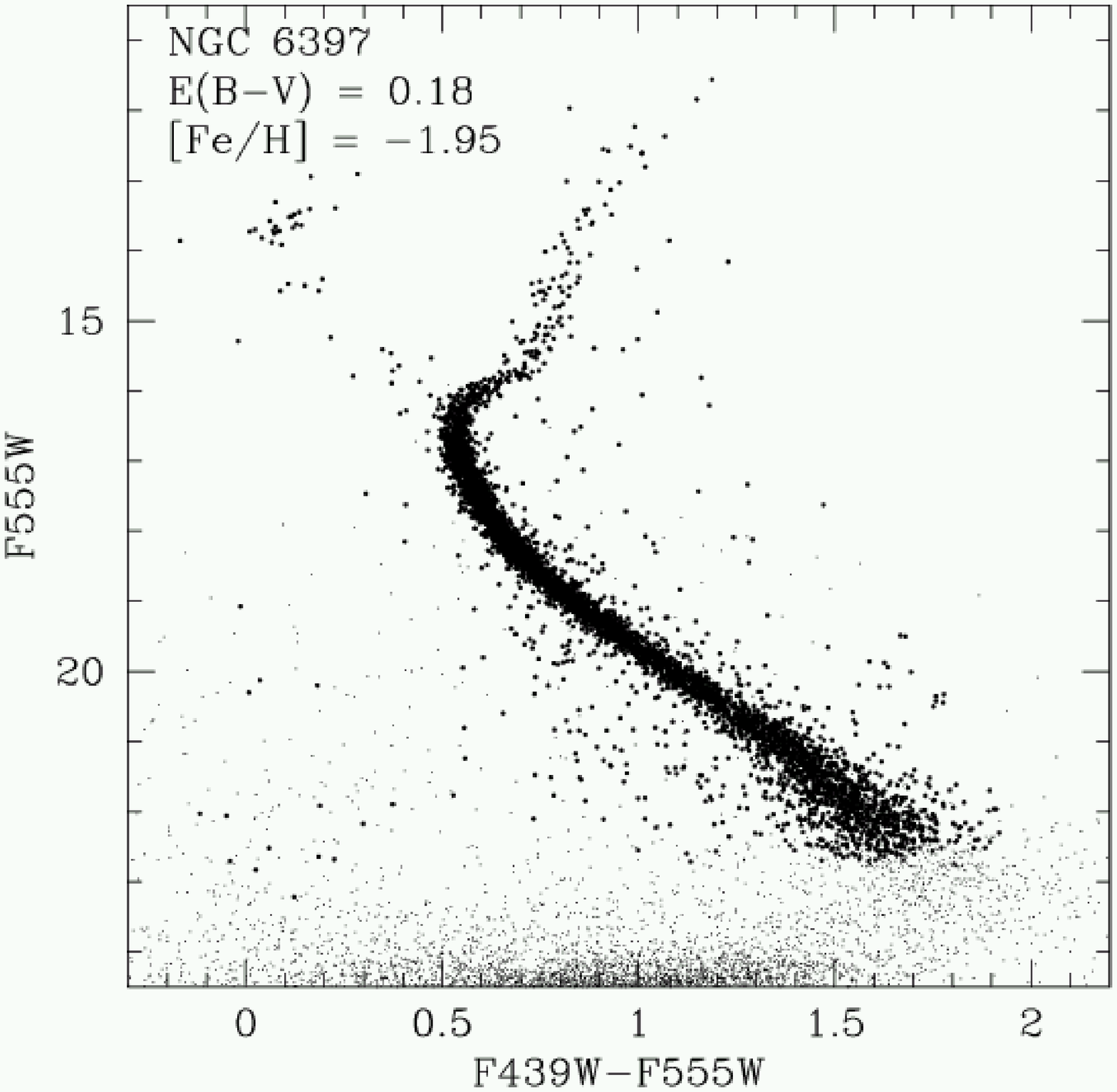}} &
\resizebox*{0.9\columnwidth}{0.36\height}{\includegraphics{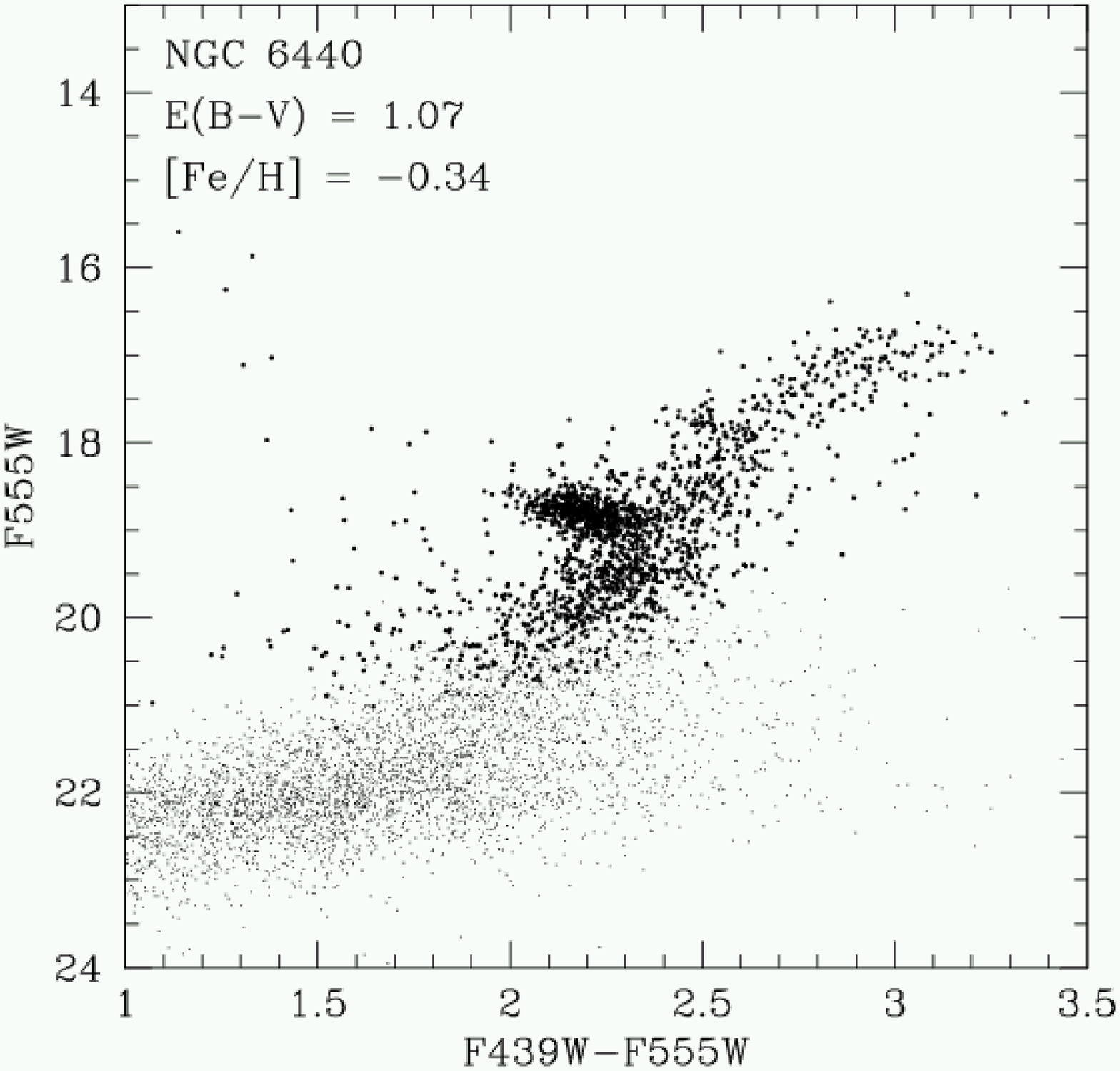}} \\
\end{tabular}
\caption{The color magnitude diagrams (cont.).}
\end{figure*}

\begin{figure*}
\begin{tabular}{cc}
\resizebox*{0.9\columnwidth}{0.36\height}{\includegraphics{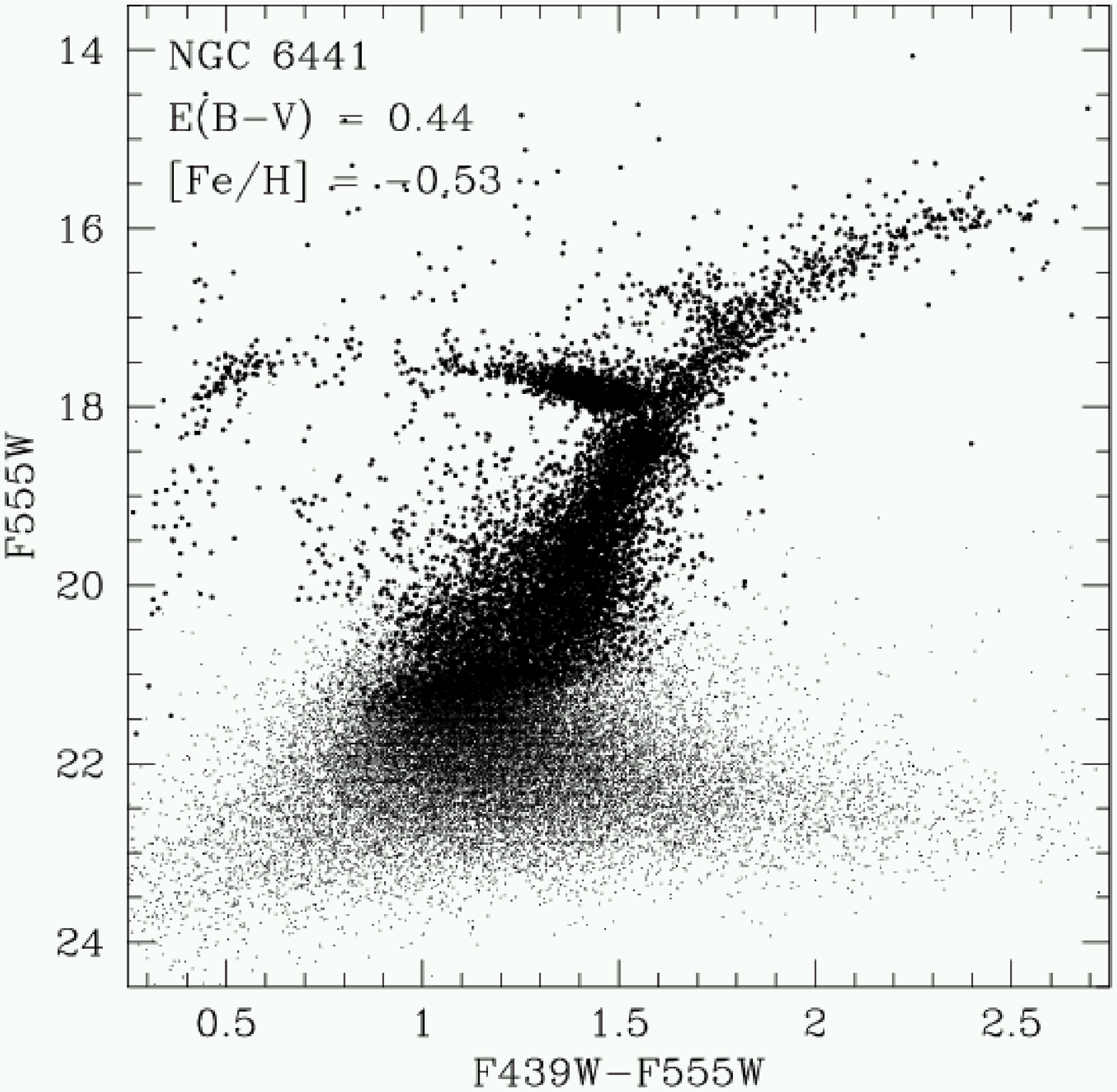}} &
\resizebox*{0.9\columnwidth}{0.36\height}{\includegraphics{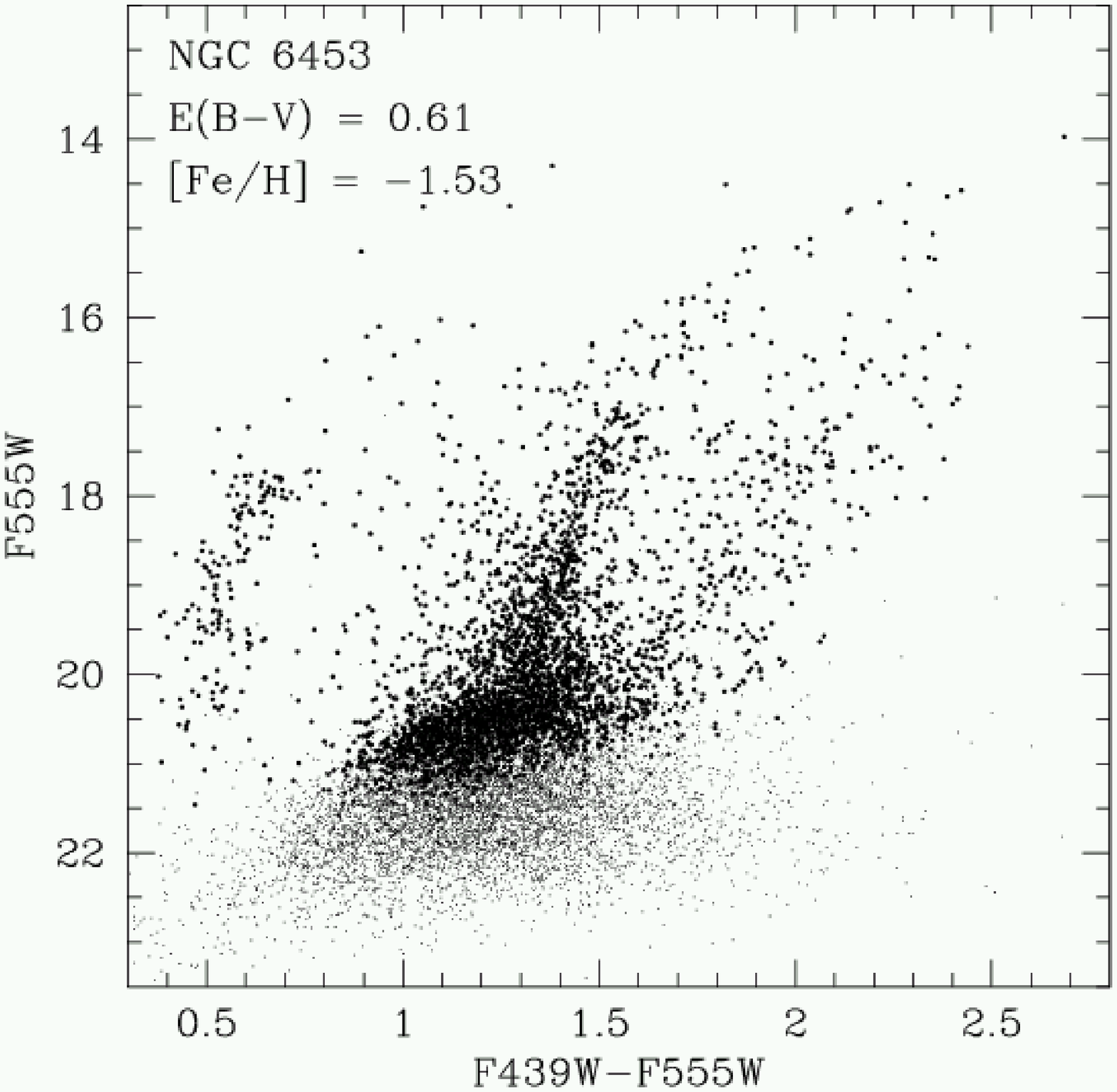}} \\
\resizebox*{0.9\columnwidth}{0.36\height}{\includegraphics{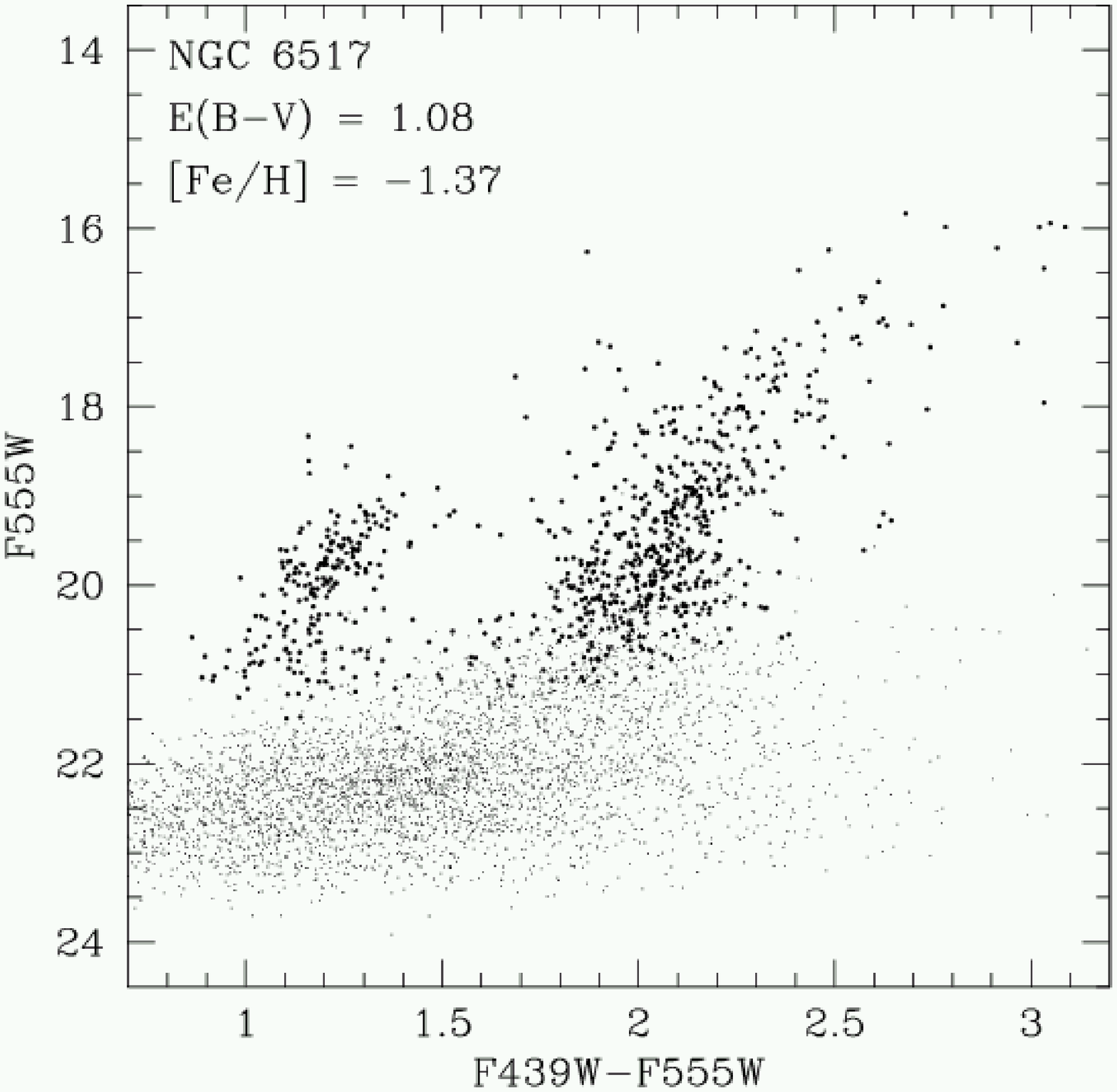}} &
\resizebox*{0.9\columnwidth}{0.36\height}{\includegraphics{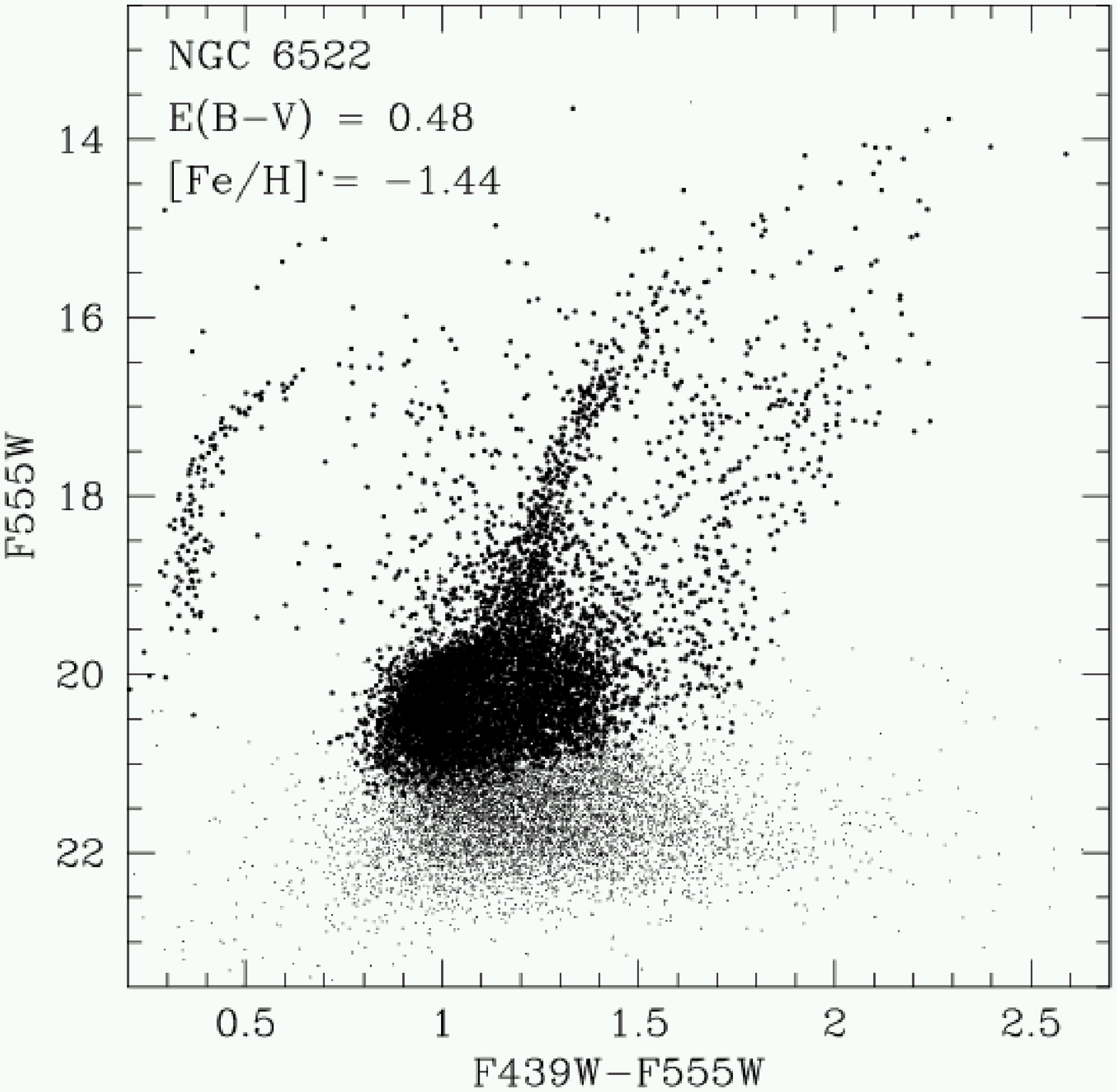}} \\
\resizebox*{0.9\columnwidth}{0.36\height}{\includegraphics{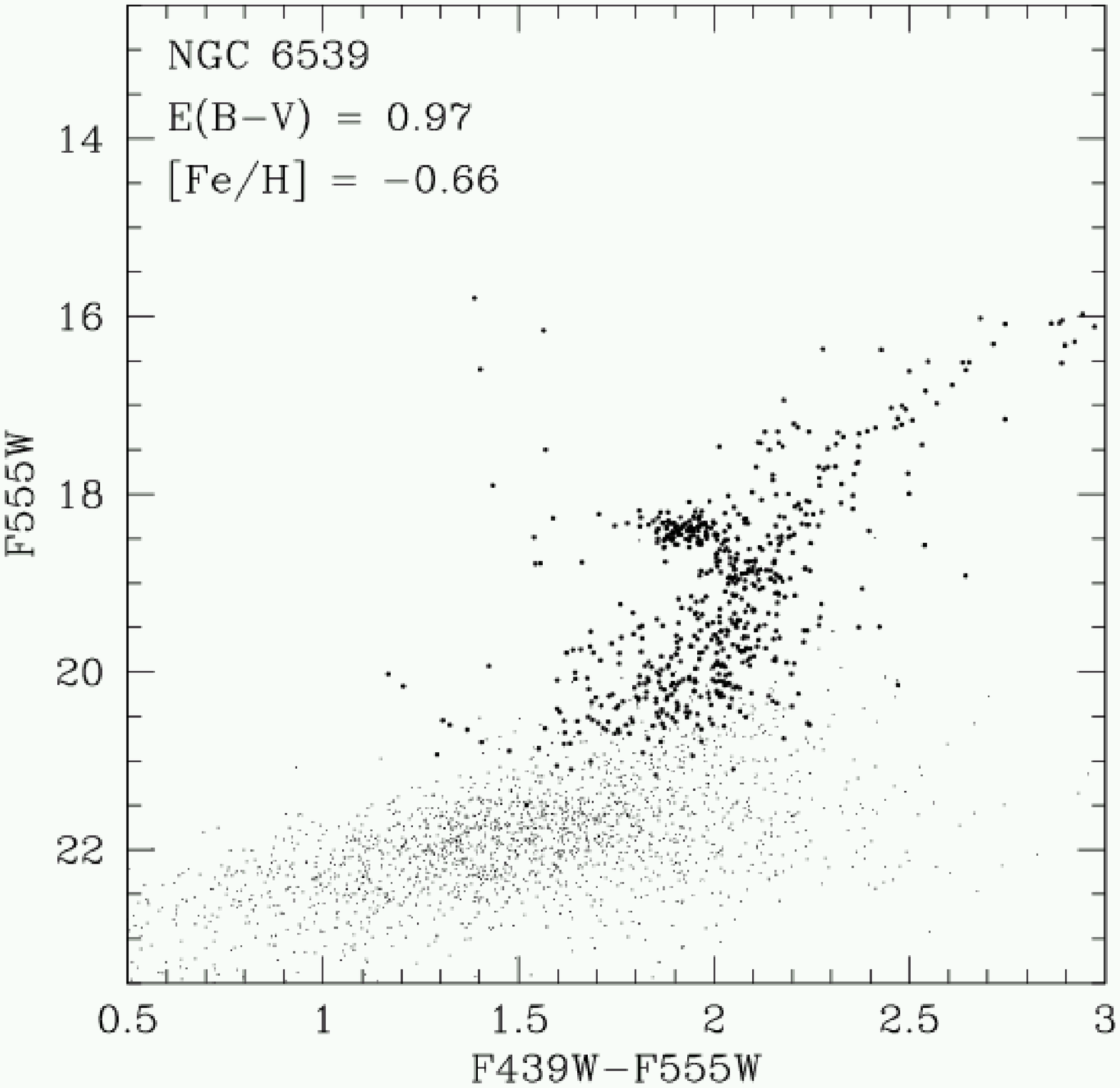}} &
\resizebox*{0.9\columnwidth}{0.36\height}{\includegraphics{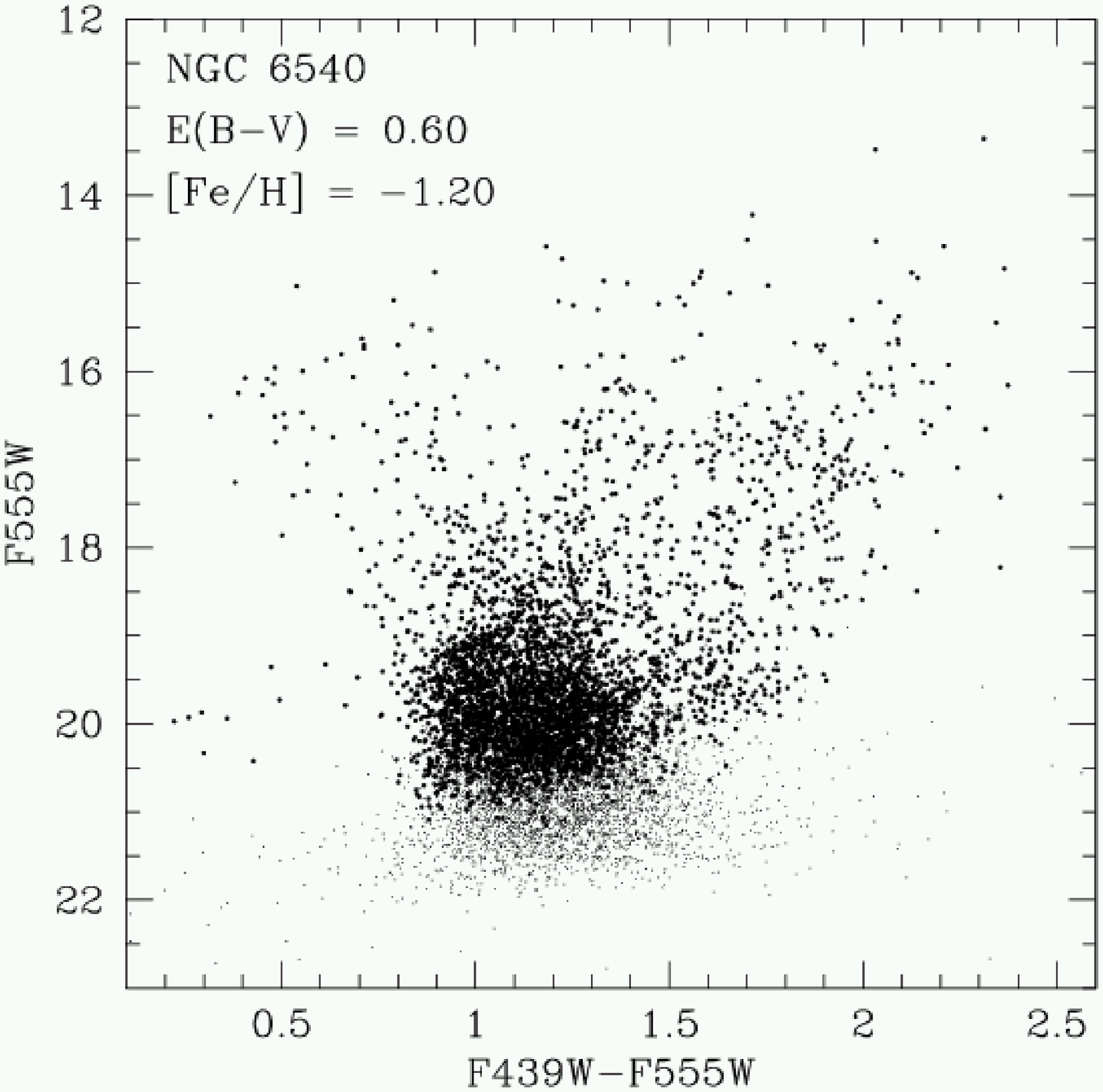}} \\
\end{tabular}
\caption{The color magnitude diagrams (cont.). Note that the magnitude
range of the CMD for NGC~6397 differs from the other cases.}
\end{figure*}

\begin{figure*}
\begin{tabular}{cc}
\resizebox*{0.9\columnwidth}{0.36\height}{\includegraphics{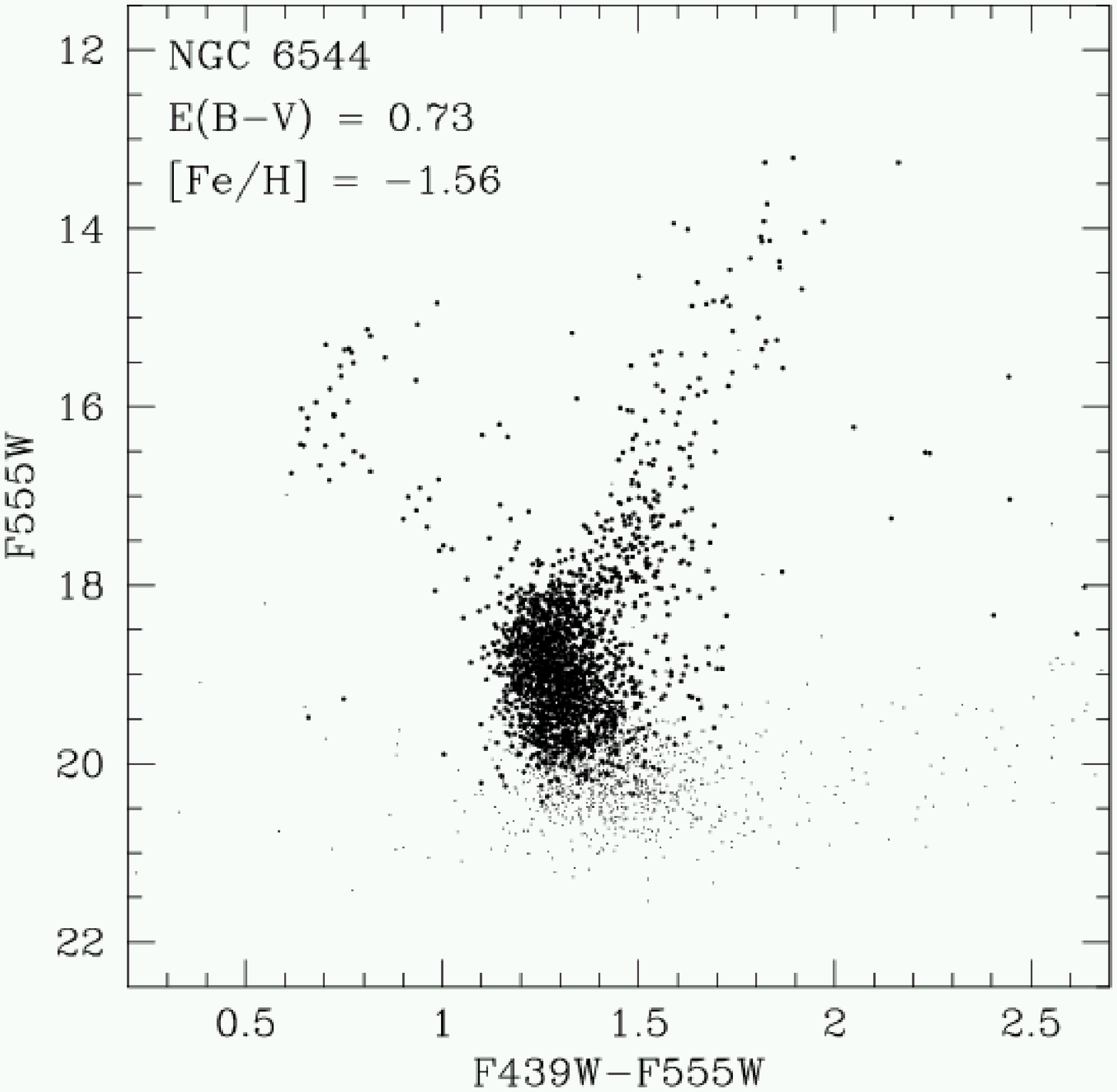}} &
\resizebox*{0.9\columnwidth}{0.36\height}{\includegraphics{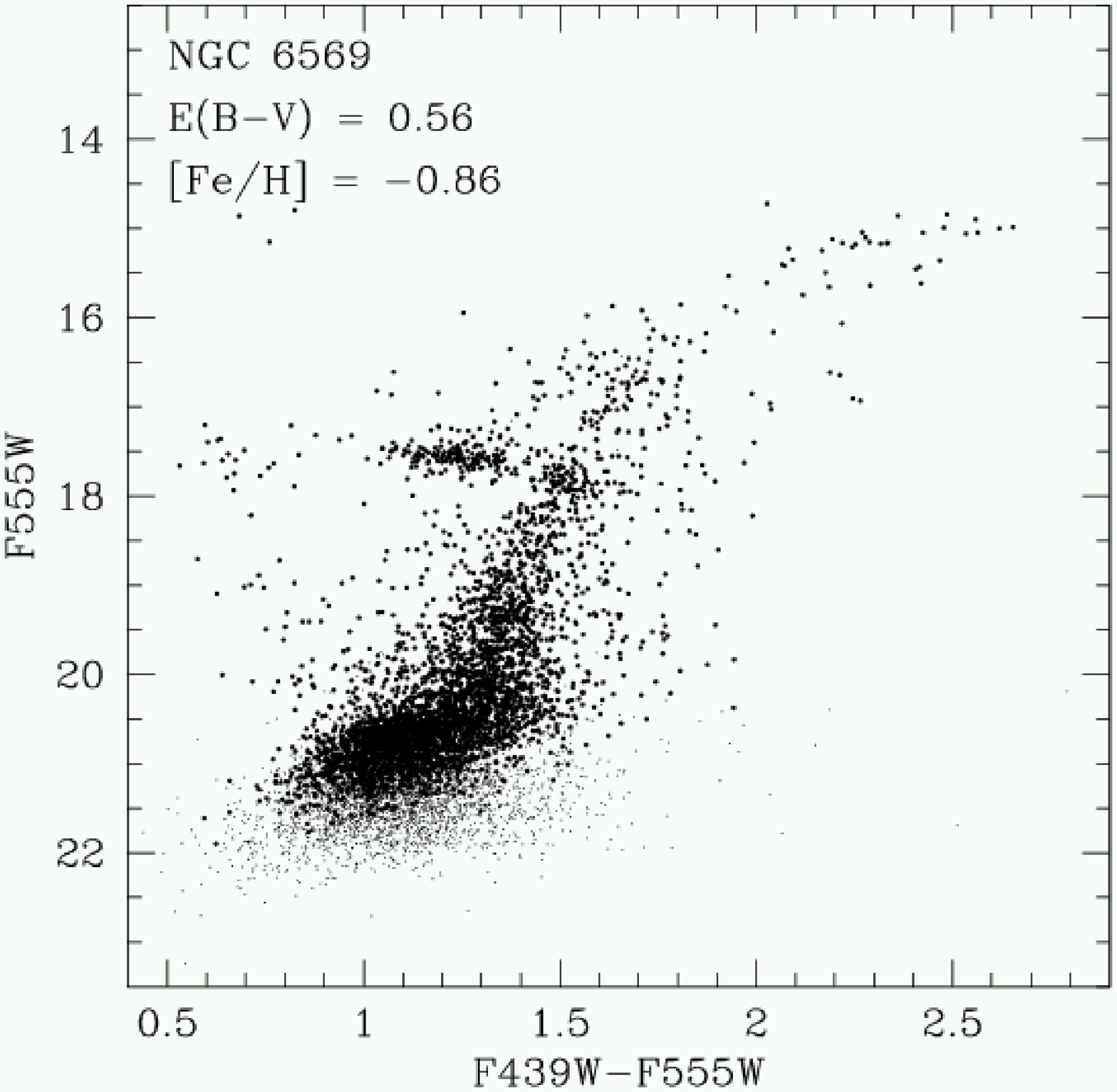}} \\
\resizebox*{0.9\columnwidth}{0.36\height}{\includegraphics{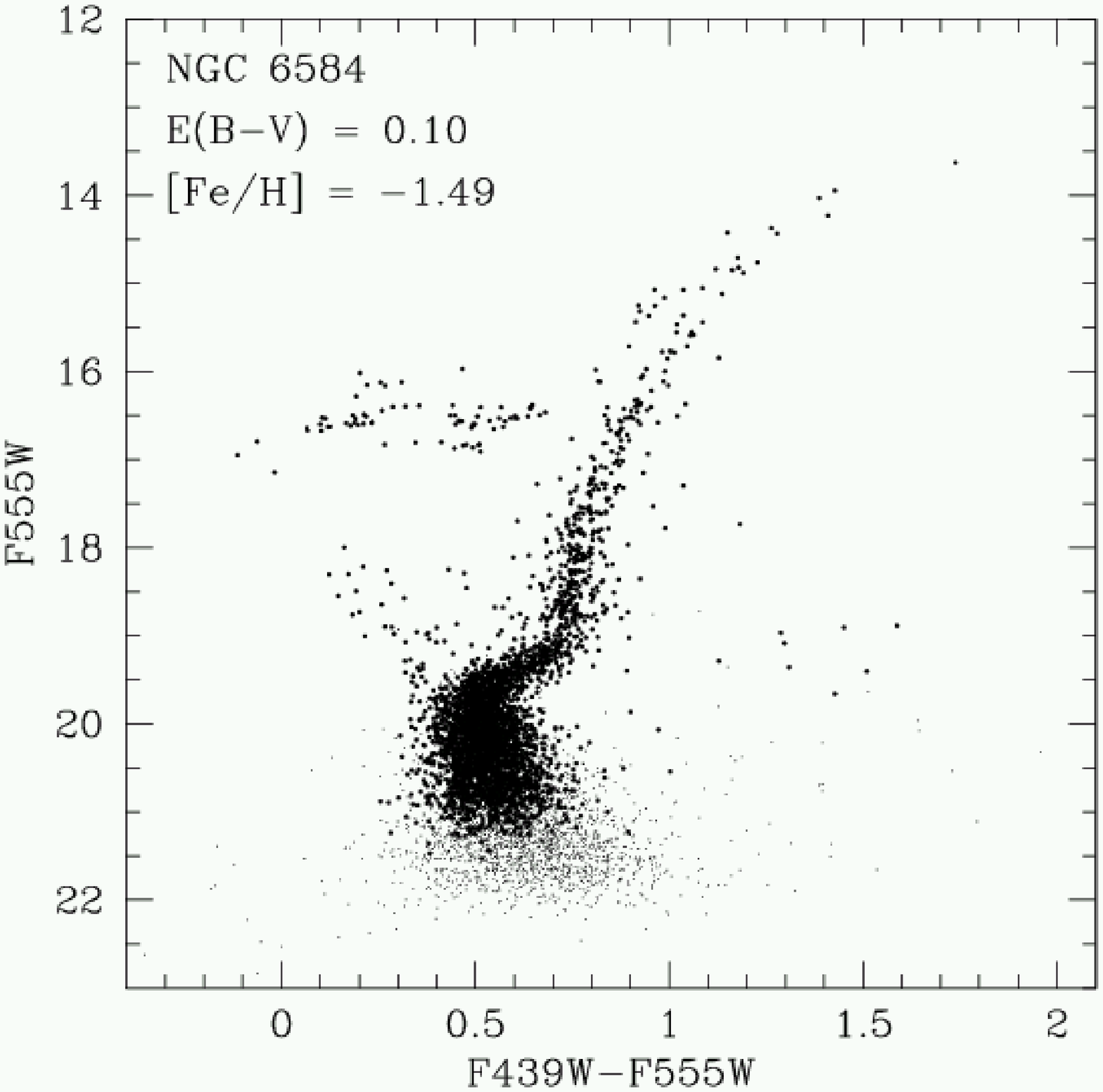}} &
\resizebox*{0.9\columnwidth}{0.36\height}{\includegraphics{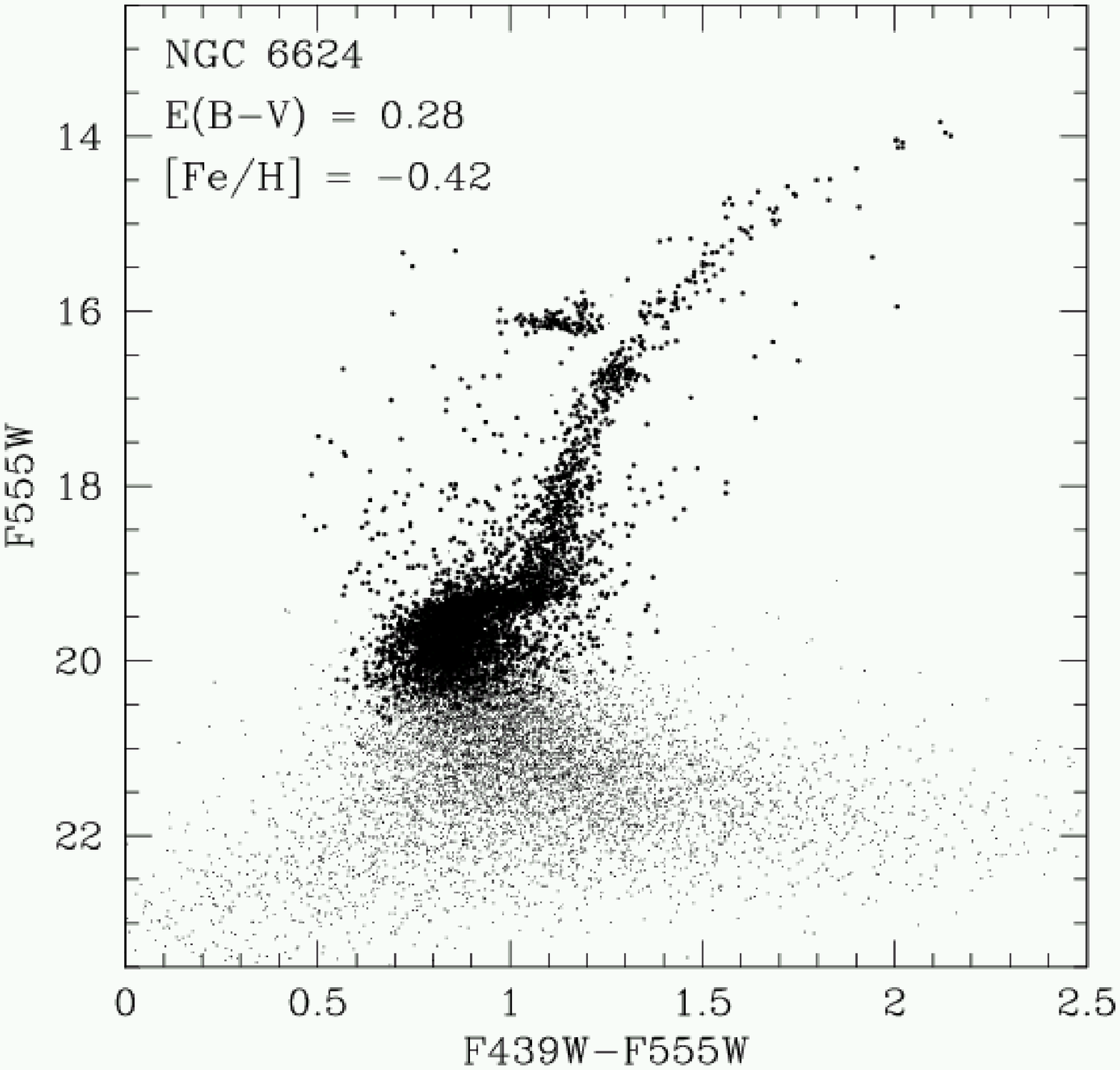}} \\
\resizebox*{0.9\columnwidth}{0.36\height}{\includegraphics{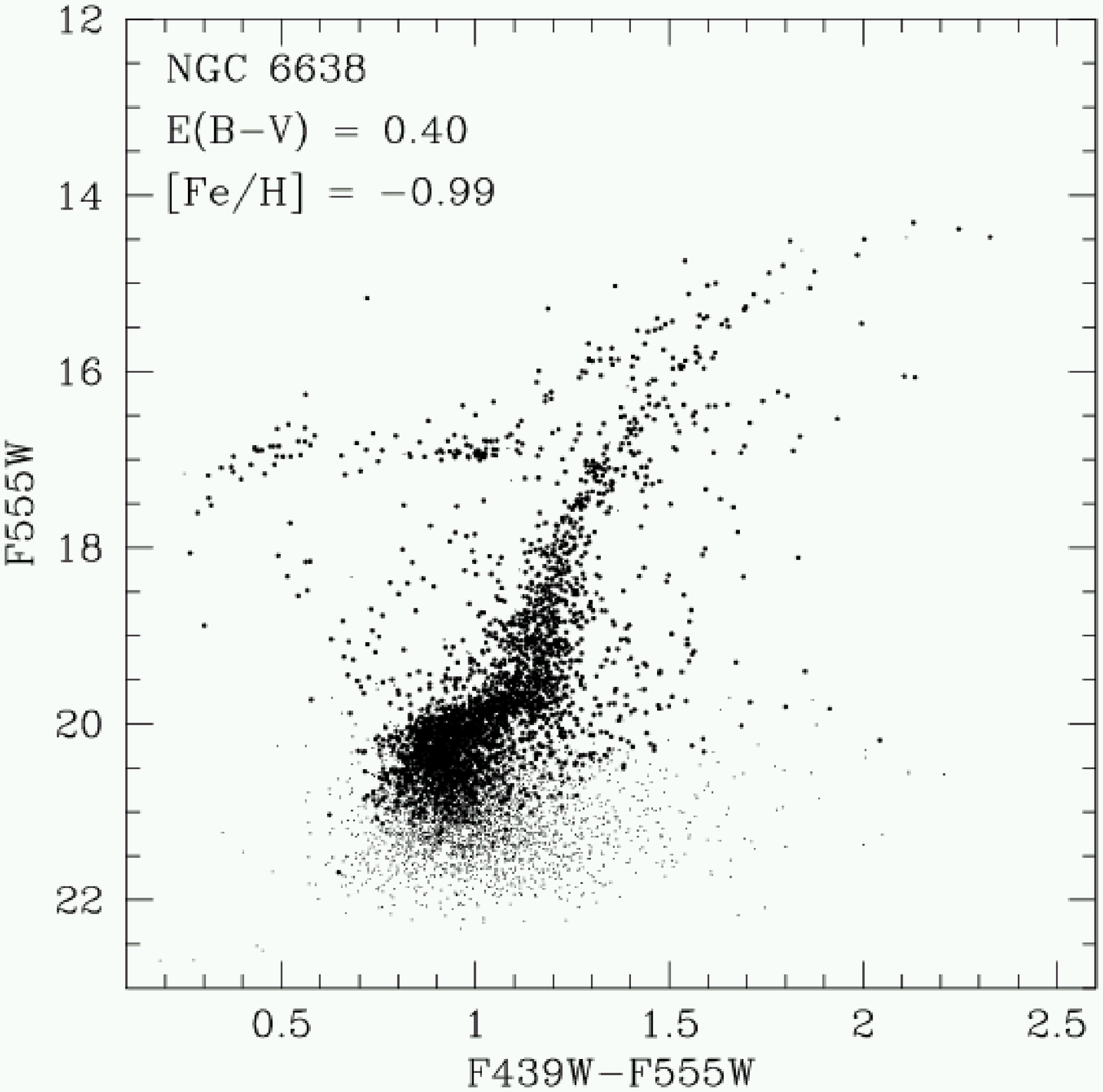}} &
\resizebox*{0.9\columnwidth}{0.36\height}{\includegraphics{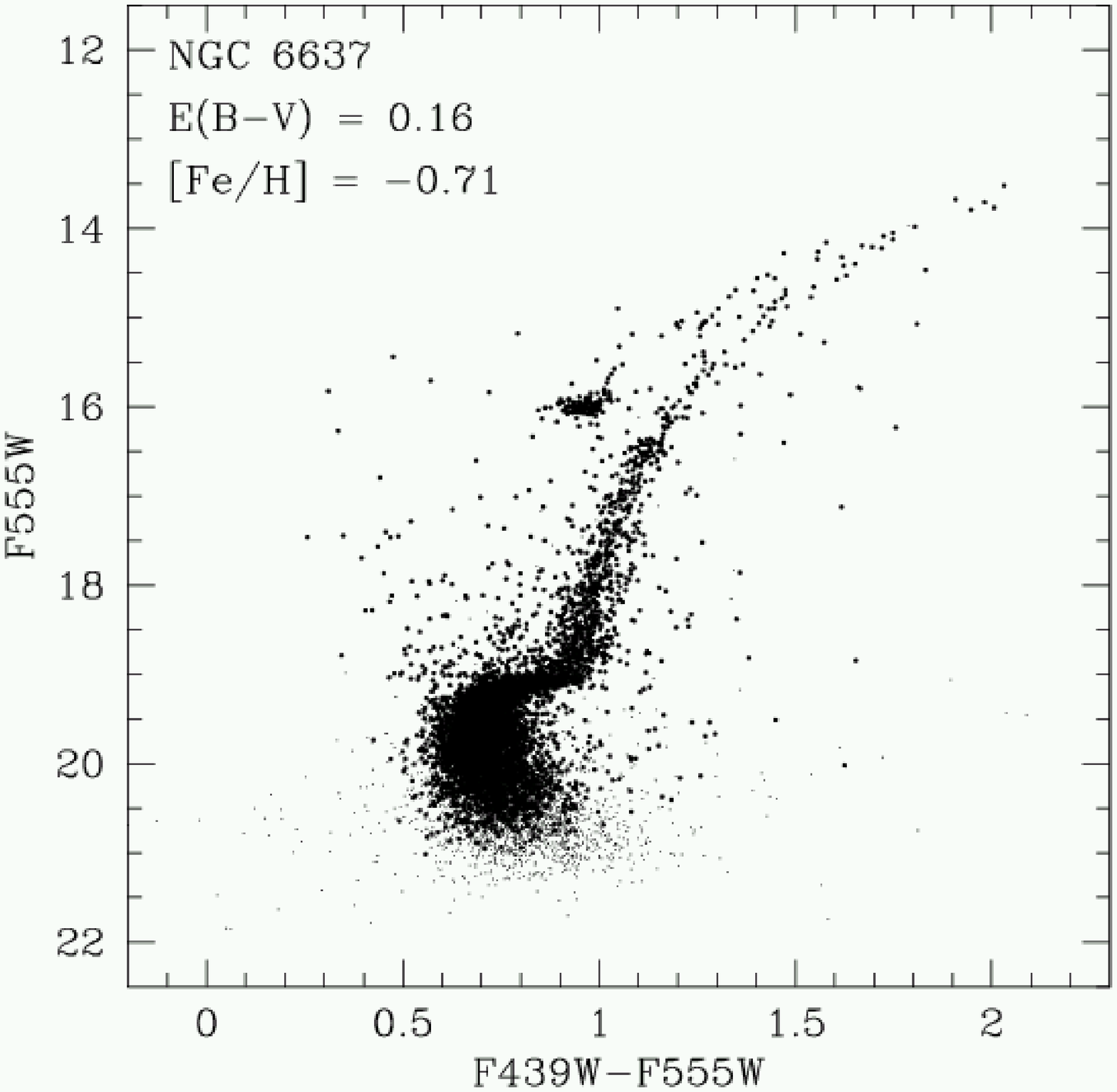}} \\
\end{tabular}
\caption{The color magnitude diagrams (cont.).}
\end{figure*}

\begin{figure*}
\begin{tabular}{cc}
\resizebox*{0.9\columnwidth}{0.36\height}{\includegraphics{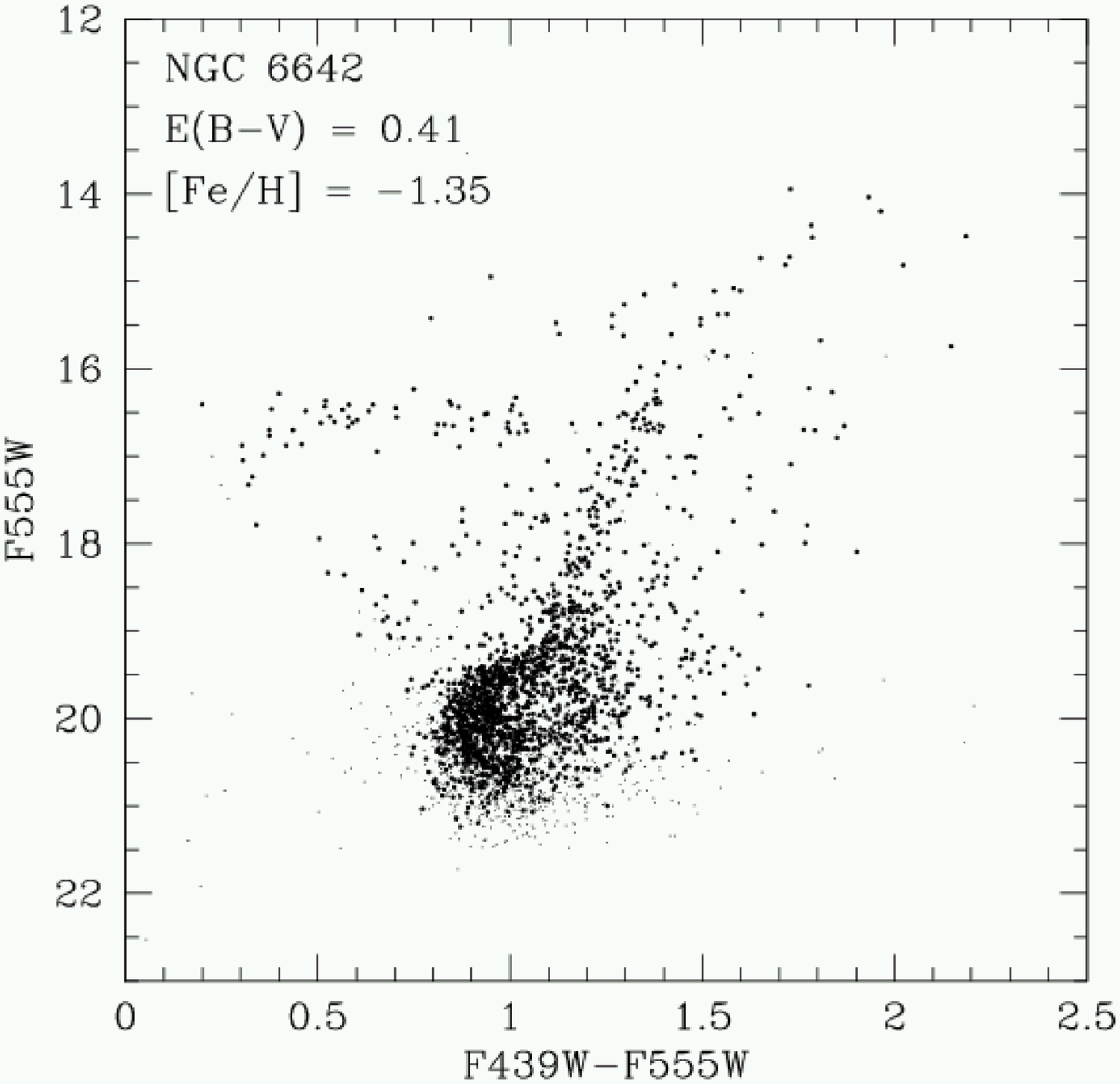}} &
\resizebox*{0.9\columnwidth}{0.36\height}{\includegraphics{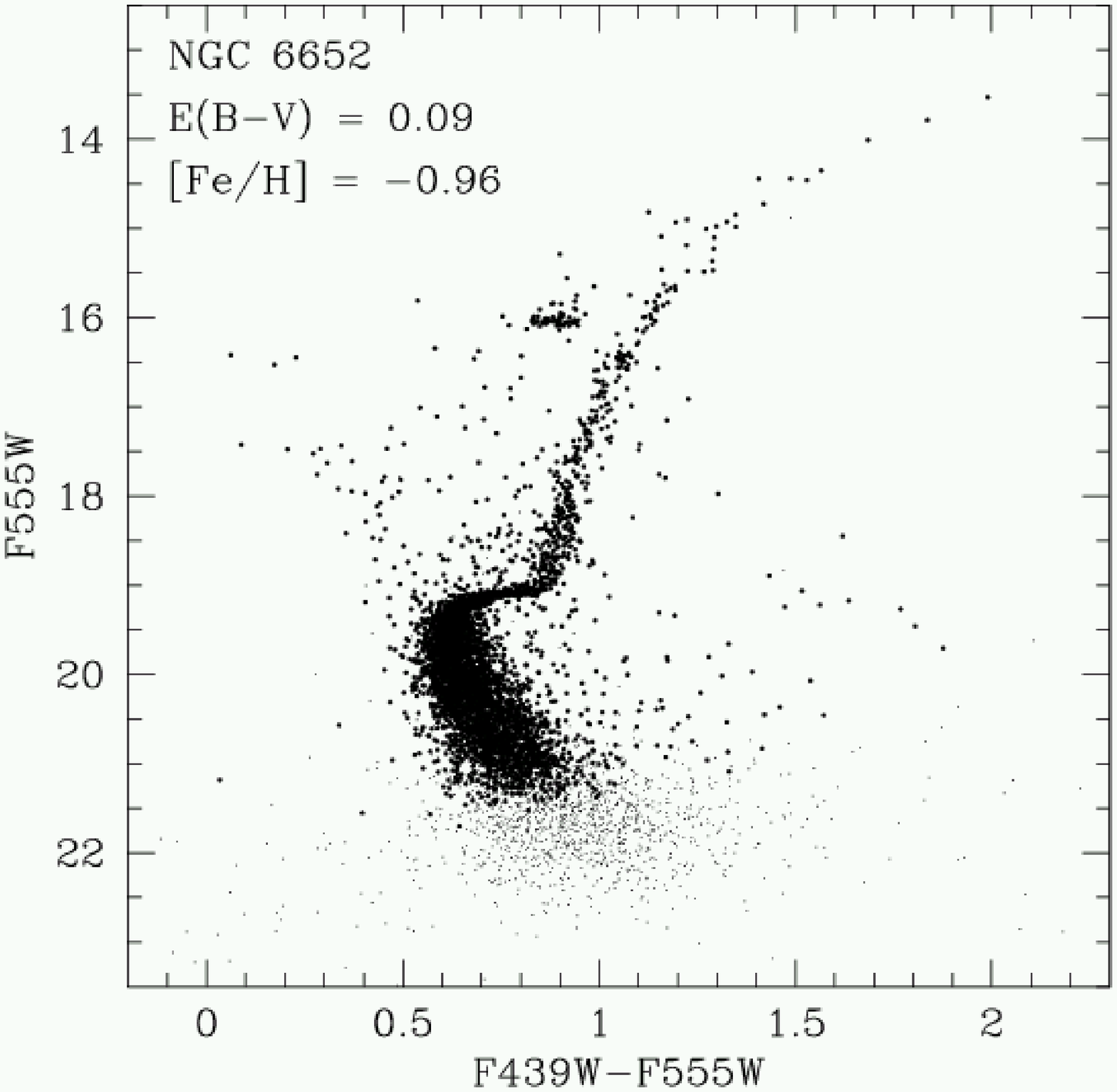}} \\
\resizebox*{0.9\columnwidth}{0.36\height}{\includegraphics{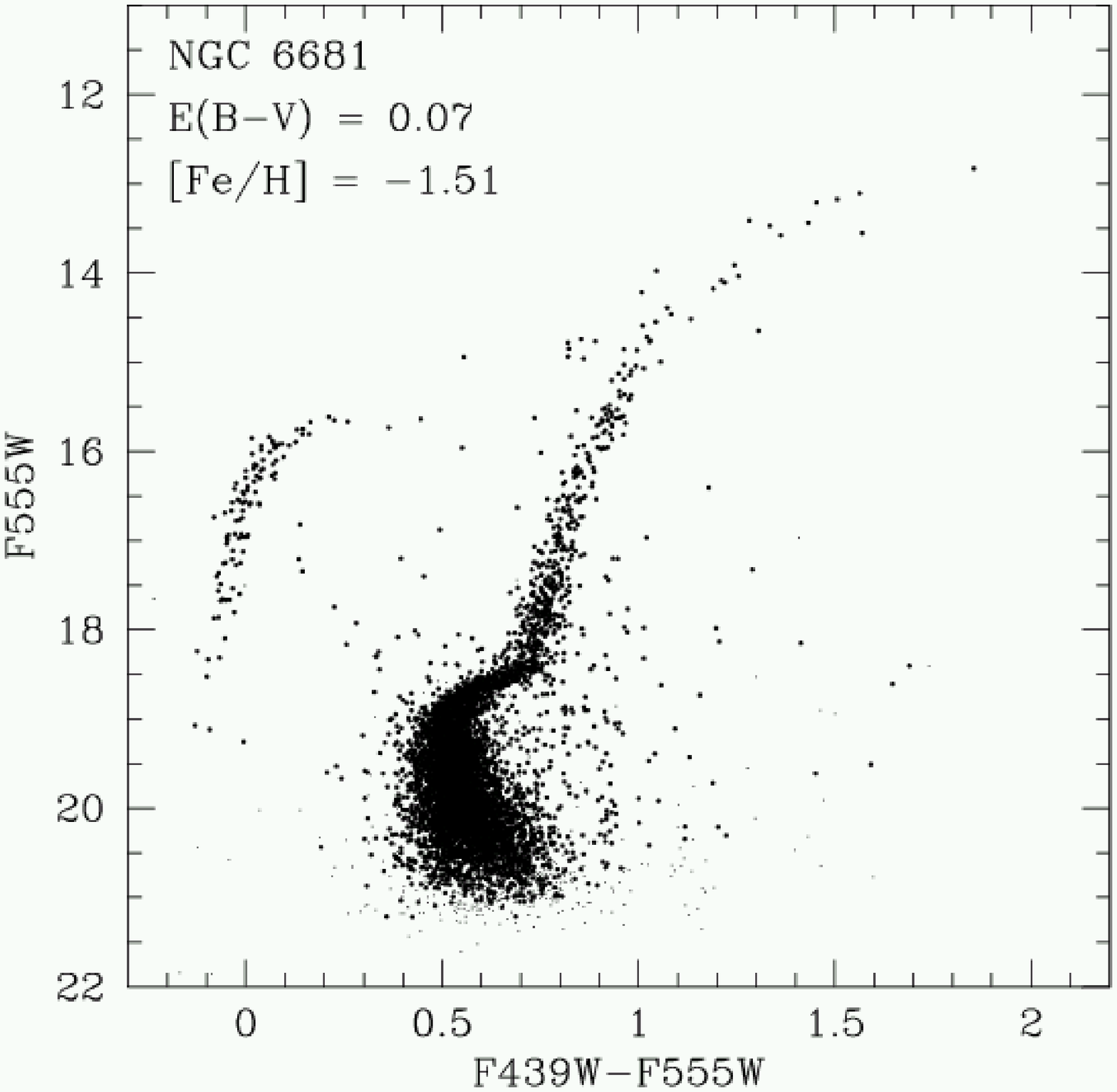}} &
\resizebox*{0.9\columnwidth}{0.36\height}{\includegraphics{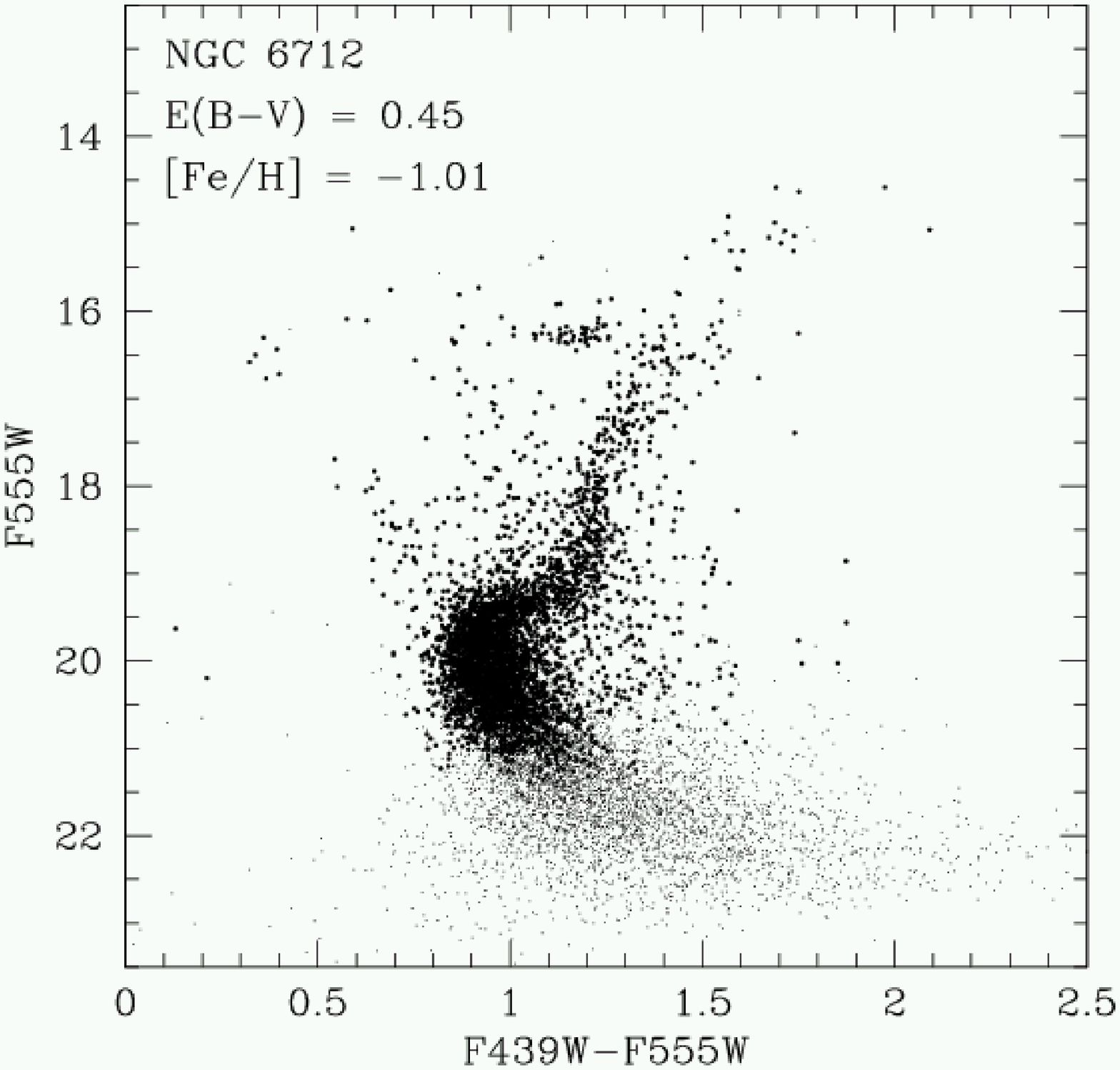}} \\
\resizebox*{0.9\columnwidth}{0.36\height}{\includegraphics{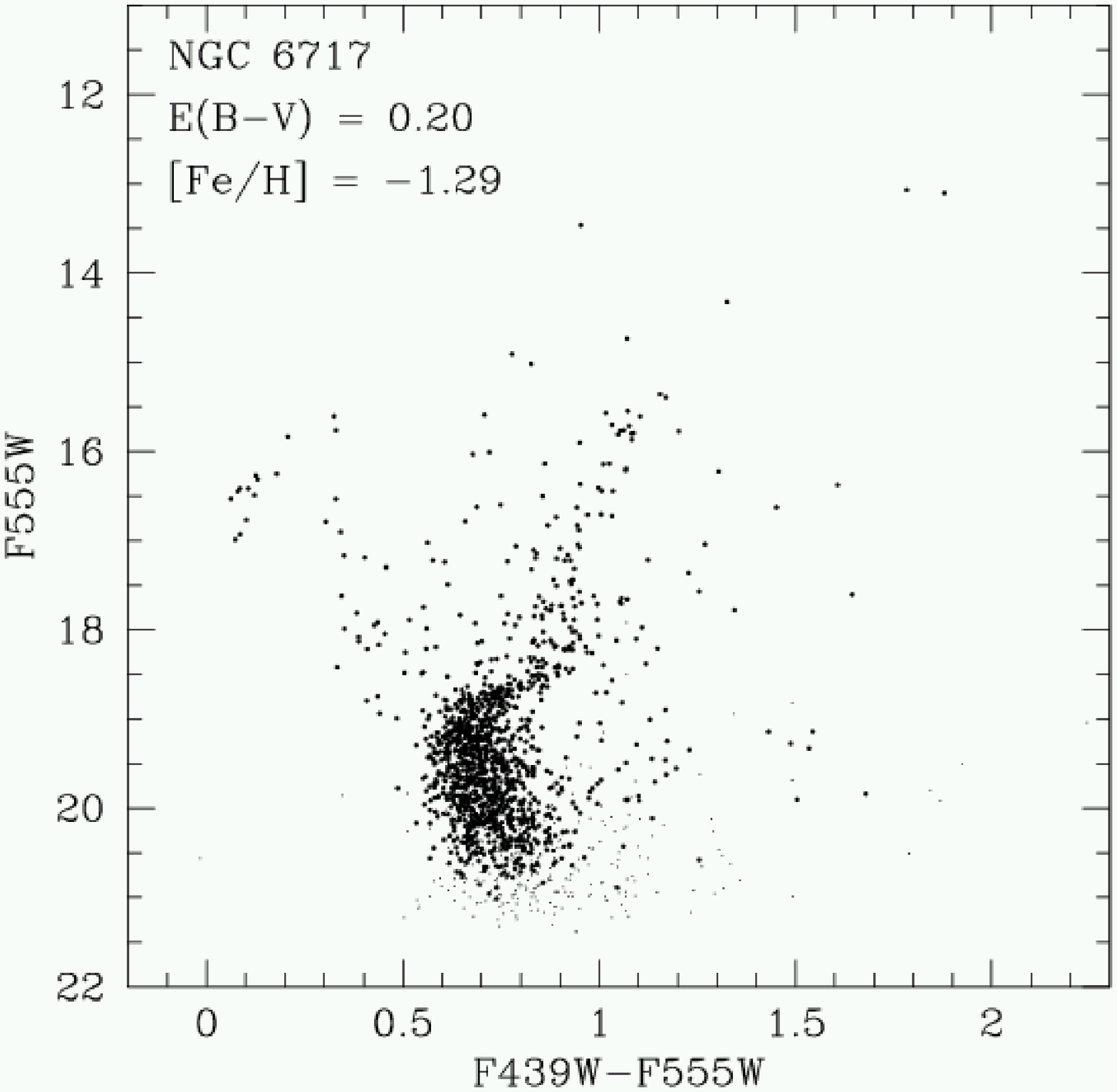}} &
\resizebox*{0.9\columnwidth}{0.36\height}{\includegraphics{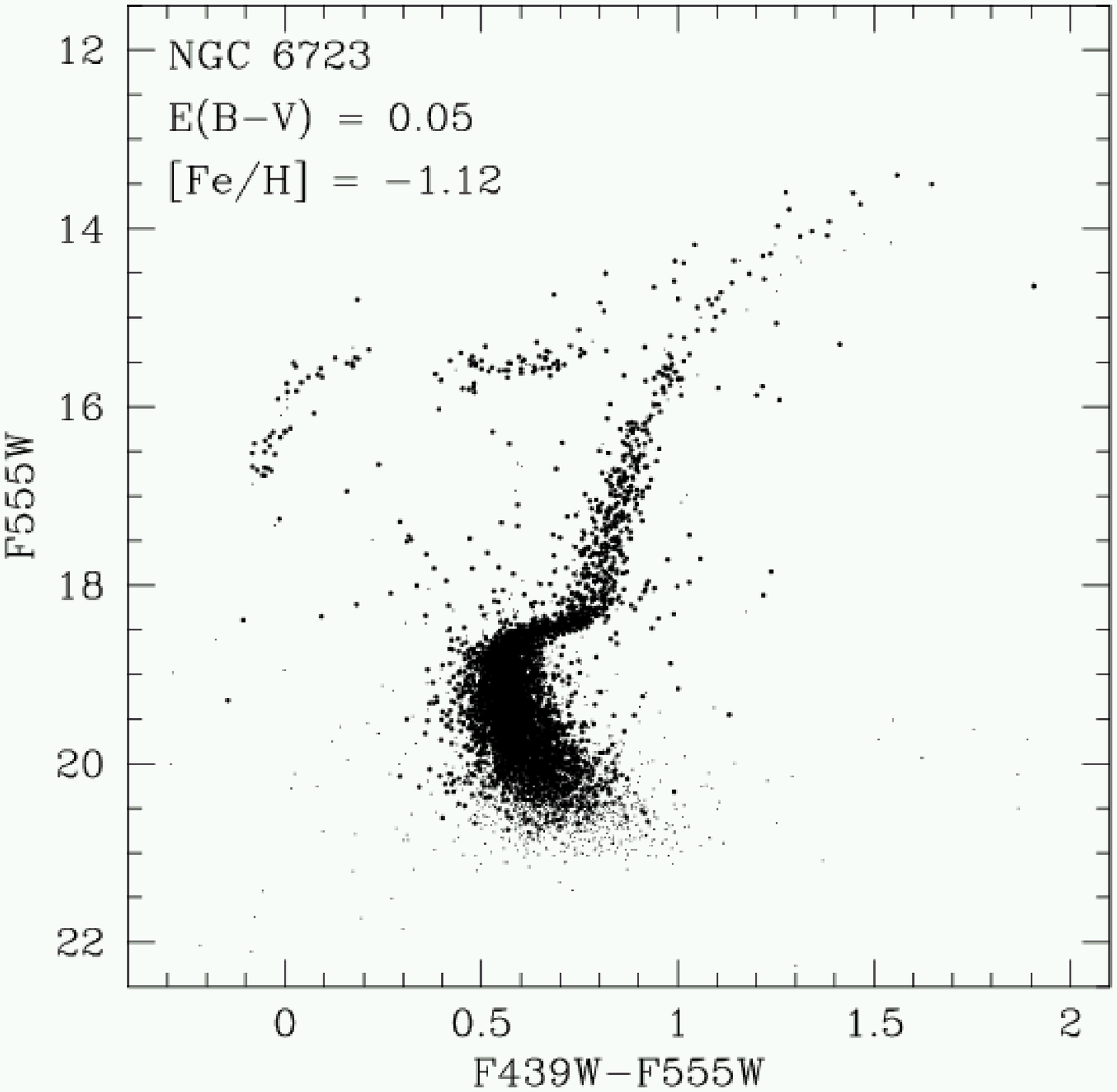}} \\
\end{tabular}
\caption{The color magnitude diagrams (cont.).}
\end{figure*}

\begin{figure*}
\begin{tabular}{cc}
\resizebox*{0.9\columnwidth}{0.36\height}{\includegraphics{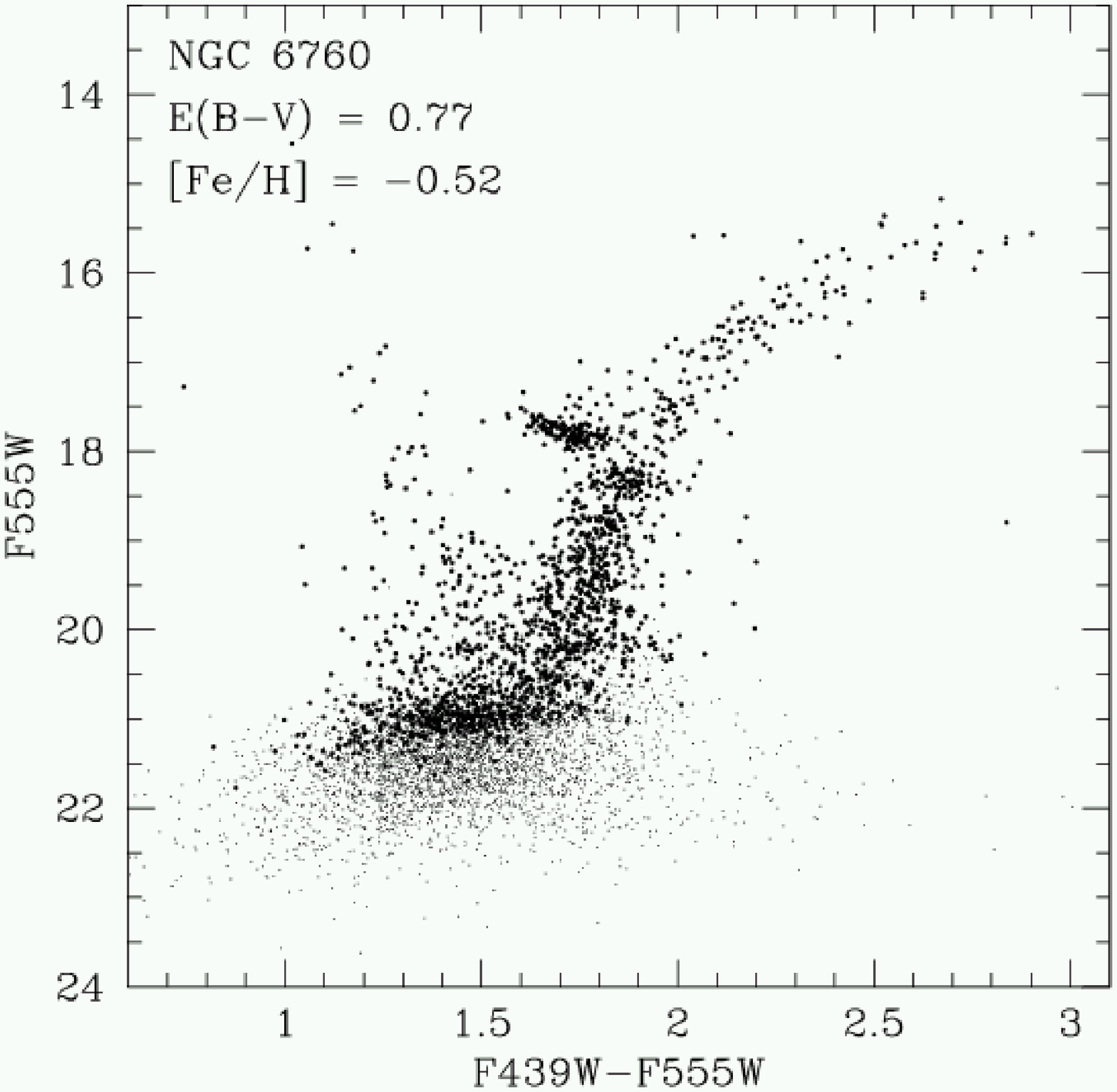}} &
\resizebox*{0.9\columnwidth}{0.36\height}{\includegraphics{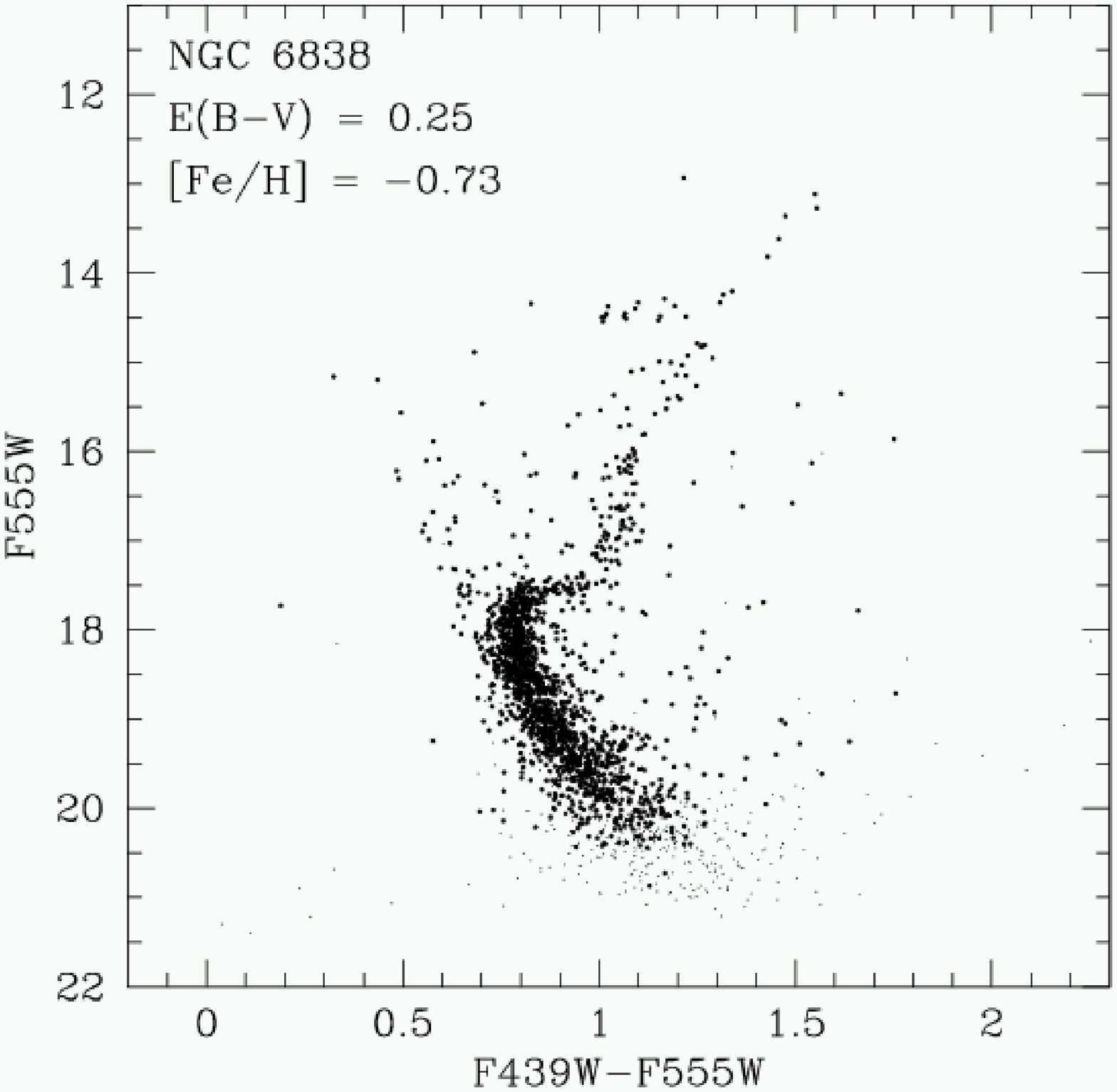}} \\
\resizebox*{0.9\columnwidth}{0.36\height}{\includegraphics{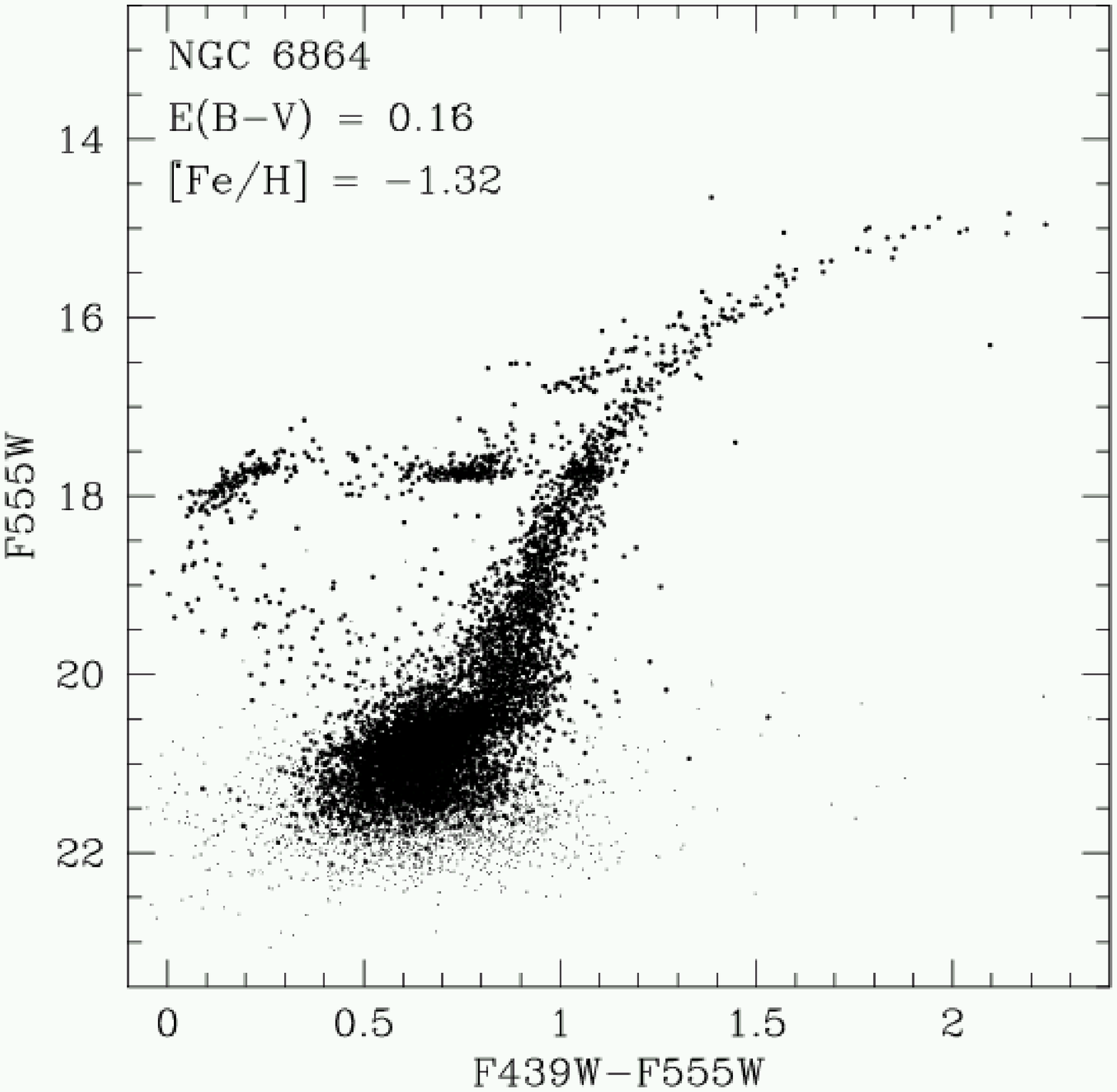}} &
\resizebox*{0.9\columnwidth}{0.36\height}{\includegraphics{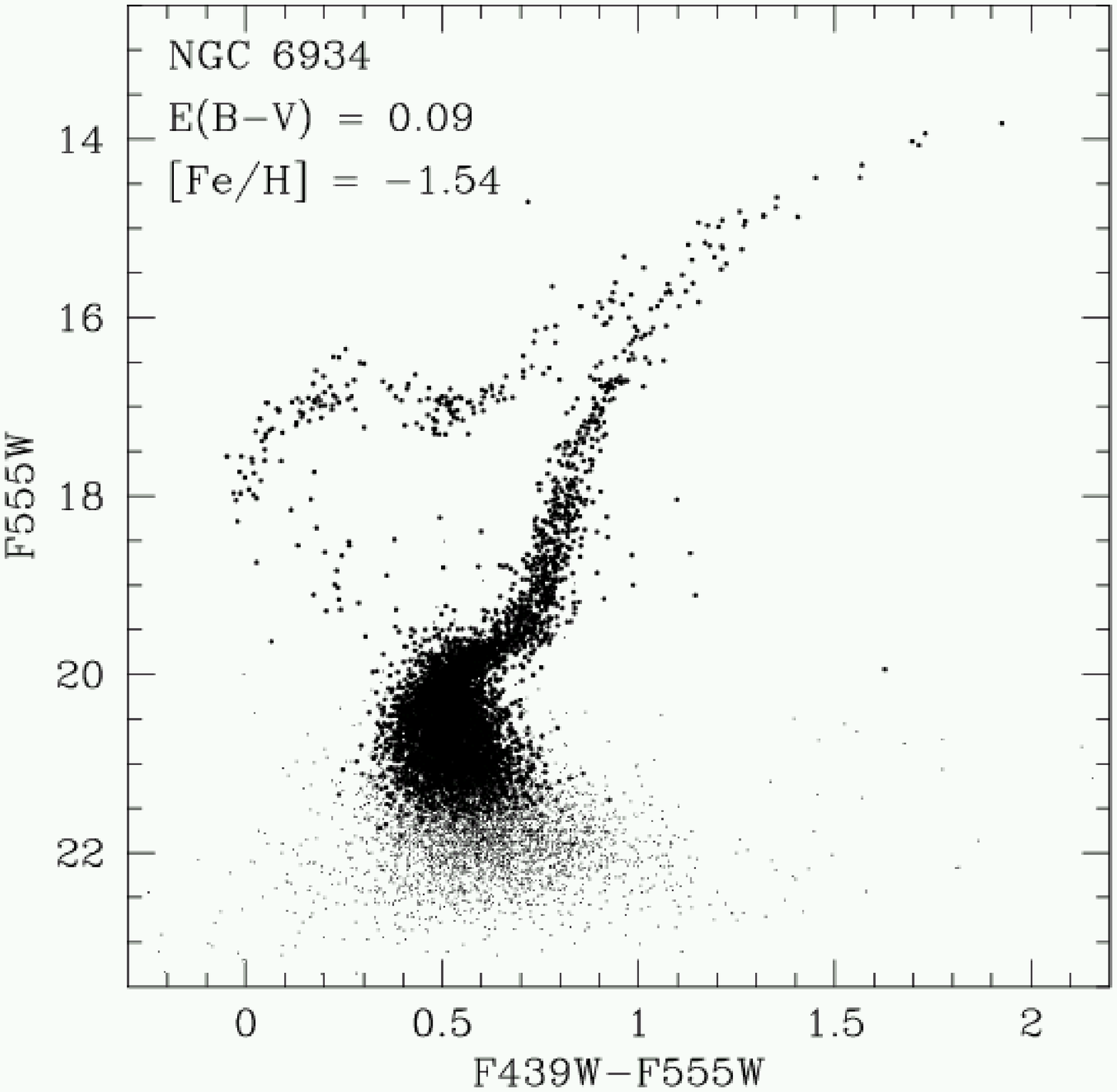}} \\
\resizebox*{0.9\columnwidth}{0.36\height}{\includegraphics{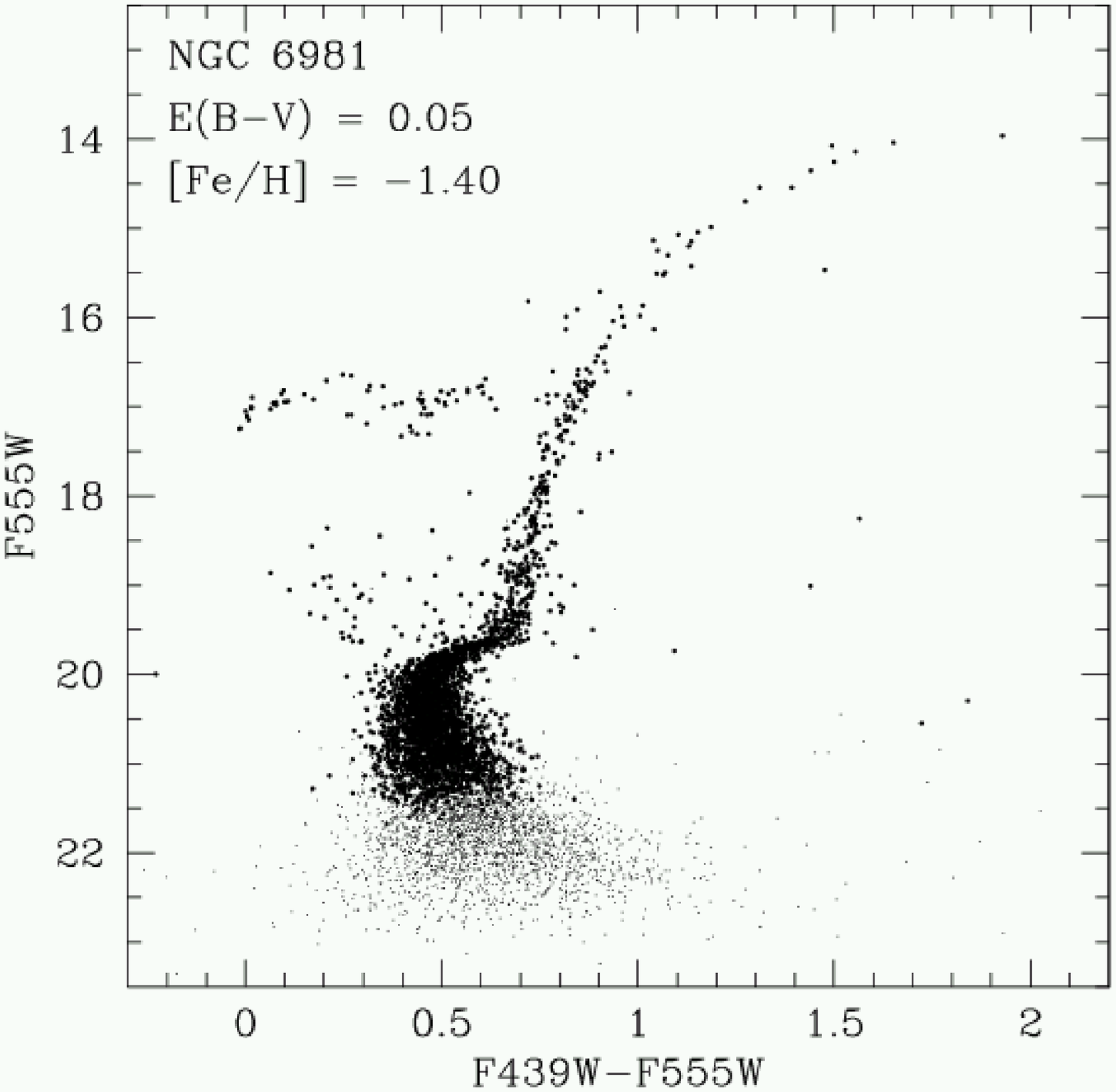}} &
\resizebox*{0.9\columnwidth}{0.36\height}{\includegraphics{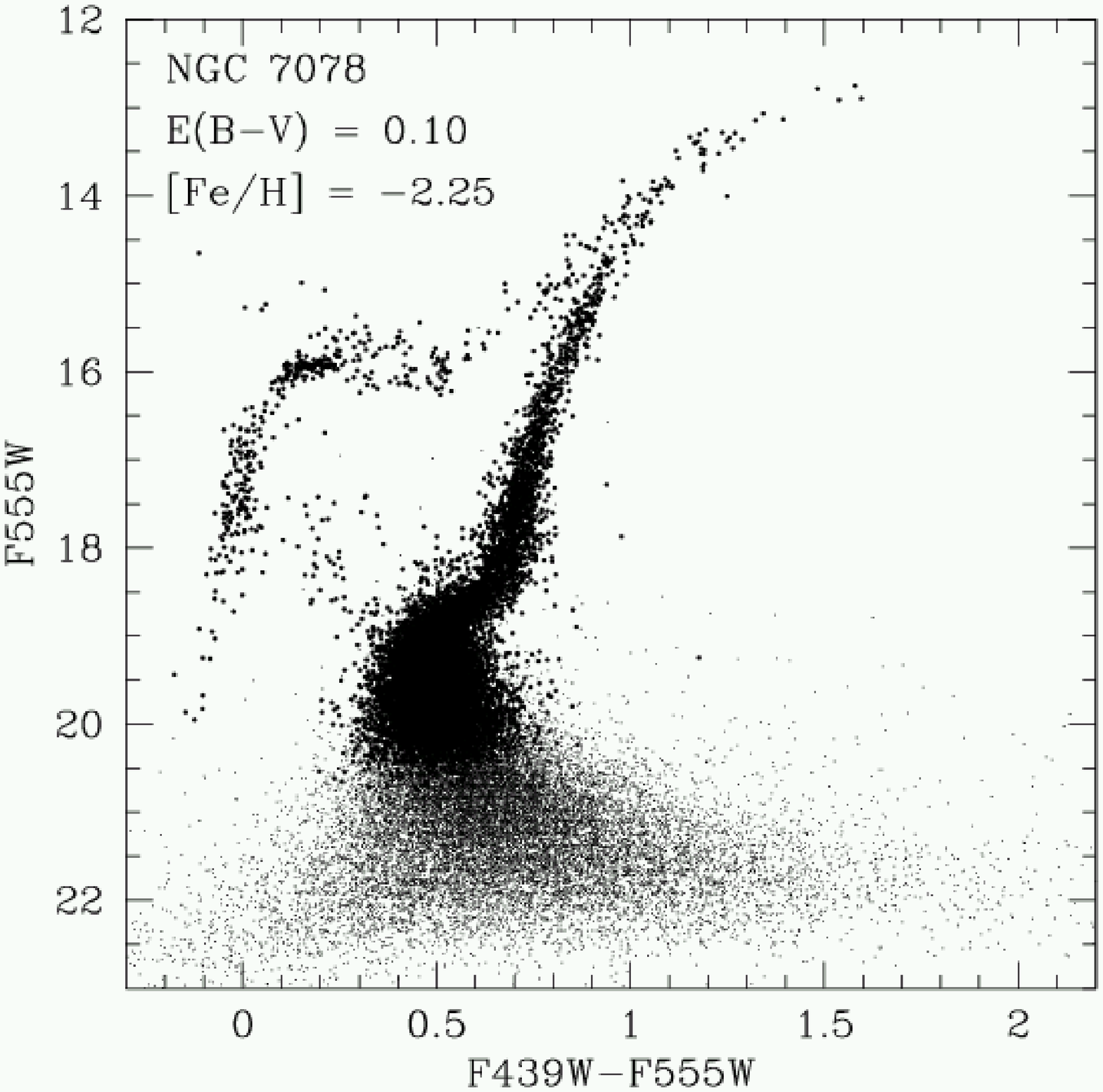}} \\
\end{tabular}
\caption{The color magnitude diagrams (cont.).}
\end{figure*}

\begin{figure*}
\begin{tabular}{cc}
\resizebox*{0.9\columnwidth}{0.36\height}{\includegraphics{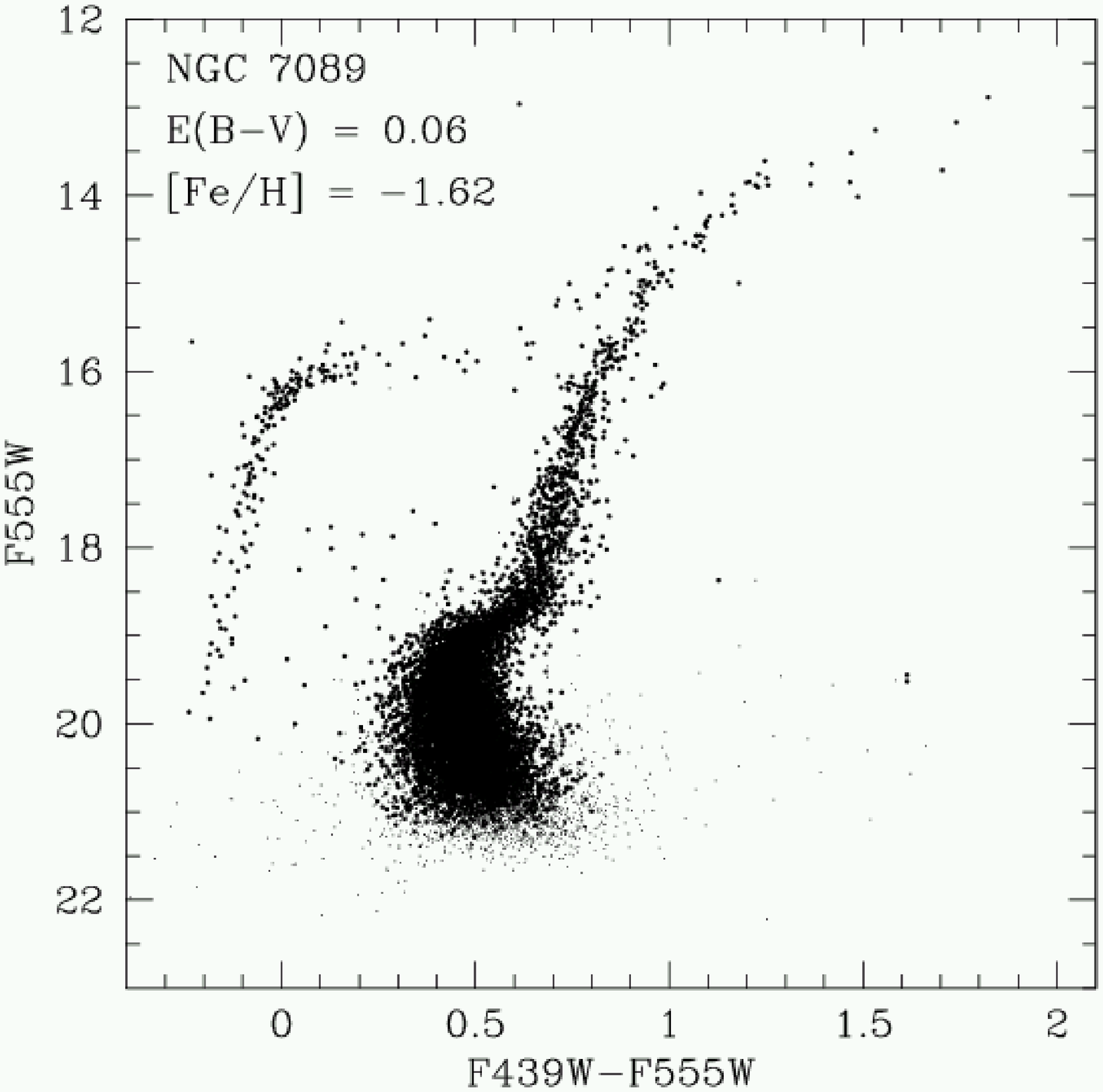}} &
\resizebox*{0.9\columnwidth}{0.36\height}{\includegraphics{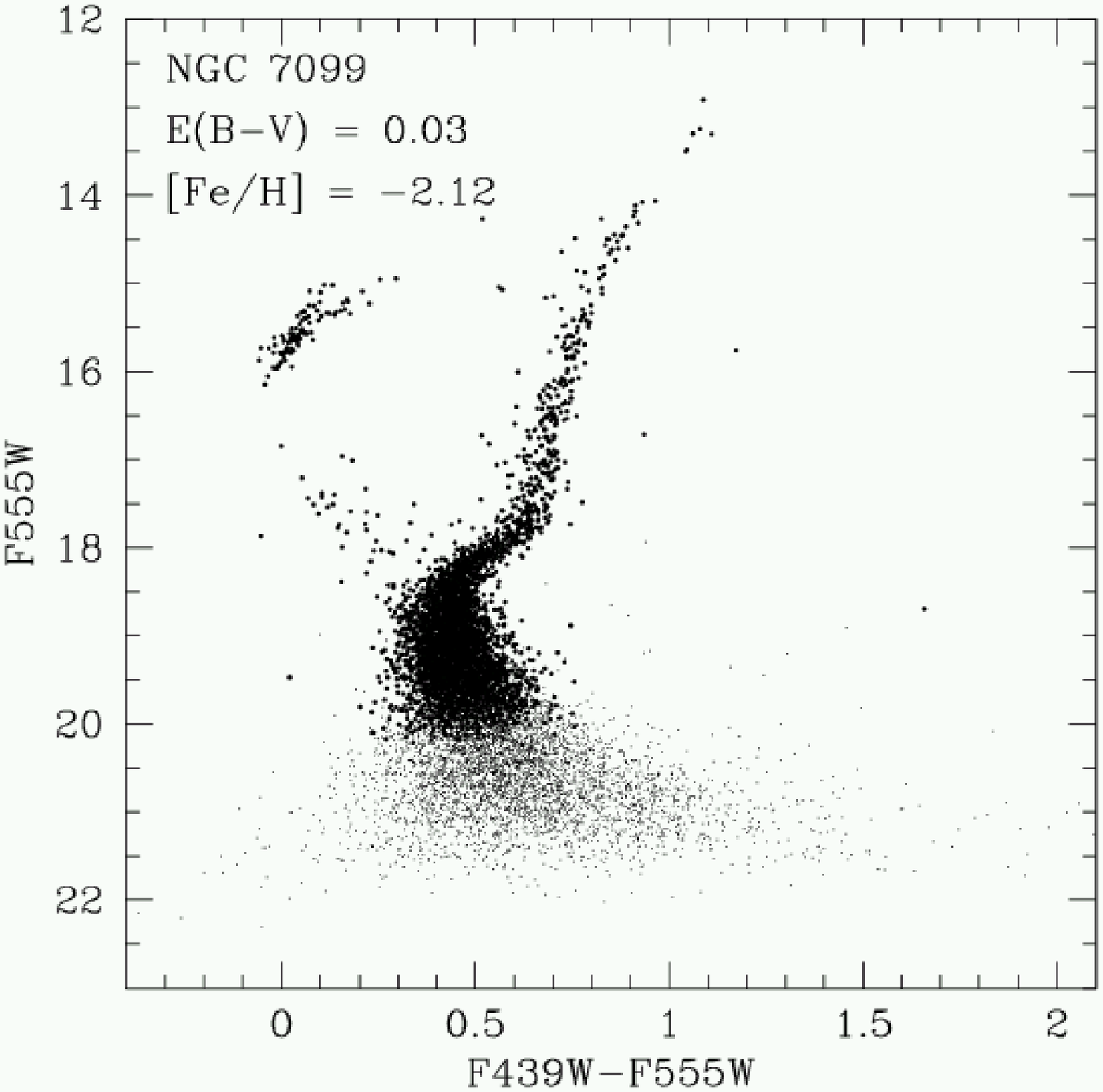}} \\
\end{tabular}
\caption{The color magnitude diagrams (cont.).}
\end{figure*}

\section{The Database}

Figures 2--13 show the final CMDs for the 74 GGCs of our database in
the F439W and F555W HST flight system. Note that the magnitude and
color ranges covered by each figure are always of the same size
(though magnitude and color intervals are different), with the exception
only of NGC~6397.  Typically from a few thousands to about 47,000
(e.g. in NGC~6388) stars have been measured in each cluster.

Table 5 shows an example of the final photometric file.
% K: No need to repeat this in the text.
%: Col. 1 gives
%an identification number, Col. 2 and 3 the position on the chip, Col.
%4 and 5 the \( V \) and \( B \) standard magnitudes (reddening corrected), Col. 6
%and 7 the F555W and F439W magnitudes in the HST flight system
%(reddening corrected), Col. 8 and 9 the photometric errors given by
%\noun{allframe}, Col. 10 and 11 the \( V \) and \( B \) standard magnitudes
%(before the reddening correction), Col. 12 and 13 the F555W and F439W
%flight magnitudes (before the reddening correction), Col. 14 and 15
%the $\chi$ and sharp parameters as given by \noun{allframe}, and
%Col. 16 the chip number (1 for PC, and 2, 3, 4 for WF2, WF3, WF4
%respectively).

After the publication of the present paper, all the CMDs, the star
positions, the magnitudes in both the F439W and F555W flight system
and the \( B \) and \( V \) standard Johnson system can be retrieved from the
Padova Globular Cluster Group Web pages at 
http://dipastro.pd.astro.it/globulars
We will also make available the star-count completeness results of the
single clusters, upon request to the first author. Anyone who uses the
photometric data retrieved from the Web pages and the completeness
corrections is kindly asked to cite the present paper.

%In the file with the photometry we have included the magnitudes in 4
%photometric systems:\ the magnitudes in the HST flight system (F439W,
%F555W) and Johnson \( B \) and \( V \) system resulting from the
%iterative procedure with the Harris (1996) reddening subtraction
%described in Section 2.4, and the magnitudes in the HST flight system
%(F439W$_{\rm nr}$, F555W$_{\rm nr}$) and Johnson \( B$_{\rm nr}$ \)
%and \( V$_{\rm nr}$ \) before the reddening correction. We would
%advise the reader to use the F439W$_{\rm nr}$ F555W$_{\rm nr}$
%(Cols.\ 12 and 13) magnitudes. If one needs to work in the Johnson
%system, for highly reddened clusters, the only magnitudes which can be
%safely used are the \( B \) and \( V \) magnitudes resulting from the
%iterative calibration procedure that takes into account the reddening
%effect (Cols.\ 4 and 5), though we cannot guarantee that the adopted
%reddening is correct.

% IRK revision: (GP -- Please read carefully and see if I have correctly
% conveyed your meaning.)
% (Note: from this I saw that 4 of the headings in Table 5 needed to be
% fixed.  I did so.)
In the file that gives the photometry we have included two sets of
magnitudes in each of four photometric bands.  We give magnitudes in the
HST flight system (F439W, F555W), and the Johnson \( B \) and \( V \)
magnitudes that result from the iterative procedure that uses the Harris
(1996) reddening subtraction, as described in Section 2.4.  We also give
magnitudes in both systems before the reddening correction: 
(F439W$_{\rm nr}$, F555W$_{\rm nr}$) and ($B_{\rm nr}$, $V_{\rm
nr}$).  In general we would advise the reader to use the F439W$_{\rm
nr}$ and F555W$_{\rm nr}$ magnitudes (Cols.\ 12 and 13). If one needs to
work in the Johnson system for highly reddened clusters, however, the
only magnitudes which can be safely used are the \( B \) and \( V \)
magnitudes resulting from the iterative calibration procedure (Cols.\ 4
and 5), though we cannot guarantee that the adopted reddening is
correct.

\begin{acknowledgements}
The present work has been supported by the Italian Ministero della
Universit\`a e della Ricerca under the program {\it Stellar Dynamics
and Stellar Evolution in Globular Clusters} and by the Agenzia
Spaziale Italiana. ARB wishes to thank the Istituto Nazionale di
Astrofisica for support.  IRK, SGD, CS, and RMR acknowledge support
from the STScI grants GO-6095, GO-7470, GO-8118, and GO-8723.
% K: Be sure to get something from SGD, RMR, and GM.
% GP Done.
\end{acknowledgements}

%\twocolumn

\end{document}